\documentclass[letterpaper,12pt,final,twoside]{thesis-umd}
\usepackage[colorlinks, letterpaper, pdfpagelayout=OneColumn, pdfpagemode=UseOutlines, bookmarksnumbered, pdfstartview=FitH, linkcolor=black, anchorcolor=black, citecolor=black, filecolor=black, menucolor=black, pagecolor=black, urlcolor=black]{hyperref}
\usepackage[tbtags]{amsmath}
\usepackage{amsfonts}
\usepackage{amssymb}

\def\a{\alpha}
\def\b{\beta}
\def\d{\delta}
\def\e{\varepsilon}
\def\f{\phi}
\def\g{\gamma}
\def\h{\eta}

\def\k{\kappa}
\def\l{\lambda}
\def\m{\mu}
\def\n{\nu}
\def\o{\omega}

\def\q{\theta}
\def\r{\rho}
\def\s{\sigma}
\def\t{\tau}

\def\F{\Phi}
\def\G{\Gamma}

\def\S{\Sigma}

\def\IR{\relax{I\kern-.35em R}}
\allowdisplaybreaks[1]

\begin{document}


\renewcommand{\baselinestretch}{2}
\small\normalsize


\pagestyle{empty}


\hfill\href{http://www.arxiv.org/abs/hep-th/0507104}{hep-th/0507104}\\
\vspace{1in}

\renewcommand{\baselinestretch}{1}
\small\normalsize


\begin{center}
\large{{\bf ABSTRACT}}
\end{center}

\vspace{3em}

\hspace{-.5in}\begin{tabular}{ll}
Title of dissertation:      & {FLUX COMPACTIFICATION OF M-THEORY}\\
                            & {ON COMPACT MANIFOLDS WITH SPIN(7)}\\
                            & {HOLONOMY}\\
\ \\
                            & {Drago\c{s} Eugeniu Constantin,}\\
                            & {Doctor of Philosophy, 2005}\\
\ \\
Dissertation directed by:   & {Professor Melanie Becker}\\
                            & {Department of Physics } \\
\end{tabular}


\vspace{3em}

\renewcommand{\baselinestretch}{2}
\small\normalsize


At the leading order, M-theory admits minimal supersymmetric compactifications if the internal manifold has exceptional holonomy. The inclusion of non-vanishing fluxes in M-theory and string theory compactifications induce a superpotential in the lower dimensional theory, which depends on the fluxes. In this work, we check the conjectured form of this superpotential in the case of warped M-theory compactifications on Spin(7) holonomy manifolds. We perform a Kaluza-Klein reduction of the eleven-di\-men\-sio\-nal supersymmetry transformation for the gravitino and we find by direct comparison the superpotential expression. We check the conjecture for the heterotic string compactified on a Calabi-Yau three-fold as well. The conjecture can be checked indirectly by inspecting the scalar potential obtained after the compactification of M-theory on Spin(7) holonomy manifolds with non-vanishing fluxes. The scalar potential can be written in terms of the superpotential and we show that this potential stabilizes all the moduli fields describing deformations of the metric except for the radial modulus.

All the above analyses require the knowledge of the  minimal supergravity action in three dimensions. Therefore we calculate the most general causal ${\cal N}=1$ three-di\-men\-sio\-nal, gauge invariant action coupled to matter in superspace and derive its component form using Ectoplasmic integration theory. We also show that the three-di\-men\-sio\-nal theory which results from the compactification is in agreement with the more general supergravity construction.

The compactification procedure takes into account higher order quantum correction terms in the low energy effective action. We analyze the properties of these terms on a Spin(7) background. We derive a perturbative set of solutions which emerges from a warped compactification on a Spin(7) holonomy manifold with non-vanishing flux for the M-theory field strength and we show that in general the Ricci flatness of the internal manifold is lost, which means that the supergravity vacua are deformed away from the exceptional holonomy. Using the superpotential form we identify the supersymmetric vacua out of this general set of solutions.



\hbox{\ }
\vspace{1in}

\renewcommand{\baselinestretch}{1}
\small\normalsize


\begin{center}

\large{{\bf FLUX COMPACTIFICATION OF M-THEORY ON COMPACT MANIFOLDS WITH SPIN(7) HOLONOMY}}\\
\ \\
\ \\
\large{by} \\
\ \\
\large{Drago\c{s} Eugeniu Constantin}
\ \\
\ \\
\ \\
\ \\
\normalsize
Dissertation submitted to the Faculty of the Graduate School of the \\
University of Maryland, College Park in partial fulfillment \\
of the requirements for the degree of \\
Doctor of Philosophy \\
2005
\end{center}

\vspace{8em}

\noindent Advisory Committee:\vspace{2ex}

Professor Melanie Becker, Chair/Advisor

Professor Andrew Baden

Professor Sylvester James Gates, Jr.

Professor Rabindra Nath Mohapatra

Professor Jonathan Micah Rosenberg



\hbox{\ }
\vspace{3.5in}

\renewcommand{\baselinestretch}{1.5}
\small\normalsize


\begin{center}
\large{\copyright \hbox{ }Copyright by\\
Drago\c{s} Eugeniu Constantin\\
2005}
\end{center}



\pagestyle{plain}
\pagenumbering{roman}
\setcounter{page}{2}


\hbox{\ }
\vspace{3.5in}


\begin{center}
\large{\em For my parents, Dumitra and Gabriel Constantin}
\end{center}



\hbox{\ }
\vspace{1in}

\renewcommand{\baselinestretch}{2}
\small\normalsize


\begin{center}
\large{{\bf ACKNOWLEDGMENTS}} \\[0.5in]
\end{center}

First of all, I would like to thank my advisor, Professor Melanie Becker, for allowing me to be part of her research group. I warmly thank her for her support and advice during my time as a graduate student in the Department of Physics at University of Maryland. I have enjoyed the extraordinary collaboration with Professor Sylvester James Gates, Jr. and I would like to thank him for all his advices and interesting discussions that we had about theoretical physics. I would also like to thank Professor Michael Fisher for his excellent advices and kind support and for opening my appetite for biophysics. I would like to thank Professor Theodore Jacobson for his thoughtful courses.

Many thanks are due to Professor Andrew Baden, Professor Rabindra Nath Mohapatra and Professor Jonathan Rosenberg for agreeing to serve on my thesis committee and for sparing their invaluable time reviewing the manuscript.

Also I would like to thank my collaborators, William Linch, Willie Merrell and Joseph Phillips for crystalizing the ideas about the minimal three-di\-men\-sio\-nal supergravity. I would like to acknowledge useful discussions about various aspects of my thesis with Sergei Gukov, Michael Haack, Dominic Joyce and Axel Krause.

It is a great pleasure to acknowledge all my former professors who guided my first steps of my career: Ileana Cucu, Rozalia Dinu, Maria Magdalena Mihail, Emilian Mihail, Aurora Nichifor, Victor Nichifor, Constantin Vrejoiu, Radu Lungu, Voicu Grecu, Gheorghe Ciobanu, Mihai Visinescu and Irinel Caprini.

My first two years of graduate studies at Carnegie Mellon University will always be warmly remembered for being filled with a friendly and scientific atmosphere. I would like to thank my former professors from the Department of Physics at Carnegie Mellon University, Ling-Fong Li, Anthony Duncan, Gregg Franklin, Colin Morningstar, Michael Widom, Ira Rothstein, Richard Holman and Robert Swendsen for their great lectures and support during that time.

Many, many thanks are due to my brother Tudor Constantin and my best friend Alexandru Popa. The discussions with them were always refreshing and optimistic and helped me regain my tonus.

Sincere thanks are due to my parents for being a real support in the early stages of my career and their unbound love. I am grateful for everything that they have done for me.

Finally but not least, I would like to thank my beloved wife, Magdalena Constantin, for her continuous and unconditional support, for her lucid advices and for her unlimited patience. Her encouragements helped me finalize this thesis. I also want to thank my son, Gabriel Constantin, for being so patient with me in the last stages of writing this thesis.\vspace{3ex}

Thank all of you!



\renewcommand{\baselinestretch}{1}
\small\normalsize

\tableofcontents
\newpage
\hbox{\ }
\newpage


\setcounter{page}{1}
\pagenumbering{arabic}

\renewcommand{\baselinestretch}{2}
\small\normalsize


\chapter[Introduction]{INTRODUCTION}
\label{chapter-introduction}

\setcounter{equation}{0}
\renewcommand{\theequation}{\thechapter.\arabic{equation}}


Low dimensional compactifications of M-theory have previously been discussed in the literature. The amount of supersymmetry obtained in the low-di\-men\-sio\-nal effective theory is directly related to the holonomy of the internal manifold. Compactifications on Riemannian manifolds of exceptional holonomy are of special interest because they allow us to obtain theories with less supersymmetry and in a different number of space-time dimensions. In particular, M-theory compactifications on Spin(7) holonomy manifolds\footnote{For a mathematical introduction into the subject of manifolds with exceptional holonomy the reader can consult the book by Dominic Joyce \cite{Joyce}.} lead to a minimal supersymmetric theory in three dimensions. Early papers which have considered compactification of M-theory on exceptional holonomy backgrounds are \cite{Shatashvili:1994zw} and \cite{Papadopoulos:1995da}.

Recall that there is a close connection between the theory of Riemannian manifolds with reduced holonomy and the theory of calibrated geometry \cite{Harvey:1982xk}. Calibrated geometry is the theory which studies calibrated submanifolds, a special kind of minimal submanifolds of a Riemannian manifold, which are defined using a closed form called the calibration. Riemannian manifolds with reduced holonomy usually come equipped with one or more natural calibrations. Based on this close relation to calibrated geometry and generalizing the result for the superpotential found in \cite{Gukov:1999ya}, Gukov made a conjecture about the form of the superpotential appearing in string theory compactifications with non-vanishing Ramond-Ramond fluxes on a manifold $X$ of reduced holonomy \cite{Gukov:1999gr}
\begin{equation}
\label{01-avii}
W=\sum \int_X (\mathrm{Fluxes}) \wedge (\mathrm{Calibrations}) \,.
\end{equation}
In this formula the sum is over all possible combinations of fluxes and calibrations. This conjecture has been previously checked by computing the scalar potential from a Kaluza-Klein reduction of the action for a certain type of theories. This, in turn, determines the superpotential. We want to emphasize that this procedure is an indirect verification of \eqref{01-avii}. In this thesis we present a direct computation based on the general observation that the gravitino supersymmetry transformation contains a term proportional to $W$. For the Type IIB theory these potentials have been computed in \cite{Taylor:1999ii} and \cite{Giddings:2001yu}. The superpotentials for Type IIA compactifications on Calabi-Yau four-folds were derived in \cite{Gates:2000fj, Haack:2000di, Gukov:2002iq}, while the scalar potential for M-theory on ${\mathrm G}_2$ holonomy manifolds has been computed in \cite{Beasley:2002db}. One of our main goals will be to compute directly the superpotential for the three-dimensional theory obtained from compactification of M-theory on Spin(7) holonomy manifolds. Having the form of $W$ we can then determine the concrete form of the scalar potential which arises in the low energy effective action. This is another important problem addressed in this thesis.

It is well known that for a conventional compactification of the heterotic string on a Calabi-Yau three-fold, i.e. without taking warp factors into account, turning on an expectation value for the heterotic three-form will induce a superpotential, which breaks supersymmetry without generating a cosmological constant \cite{Dine:1985rz}. In the context of Gukov's conjecture \cite{Gukov:1999gr}, it was argued in \cite{Behrndt:2000zh} that this superpotential can be written as in \eqref{01-avii}, generalizing the original proposal \cite{Gukov:1999gr} to fluxes of Neveu-Schwarz type. For an earlier discussion on the superpotential one can consult \cite{Dine:1985rz}. Due to the fact that this is a rather important result, we have included in our thesis the superpotential derivation for the heterotic theory alongside with the derivation for M-theory on Spin(7) manifolds. We shall check the above conjecture for both theories in section \ref{01-gravitino} by computing the superpotential explicitly from a Kaluza-Klein reduction of the gravitino supersymmetry transformation.

Of great importance are the compactifications of M-theory and string theory with non-vanishing expectation values for tensor fields. The previously mentioned procedures play a very special role when trying to find a realistic string theory model that could describe our four-di\-men\-sio\-nal world. Especially interesting are the so called warped compactifications. Such compactifications were first discovered for the heterotic string in \cite{Strominger:1986uh} and \cite{deWit:1986xg} and were later generalized to warped compactifications of M-theory and F-theory in \cite{Becker:1996gj, Sethi:1996es, Dasgupta:1999ss}. In these compactifications tensor fields acquire non-vanishing expectation values, while leaving supersymmetry unbroken. The compactification generates scalar fields in the low-energy effective supergravity theory, the so called moduli fields. The vacuum expectation values of the moduli fields characterize the vacuum. If the compactified theory contains no scalar potential, the moduli fields can take any possible values and the theory loses its predictive power because the vacuum is undetermined. However, it was realized in \cite{Gukov:1999ya, Taylor:1999ii, Giddings:2001yu, Dasgupta:1999ss, Haack:2001jz} and \cite{Berg:2002es} that for string theory and M-theory compactifications with non-vanishing fluxes a scalar potential emerges, which stabilizes many of the moduli fields. More specifically, the restrictions imposed by supersymmetry on the fluxes lead to constraints on the moduli fields of the theory and most of these moduli fields will be stabilized, hence the number of possible vacua is reduced.

In this thesis we would like to consider warped compactifications of M-theory on a smooth and compact Spin(7) holonomy manifold. As we have mentioned before the resulting action has an ${\cal N}=1$ supersymmetry in three dimensions and it is interesting from several reasons. First of all these theories are closely related to four-di\-men\-sio\-nal counterparts with completely broken supersymmetry. This is because they can not be obtained by a dimensional reduction from a supersymmetric four-di\-men\-sio\-nal theory\footnote{The minimal supersymmetric theory in four dimensions compactified on $S^1$ produces a three-di\-men\-sio\-nal theory with ${\cal N}=2$ supersymmetry.}, thus one might understand the mechanism of ${\cal N}=1$ supersymmetry breaking in four dimensions by studying the three-di\-men\-sio\-nal theory with ${\cal N}=1$ supersymmetry. Also, because the string world-sheet is two-di\-men\-sio\-nal one expects to observe interesting phenomena upon compactification of string theory to two dimensions \cite{Gates:2000fj} and for this reason three-di\-men\-sio\-nal compactifications of M-theory are attractive. Another strong reason to pursue a serious analysis of M-theory on such a background is the close relation which exists between manifolds with Spin(7) holonomy and manifolds with ${\mathrm G}_2$ holonomy. We would like to remind the reader that M-theory compactified on manifolds with ${\mathrm G}_2$ holonomy generates a minimal supersymmetric theory in four dimensions which is appealing from a phenomenological point of view.

As well, models with ${\cal N}=1$ supersymmetry in three dimensions are interesting in connection to the solution of the cosmological constant problem along the lines proposed by Witten in \cite{Witten:1994cg} and \cite{Witten:1995rz} and exemplified in the three-di\-men\-sio\-nal case in \cite{Becker:1995sp}. The basic idea of this proposal is that in three dimensions supersymmetry can ensure the vanishing of the cosmological constant, without implying the unwanted Bose-Fermi degeneracy. However, this mechanism does not explain why the cosmological constant of our dimensional world is so small, unless there is a duality between a three-di\-men\-sio\-nal supersymmetric theory and a four-di\-men\-sio\-nal non-supersymmetric theory of the type that we are discussing. So, M-theory compactifications on Spin(7) holonomy manifolds allow us to address the cosmological constant problem from a three-di\-men\-sio\-nal perspective.

In general, due to membrane anomaly \cite{Vafa:1995fj, Duff:1995wd, Witten:1997md} and the global tadpole anomaly \cite{Sethi:1996es}, the compactification of M-theory on eight-di\-men\-sio\-nal manifolds involves the presence of a non-vanishing flux for the field strength \cite{Becker:1996gj}. The supersymmetry imposes restrictions on the form of the field strength flux. In the Spin(7) holonomy case the restrictions imposed to the flux were derived in \cite{Becker:2000jc}. It was later shown in \cite{Acharya:2002vs} and \cite{Becker:2002jj} that these constraints can be derived from certain equations which involve the superpotential
\begin{equation}
\label{03-w-conditions}
W=D_A W=0 \,,
\end{equation}
where $D_A W$ indicates the covariant derivative of $W$ with respect to the moduli fields which correspond to the metric deformations of the Spin(7) holonomy manifold\footnote{In the previously mentioned papers the external space is considered to be Minkowski.}. We want to note that the compactness of the internal manifold was essential in the analysis performed in \cite{Acharya:2002vs} and \cite{Becker:2002jj}. In the present paper we restrict ourselves to manifolds with Spin(7) holonomy which are smooth and compact\footnote{A compact manifold with Spin(7) holonomy is simply connected. For details see \cite{Joyce}.}. However, as stated in \cite{Acharya:2002vs}, the result obtained using \eqref{03-w-conditions} is valid for non-compact manifolds as well but the proof does not involve the above equations. The existence of a Ricci flat metric for such manifolds is not guaranteed as in the Calabi-Yau case, therefore we will tacitly suppose that there are such metrics and we will perform all the derivations under this assumption. Even if we will be concerned only with compact manifolds which have Spin(7) holonomy we would like to mention a few papers, such as \cite{Gukov:2001hf, Cvetic:2001pg}, where non-compact examples of such manifolds have been constructed and analyzed. Also in \cite{Gukov:2002zg} aspects of topological transitions on non-compact manifolds with Spin(7) holonomy and phase transitions have been considered. A more complete list of papers regarding M-theory on singular manifolds with exceptional holonomy can be found in \cite{Acharya:2004qe} which is a recent review of the subject.

In this thesis, we calculate the Kaluza-Klein compactification of M-theory on a Spin(7) holonomy manifold with non-vanishing fluxes. Our calculation is similar to that of \cite{Haack:2001jz}, which has been done in the context of M-theory compactifications on conformally Calabi-Yau four-folds. We will see that the resulting scalar potential leads to the stabilization of all the moduli fields corresponding to deformations of the internal manifold, except the radial modulus. This scalar potential can be expressed in terms of the superpotential which has appeared previously in the literature \cite{Gukov:1999gr, Acharya:2002vs} and \cite{Becker:2002jj}.

This thesis is based on our recent results published in \cite{Becker:2002jj, Becker:2003wb} and \cite{Constantin:2004eg}. However several additions were necessary in order to present the results in a logical fashion. In what follows we present the structure of this thesis.

In chapter \ref{chapter-vacua} we study the possible solutions of the equations of motion of M-theory on a warped geometry with a Spin(7) holonomy internal manifold. We start by introducing in section \ref{03-effective-action} the M-theory action and we define the quartic polynomials which define the quantum correction terms. In section \ref{03-eq-motion} we derive perturbatively the form of the equations of motion and we discuss the Ricci flatness problem of the internal manifold. In section \ref{03-quartic-polynomials} we have collected some of the most important properties of the above mentioned quartic polynomials.

Chapter \ref{chapter-compactification} is devoted to the derivation of the low energy effective action that emerges from M-theory compactification on Spin(7) holonomy manifolds in the presence of non-zero background flux for the field strength. We start in section \ref{02-zero-flux} with a simpler situation with the compacification of the theory without background fluxes. In section \ref{02-non-zero-flux}, we take the fluxes into account and derive the complete form of the bosonic part of the action. In this way we are able to identify the scalar potential which arises in the compactified theory due to the inclusion of fluxes.

Some of the vacua emerged from compactification which were found in chapter \ref{chapter-vacua} are candidates for supersymmetric solutions and they correspond to a minimal supergravity theory in three dimensions. The conditions which lead to a supersymmetric background can be derived by knowing the three-dimensional supergravity theory. Also, the analysis of the properties of these solution requires the knowledge of the above mentioned supergravity. Hence, chapter \ref{chapter-3dsugra} is dedicated to the derivation of the most general off-shell three-di\-men\-sio\-nal ${\cal N}=1$ supergravity action coupled to an arbitrary number of scalars and $U(1)$ gauge fields. In section \ref{02-supergeometry}, we present the algebra of supercovariant derivatives which describes the superspace geometry. We then discuss Ectoplasmic integration, the technique used to calculate the density projector, which is required to integrate over curved supermanifolds. In section \ref{02-super-3forms}, we solve the Bianchi identities for a super three-form subject to the given algebra required for Ectoplasmic integration. In section \ref{02-ectoplasmic}, we detail the use of Ectoplasm to calculate the density projector. In section \ref{02-component-formalism}, we complete the supergravity analysis by first deriving the component fields and then calculating the component action. We end the analysis by giving the supersymmetry transformations for the component fields and putting the component action on shell, i.e., we remove the auxiliary fields by their algebraic equations of motion.

Chapter \ref{chapter-superpotential} contains our main analyses of the topic. Rather than computing first the scalar potential and from there obtaining the superpotential, we compute in section \ref{01-gravitino} the superpotential directly by a Kaluza-Klein compactification of the gravitino supersymmetry transformation. We illustrate this idea in section \ref{01-spin} in the case of M-theory compactified on a Spin(7) holonomy manifold and in section \ref{01-heterotic} we compute the superpotential for the heterotic string compactified on a Calabi-Yau three-fold. In section \ref{02-moduli} we determine the form of the scalar potential generated for the moduli fields by the field strength flux. We conclude that most of the moduli fields are stabilized but the radial modulus remained unconstrained. We also show that the result obtained at the end of chapter \ref{chapter-compactification} is a particular case of the more general construction of chapter \ref{chapter-3dsugra}. Based on \eqref{03-w-conditions} and using the form \eqref{01-avii} for $W$, we derive in section \ref{03-susy-solutions} the conditions imposed on the internal flux by a supersymmetric solution and we investigate the conditions under which the supersymmetry is broken dynamically by the internal flux.

Our concluding remarks are presented in chapter \ref{chapter-conclusions}. We give a summary of our results and comment on the physics implied by the explicit form of the scalar potential. We conclude this section with some open questions and directions for future investigations suggested by our findings.

Finally, details related to our calculations are contained in the appendices. In appendix \ref{app-compact}, we provide the conventions used in this thesis as well as some useful identities and small derivations of results which were used in different sections of this work. In appendix \ref{app-3d}, we provide the conventions used in chapter \ref{chapter-3dsugra} and we provide various derivations and check procedures needed in the previously mentioned chapter. In appendix \ref{app-spin7}, we provide relevant aspects related to manifolds with Spin(7) holonomy which were used in some parts of our computation.


\chapter[M-theory Vacua]{M-THEORY VACUA}
\label{chapter-vacua}

\setcounter{equation}{0}
\renewcommand{\theequation}{\thechapter.\arabic{equation}}


In this chapter we find all the vacua generated by a warped compactification of M-theory on compact eight-di\-men\-sio\-nal manifolds with Spin(7) holonomy in the presence of a non-zero flux for the field strength. We will take into consideration all the known terms in the low energy effective action up to the order $\kappa_{11}^{-2/3}$, where $\kappa_{11}$ is the e\-le\-ven-di\-men\-sio\-nal gravitational coupling constant. Not all the terms in the effective action are known to this order. Therefore we will need a criterium to consistently eliminate the contribution to the equations of motion that comes from these unknown terms. Terms like $F^2 R^3$ are known to appear in the $\kappa_{11}^{-2/3}$ order \cite{Strominger:1997eb} but they are suppressed in the large volume limit \cite{Peeters:2003ys}, which is the most realistic compactification scenario. In this limit the ``radius'' of the internal manifold\footnote{The ``radius'' of the manifold is nothing else but the characteristic length (the average size) of the manifold.} is much bigger than the e\-le\-ven-di\-men\-sio\-nal Planck length and because of this property their ratio  generates a big number. It is natural to consider as the key ingredient for our analysis a perturbative series expansion in terms of the above defined ratio. The most obvious ansatz is to consider the leading order of the internal metric to be proportional to the square of  this dimensionless parameter. If we restrict ourselves to the first few orders of this perturbative expansion we can exclude the contribution that comes from the above mentioned unknown terms. We would like to mention that such a large ``radius'' expansion for the case of a Calabi-Yau manifolds was previously considered in \cite{Witten:1987kg} and \cite{Becker:2001pm}.

This chapter is organized as follows. In section \ref{03-effective-action} we introduce the low energy effective action of M-theory with all the known leading quantum correction terms. Also we carefully define the quartic polynomials in the Riemann tensor, that enter in the definition of the quantum correction terms. In section \ref{03-eq-motion} we analyze perturbatively the equations of motion and we derive conditions that have to be satisfied by the internal background flux in order to have a valid solution. Also at the end of section \ref{03-eq-motion} we argue that in general the internal manifold looses its Ricci flatness once the quantum correction terms are taken into account. However we show that the manifold remains Ricci flat if a certain condition is satisfied by the warp factors. This relation is rather important because it shows under what conditions we obtain a supersymmetric solution after compactification. In section \ref{03-quartic-polynomials} we discuss some of the properties of the quartic polynomials $E_8$, $J_0$ and $X_8$. These properties are used throughout our analysis presented in section \ref{03-eq-motion} and we have considered it is useful to have them listed in a separate section. In particular in sub-section \ref{03-j0-computation} we prove that $J_0$ vanishes on a Spin(7) background and we also derive a compact expression for the first variation of $J_0$ with respect to the internal metric. At the end of sub-section \ref{03-j0-computation} we compute an elegant formula for the trace of the first variation of $J_0$.


\section{The Low Energy Effective Action}
\label{03-effective-action}

\setcounter{equation}{0}
\renewcommand{\theequation}{\thesection.\arabic{equation}}


For completeness we introduce in this section the bosonic truncation of the e\-le\-ven-di\-men\-sio\-nal supergravity action along with its known correction terms. The effective action for M-theory has the following structure
\begin{equation}
\label{03-start-point}
S=S_0+S_1 + \ldots \,.
\end{equation}
In the above expression $S_0$ represents the bosonic truncation of e\-le\-ven-di\-men\-sio\-nal supergravity \cite{Cremmer:1978km} and $S_1$ represents the leading quantum corrections term. $S_0$ is of order $\kappa_{11}^{-2}$, $S_1$ is of order $\kappa_{11}^{-2/3}$ and the ellipsis denotes higher order terms in $\kappa_{11}$. The explicit expressions of $S_0$ and $S_1$ are
\begin{subequations}
\begin{align}
S_0 &= \frac{1}{2\kappa_{11}^2}\, \int_{M_{11}} d^{11}x\sqrt{-g_{11}}R - \frac{1}{4\kappa_{11}^2} \int_{M_{11}} \left(F \wedge \star F + \tfrac{1}{3} \,C \wedge F \wedge F \right)\,, \label{03-zero} \\
S_1 & = -T_2 \int_{M_{11}} {C \wedge X_8} + b_1 T_2 \int_{M_{11}} {d^{11}x \sqrt{-g_{11}} \left( J_0 - \tfrac{1}{2} \, E_8 \right)} + \ldots \,, \label{03-pppS_1}
\end{align}
\end{subequations}
where $g_{11}$ is the determinant of the e\-le\-ven-di\-men\-sio\-nal metric of $M_{11}$, $F=dC$ is the four-form field strength of the three-form potential $C$ and $b_1$ is a constant
\begin{equation}
b_1 = \frac{1}{(2 \pi)^4 3^2 2^{13}} \,.
\end{equation}
$T_2$ is the membrane tension and it is related to the e\-le\-ven-di\-men\-sio\-nal gravitational coupling constant by
\begin{equation}
T_2 = \left( \frac{2 \pi^2}{\kappa_{11}^2} \right)^{1/3} \,.
\end{equation}
$X_8$ is a differential form of order eight whose components are quartic polynomials in the e\-le\-ven-di\-men\-sio\-nal Riemann tensor
\begin{equation}
\label{03-x8}
X_8(M_{11}) = \frac{1}{192 \,(2 \pi)^4} \Big[ {\rm Tr} \, {\cal R}^4 - \tfrac{1}{4} \, ({\rm Tr}\,  {\cal R}^2)^2 \Big] \,,
\end{equation}
where ${\cal R}_{ij} = \frac{1}{2} R_{ijkl} \, e^k \wedge e^l$ is the curvature two-form written in some orthonormal frame $e^i$. Furthermore, $E_8$ and $J_0$ are also quartic polynomials of the e\-le\-ven-di\-men\-sio\-nal Riemann tensor and have the following expressions \cite{Tseytlin:2000sf}
\begin{align}
E_8(M_{11}) &= - \, \tfrac{1}{3!} \, \delta^{ABCM_1 N_1 \ldots M_4 N_4}_{ABCM_1' N_1' \ldots M_4' N_4'} \, {R^{M_1' N_1'}}_{M_1 N_1} \ldots {R^{M_4' N_4'}}_{M_4 N_4} \,, \label{03-pppE_8} \\
\begin{split}
J_0(M_{11}) &= t^{M_1 N_1 \ldots M_4 N_4} \, t_{M_1' N_1' \ldots M_4' N_4'} \, {R^{M_1' N_1'}}_{M_1 N_1} \ldots {R^{M_4' N_4'}}_{M_4 N_4}\\
& \quad + \tfrac{1}{4} \,E_8(M_{11})\,.
\end{split}\label{03-pppJ_0}
\end{align}
The tensor $t$ is defined by its contraction with some antisymmetric tensor $A$
\begin{equation}
t^{M_1 \ldots M_8} A_{M_1 M_2} \ldots A_{M_7 M_8}
  = 24 {\rm Tr} A^4 - 6 ({\rm Tr} A^2)^2 \,.
\end{equation}
More details regarding the properties of polynomials $E_8$, $J_0$ and $X_8$ can be found in section \ref{03-quartic-polynomials}.


\section{The Equations of Motion}
\label{03-eq-motion}

\setcounter{equation}{0}
\renewcommand{\theequation}{\thesection.\arabic{equation}}


In this section we perform a perturbative analysis of the equations of motion and we derive the conditions that the internal flux has to satisfy in order to have a valid solution. We conclude this section with a discussion about the way the internal manifold gets deformed under the influence of higher order correction terms.

The equation of motion which follows from the variation of action \eqref{03-start-point} with respect to the metric is
\begin{equation}
\label{03-eom-g}
\begin{split}
R_{MN}(M_{11}) &- \frac{1}{2} g_{MN} R(M_{11}) - \frac{1}{12} T_{MN}\\
&= - \beta \frac{1}{\sqrt{-g}} \frac{\delta}{\delta g^{MN}} \left[ \sqrt{-g} (J_0- \frac{1}{2} E_8) \right] \,,
\end{split}
\end{equation}
where $T_{MN}$ is the energy momentum tensor of $F$ given by
\begin{equation}
\label{03-en-mom-tensor}
T_{MN}=F_{MPQR} \, {F_N}^{PQR}- \tfrac{1}{8} \, g_{MN} \, F_{PQRS} \, F^{PQRS} \,,
\end{equation}
and we have set $\beta=2 \kappa_{11}^2 b_1 T_2$. We have listed in appendix \ref{03-scaling} the expressions for the external and internal e\-ner\-gy-mo\-men\-tum tensor. Also in the above mentioned appendix we provide the results obtained for the external and internal components of the term in right hand side of the Einstein equation \eqref{03-eom-g}.

Without sources the field strength obeys the Bianchi identity
\begin{equation}
dF=0 \,,
\end{equation}
and the equation of motion
\begin{equation}
\label{03-eom-F}
d*F= \frac{1}{2} F \wedge F+ \frac{\beta}{b_1} X_8 \,.
\end{equation}
The metric ansatz is a warped product
\begin{equation}
\label{03-full-metric}
\begin{split}
ds^2 &= \widetilde{g}_{MN} \, dX^M \, dX^N = \widetilde{\eta}_{\mu \nu }(x,y) \, dx^{\mu} dx^{\nu} + \widetilde{g}_{mn}(y) \, dy^m dy^n \\
&= e^{2A(y)} \, \eta_{\mu \nu } (x) \, dx^{\mu} dx^{\nu} + e^{2B(y)} \, g_{mn}(y) \, dy^m dy^n \,,
\end{split}
\end{equation}
where $\eta_{\mu\nu}$ describes a three-di\-men\-sio\-nal external space $M_3$ and $g_{mn}$ is a Spin(7) holonomy metric of a compact manifold $M_8$. As usual the big Latin indices $M$, $N$ take values between $0$ and $10$, the Greek indices $\mu$, $\nu$ take values between $0$ and $2$ and small Latin indices $m$, $n$ take values between $3$ and $10$. Also, $X^M$ refers to the coordinates on the whole e\-le\-ven-di\-men\-sio\-nal manifold $M_{11}$, $x^\mu$ are the coordinates on $M_3$ and $y^m$ are the coordinates on $M_8$. We want to note that $M_{11}$ is the direct product between $M_3$ and $M_8$ only in the leading order approximation.

In what follows we introduce a dimensionless parameter ``$t$'' defined as the square of the ratio between $l_8$, the characteristic size of the internal manifold $M_8$, and $l_{11}$ which denotes the e\-le\-ven-di\-men\-sio\-nal Planck length
\begin{equation}
\label{03-t-def}
t=\left( \frac{l_8}{l_{11}} \right)^2 \gg 1 \,,
\end{equation}
where $l_8$ is given by
\begin{equation}
(l_8)^8 = \int_{M_8} d^8y \, \sqrt{g} \,,
\end{equation}
and we have considered the large volume limit for $M_8$, which means that $l_8 \gg l_{11}$. We will suppose that all the fields of the theory have a series expansion in ``$t$'' and we will analyze the equations of motion order by order in the ``$t$'' perturbative expansion. The main ansatz is to consider that the metric of the internal compact space $M_8$ has the following series expansion in ``$t$''
\begin{equation}
\label{03-metric-exp}
g_{mn}=t \, [g^{(1)}]_{mn} + [g^{(0)}]_{mn} + \ldots \,.
\end{equation}
Thus the inverse metric is
\begin{equation}
g^{mn} = t^{-1} [g^{(1)}]^{mn} + t^{-2} [g^{(2)}]^{mn} + \ldots \,,
\end{equation}
where the above expansion coefficients are derived in appendix \ref{03-scaling}. It is obvious now that the Riemann tensor, the Ricci tensor and the Ricci scalar of the internal manifold $M_8$ will have a series expansion in ``$t$'' of the form
\begin{subequations}
\begin{align}
&{R^a}_{mbn}(M_8) = {[R^{(0)}]^a}_{mbn} + t^{-1} {[R^{(1)}]^a}_{mbn} + \ldots \,, \\
&R_{mn}(M_8) = [R^{(0)}]_{mn} + t^{-1} [R^{(1)}]_{mn} + \ldots\,, \\
&R(M_8) = t^{-1} R^{(1)} +t^{-2} R^{(2)} + \ldots \,,
\end{align}
\end{subequations}
where the coefficients in the above expansions can be determined in terms of the expansion coefficients of $g_{mn}$ and $g^{mn}$ and their derivatives. It is not so obvious at this stage of computation that the right ansatz for the warp factors is
\begin{equation}
\label{03-warp-exp}
A=t^{-3} \, A^{(3)} + \ldots \quad \Rightarrow \quad e^{A}=1+t^{-3} \, A^{(3)}+\ldots \,,
\end{equation}
and similarly for $B$. The motivation for this ansatz comes from the fact that the internal Einstein equation receives contributions from the quantum correction terms in the $t^{-3}$ order of perturbation theory. It is natural to suppose that the effect of warping appears at the same order in the equations of motion to compensate for this extra contribution.

The Poincar\'{e} invariance restricts the form of the background flux $F$ to the following structure
\begin{equation}
\label{03-flux-form}
F = F_1 + F_2 \,,
\end{equation}
where $F_1$ is the external part of the flux
\begin{equation}
F_1 = \tfrac{1}{3!} \, \varepsilon_{\mu \nu \rho} \, [\nabla_m f(y)] \, dx^\mu \wedge dx^\nu \wedge dx^\rho \wedge dy^m \,,
\end{equation}
and $F_2$ is the internal background flux
\begin{equation}
F_2 = \tfrac{1}{4!} \, F_{mnpq}(y) \, dy^m \wedge dy^n \wedge dy^p \wedge dy^q \,.
\end{equation}
We also expand the coefficients $f$ and $F_{mnpq}$ in a power series of $t$
\begin{equation}
f=f^{(0)}+t^{-1}f^{(1)}+\ldots \,,
\end{equation}
and
\begin{equation}
F_{mnpq}=F_{mnpq}^{(0)}+t^{-1}F_{mnpq}^{(1)}+\ldots \,.
\end{equation}

Taking into account that the three-di\-men\-sio\-nal external space described by $\eta_{\mu \nu}$ is not at all influenced by the size of the eight-di\-men\-sio\-nal manifold $M_8$ described by $g_{mn}$, all the quantities that emerge from the metric $\eta_{\mu \nu}$ are of order zero in an expansion in ``$t$'', in other words all these quantities are independent of the scale ``$t$''. The external manifold suffers no change due to the deformations of the internal manifold and $\eta_{\mu \nu}$ generates the same equations of motion as in the absence of fluxes and without the quantum correction terms. The zeroth order of the external component of equation \eqref{03-eom-g} reads
\begin{equation}
R_{\mu\nu}(M_3) - \tfrac{1}{2} \, \eta_{\mu\nu} R(M_3) = 0 \,,
\end{equation}
therefore
\begin{subequations}
\begin{align}
&R_{\mu\nu}(M_3) = 0 \,, \\
&R(M_3) = 0 \,,
\end{align}
\end{subequations}
which means that the external space is Minkowski\footnote{In three dimensions the Riemann tensor is proportional to the Ricci tensor.}. However our result does not eliminate the possibility for an $AdS_3$ background in the case when membranes are included in the analysis (see e.g. \cite{Martelli:2003ki}).

A careful analysis of the internal and external Einstein equations to orders no higher than $t^{-2}$ and $t^{-3}$, respectively, reveals that the internal manifold remains Ricci flat to the $t^{-2}$ order in perturbation theory
\begin{equation}
\label{03-Ricci-t2}
R^{(0)}_{mn} = R^{(1)}_{mn}=R^{(2)}_{mn}=0 \,,
\end{equation}
and the Ricci scalar vanishes to the $t^{-3}$ order
\begin{equation}
R^{(1)} = R^{(2)} = R^{(3)} = 0 \,.
\end{equation}
These results are natural because we expect to observe deformations of the internal manifold starting at the $t^{-3}$ order since the quantum correction terms are of this order of magnitude in an expansion in ``$t$'' and in addition the warp factors were chosen to be of the same order of magnitude. As a matter of fact, to order $t^{-2}$ even the warping has no effect and the e\-le\-ven-di\-men\-sio\-nal manifold is a direct product between $M_3$ and $M_8$.

We can also derive from the equation of motion \eqref{03-eom-F} that the covariant derivative of the external flux vanishes to order $t^{-2}$
\begin{equation}
\nabla_m f^{(0)} = \nabla_m f^{(1)} = \nabla_m f^{(2)} = 0 \,.
\end{equation}
Collecting these facts we are left with the following field decomposition for $\nabla_m f$, $R_{mn}(M_8)$ and $R(M_8)$
\begin{subequations}
\begin{align}
&\nabla_m f = t^{-3} \nabla_m f^{(3)} + t^{-4} \nabla_m f^{(4)} + \ldots \,, \\
&R_{mn}(M_8) = t^{-3} \, R^{(3)}_{mn} + t^{-4} \, R^{(4)}_{mn} + \ldots \,, \\
&R(M_8) = t^{-4} R^{(4)} + \ldots \,.
\end{align}
\end{subequations}

To order $t^{-4}$ the external component of the equation of motion \eqref{03-eom-g} has the following form
\begin{equation}
\label{03-external-t3}
R^{(4)} - 4 \triangle^{(1)} A^{(3)} - 14 \triangle^{(1)} B^{(3)} - \tfrac{1}{48} \left[F_2^{(0)} \right]^2+ \tfrac{1}{2}\,\beta E_8^{(4)}(M_8) =0 \,,
\end{equation}
where we have introduced the Laplacian
\begin{equation}
\label{03-laplace-p}
\triangle^{(1)} = [g^{(1)}]^{mn} \, \nabla_m \nabla_n \,,
\end{equation}
and
\begin{equation}
\left[F_2^{(0)} \right]^2 = [g^{(1)}]^{aa'} \, [g^{(1)}]^{bb'} \, [g^{(1)}]^{mm'} \, [g^{(1)}]^{nn'} \, F^{(0)}_{abmn} \, F^{(0)}_{a'b'm'n'} \,.
\end{equation}
We note that the right hand side of \eqref{03-eom-g} has been evaluated on the un-warped background because to this order the warping is not felt by that term. To order $t^{-4}$ in perturbation theory the trace of the internal Einstein equation has the following form
\begin{equation}
\label{03-internal-trace-t4}
\begin{split}
3[R^{(4)} - 7 \triangle^{(1)} A^{(3)} -14 \triangle^{(1)} B^{(3)}] &= 2^{17} \, \beta \, \triangle^{(1)} \, E_6^{(3)}(M_8) \,.
\end{split}
\end{equation}
Eliminating the $R^{(4)}$ term from equations \eqref{03-external-t3} and \eqref{03-internal-trace-t4} we obtain an equation for the warp factor $A^{(3)}$ and the leading order term of the internal flux $F_2^{(0)}$ which is defined below
\begin{equation}
\label{03-relation-01}
3 \triangle^{(1)} A^{(3)}  - \tfrac{1}{48} \left[F_2^{(0)} \right]^2+\tfrac{1}{2}\,\beta E_8^{(4)}(M_8) - 2^{17} \, \beta \, \triangle^{(1)} \, E_6^{(3)}(M_8) = 0 \,.
\end{equation}
The equation of motion  for the external flux at the order $t^{-4}$ is \cite{Becker:2001pm}
\begin{equation}
\label{03-relation-02}
\triangle^{(1)} f^{(3)}- \tfrac{1}{48} F_2^{(0)} \star^{(1)} F_2^{(0)}+ \tfrac{1}{2}\,\beta E_8^{(4)}(M_8) = 0 \,,
\end{equation}
where the Hodge $\star^{(1)}$ operation is performed with respect to the leading order term $[g^{(1)}]_{mn}$ of the internal metric. If we subtract \eqref{03-relation-02} from \eqref{03-relation-01} and integrate\footnote{The integration is performed on a manifold which we have denoted $M'_8$, whose metric is $[g^{(1)}]_{mn}$. In some sense we can think of $[g^{(1)}]_{mn}$ as being the undeformed Spin(7) holonomy metric and the next order term $[g^{(0)}]_{mn}$ being the deformation from the exceptional holonomy metric. Hence $M'_8$ can be thought as the undeformed Spin(7) holonomy manifold. We also want to note that $E_8^{(4)}(M_8)$ is the Euler integrand of $M'_8$.} the resulting expression we obtain that $F_2^{(0)}$ is self dual with respect to $\star^{(1)}$ operation
\begin{equation}
\label{03-self-duality}
\star^{(1)} F_2^{(0)} = F_2^{(0)} \,.
\end{equation}
The leading order term of the internal flux $F_2^{(0)}$ is a four-form defined on $M'_8$
\begin{equation}
F_2^{(0)} = \tfrac{1}{4!} F_{mnpq}^{(0)} \, dx'^m \wedge dx'^n \wedge dx'^p \wedge dx'^q \,,
\end{equation}
where $x'$ are the coordinates of $M'_8$. Also, $F_2^{(0)}$ satisfies
\begin{equation}
\label{03-tadpole-anomaly}
\frac{1}{4 \kappa^2_{11}} \, \int_{M'_8} F_2^{(0)} \wedge F_2^{(0)} = \frac{T_2}{24} \, \chi_8\,' \,,
\end{equation}
where $\chi_8\,'$ is the Euler character of $M'_8$. The last relation is obtained from integrating out the equation \eqref{03-relation-02} and considering that the internal flux is self dual. The condition \eqref{03-tadpole-anomaly} is nothing else but the perturbative leading order of the global tadpole anomaly relation\footnote{We remind the reader that we have not considered space-filling membranes in our calculations.} that the internal flux has to obey when compactifications of M-theory on eight-di\-men\-sio\-nal manifolds are taken into consideration \cite{Sethi:1996es}.

The difference between equations \eqref{03-relation-02} and \eqref{03-relation-01} together with the self duality condition \eqref{03-self-duality} of the internal flux produces an equation which relates the warp factor $A$ to the external flux
\begin{equation}
\label{03-warp-eflux}
\triangle^{(1)} \, \big[ f^{(3)} - 3 A^{(3)} + 2^{17} \, \beta \, E_6^{(3)}(M_8) \big]= 0 \,.
\end{equation}
Also the self-duality of the internal flux implies the vanishing of the following expression
\begin{equation}
[F^{(0)}]_{mabc} \, {[F^{(0)}]_n}^{abc} - \tfrac{1}{8} \, [g^{(1)}]_{mn} \, [F^{(0)}]_{abcd} \, [F^{(0)}]^{abcd} = 0 \,,
\end{equation}
where the details of the derivation are provided in appendix B of \cite{deWit:1978sh}. Therefore we are left with the following form for the internal Einstein equation to the order $t^{-3}$ in perturbation theory
\begin{equation}
\label{03-internal-t3}
R^{(3)}_{mn} - \tfrac{1}{2} \, g^{(1)}_{mn} \, R^{(4)} + 3 [ \, g^{(1)}_{mn} \triangle^{(1)} - \nabla_m \nabla_n] \, (A + 2B)^{(3)} + \beta \, \left(\frac{\delta Y}{\delta g^{mn}}\right)^{(3)} = 0 \,,
\end{equation}
where ${\delta Y}/{\delta g^{mn}}$ and its trace are computed in section \ref{03-j0-computation}.
The internal manifold remains Ricci-flat only under a very specific condition. To determine this condition we replace in \eqref{03-internal-t3} the expression for the perturbative coefficient of the Ricci scalar $R^{(4)}$ obtained from \eqref{03-internal-trace-t4}
\begin{equation}
R^{(4)} = 7 \triangle^{(1)} \, (A + 2B)^{(3)} - \, \frac{\beta}{3} \, \left(g^{ab} \frac{\delta Y}{\delta g^{ab}}\right)^{(4)} \,,
\end{equation}
and we recast \eqref{03-internal-t3} in the following form
\begin{equation}
\label{03-i-t3}
\begin{split}
R^{(3)}_{mn} &+ \tfrac{1}{6} \, g^{(1)}_{mn} \, \bigg[ \beta \, \left( g^{ab} \frac{\delta Y}{\delta g^{ab}} \right)^{(4)} - 3 \triangle^{(1)} \, (A + 2B)^{(3)} \bigg] \\
&+ \bigg[ \beta \, \left(\frac{\delta Y}{\delta g^{mn}}\right)^{(3)} - 3 \nabla_m \nabla_n \, \, (A + 2B)^{(3)} \bigg] = 0 \,.
\end{split}
\end{equation}
One quick way to obtain a supersymmetric theory after compactification is to ask for Ricci flatness of the internal manifold. Of course this requirement is not the most general one but in this way we can preserve for example the exceptional holonomy of the internal manifold. Ricci flatness of $M_8$ is not the whole story and we will see in chapter \ref{chapter-superpotential} that the internal flux has to satisfy restrictive conditions as well if we want a supersymmetric solution. Therefore, it is natural to look for vanishing solutions of \eqref{03-i-t3}. Now it is easy to see that Ricci flatness to this order in perturbation theory requires that
\begin{equation}
\label{03-R-flat-condition}
\nabla_m \nabla_n \, (A + 2B)^{(3)} = \frac{\beta}{3} \left(\frac{\delta Y}{\delta g^{mn}}\right)^{(3)} \,,
\end{equation}
which is a strong constraint on the warp factors and the $Y$ polynomial. To simplify the problem and to find cases where equation \eqref{03-R-flat-condition} is satisfied one fixes
\begin{equation}
\label{warp-factors-condition}
A+2B=0 \,,
\end{equation}
which leaves us with only one warp factor. Hence we are left with the problem of finding suitable internal manifolds for which the first variation of the polynomial $Y$ vanishes. As explained in \cite{Gross:1986iv}, even if the polynomial $Y$ vanishes because of the exceptional holonomy of the internal metric its first variation does not vanish in general.

Our analysis in chapter \ref{chapter-compactification} assumes \eqref{warp-factors-condition} and takes into consideration only manifolds for which the right hand side of \eqref{03-R-flat-condition} is zero. We would like to note that the above assumptions simplify relation \eqref{03-warp-eflux} to
\begin{equation}
\label{03-warp-eflux-02}
\triangle^{(1)} \, \big[ f^{(3)} - 3 A^{(3)}  \, \big]= 0 \,.
\end{equation}
This happens because the Laplacian of $E_6$ is proportional to the trace of the variation of $Y$ which vanishes according to \eqref{03-R-flat-condition}. This shows clearly that the external flux controls the warping and with appropriate boundary conditions one obtains $f^{(3)} = 3 A^{(3)}$, which, according to \cite{Lu:2004ng}, is exactly the condition for a supersymmetric solution to this order in perturbation theory. However, we will see in chapter \ref{chapter-superpotential} that this is not the whole story and one has to impose additional conditions on the internal flux for a supersymmetric background.

We can conclude that the internal manifold gets modified at the $t^{-3}$ order in perturbation theory in the sense that in general it looses its Ricci flatness unless the very restrictive constraint \eqref{03-R-flat-condition} is satisfied.


\section{Some Properties of the Quartic Polynomials}
\label{03-quartic-polynomials}

\setcounter{equation}{0}
\renewcommand{\theequation}{\thesection.\arabic{equation}}


In this section we look at some of the properties related to the quartic polynomials in the Riemann tensor which appear in the low energy effective action of M-theory. More precisely we will derive several relations obeyed by the polynomials which appear in the definition \eqref{03-pppS_1} of $S_1$. There are three different subsections, one for each of the polynomials $E_8$, $J_0$ and $X_8$, respectively. We want to emphasize that all the properties of these polynomials are computed on an undeformed background, i.e., our background is a direct product $M_3 \times M_8$ with $M_3$ being maximally symmetric and $M_8$ being a Spin(7) holonomy manifold. Obviously the warping and the deformation of the background will correct all the relations derived in the following subsections but these corrections are of a higher order than $t^{-4}$ and we can neglect them as our analysis stops at this order in perturbation theory.


\subsection[Properties of the $E_8$ Polynomial]{Properties of the $\mathbf{E_8}$ Polynomial}
\label{03-e8-computation}


Let us focus now on the properties of the quartic polynomial $E_8$ defined in \eqref{03-pppE_8} for an e\-le\-ven-di\-men\-sio\-nal manifold. As in \cite{Tseytlin:2000sf} we generalize its definition by introducing a polynomial $E_n({M_D})$ for any even $n$ and any $D\,$-di\-men\-sio\-nal manifold $M_D$ ($n \le D$) as follows
\begin{equation}
\label{03-generalized-euler}
E_n(M_D) = \pm \, \delta^{M_1 \ldots M_n}_{K_1 \ldots K_n} \, {R^{K_1 K_2}}_{M_1 M_2} \ldots {R^{K_{n-1} K_n}}_{M_{n-1} M_n} \,,
\end{equation}
where the indices take values from $0$ to $D-1$ and the ``$+$'' corresponds to the Euclidean signature and the ``$-$'' corresponds to the Lorentzian signature. As we have mentioned at the beginning of section \ref{03-quartic-polynomials}, $E_8$ is computed on a direct product manifold $M_{11}=M_3 \times M_8$, therefore we have \cite{Haack:2001jz}
\begin{equation}
\label{03-e8-property}
E_8(M_3 \times M_8) = - E_8(M_8) - 8 R(M_3) E_6(M_8) = - E_8(M_8) \,,
\end{equation}
where $R(M_3)$ is the Ricci scalar for the external manifold which is zero in our case. If $n=D$ in formula \eqref{03-generalized-euler} then $E_n(M_n)$ is proportional to the Euler integrand of $M_n$. In particular for $E_8(M_8)$ we have that
\begin{equation}
\label{03-e8-euler}
\int_{M_8} E_8(M_8) \sqrt{g} \, d^8y = \frac{\chi_8}{12\,b_1} \,,
\end{equation}
where $\chi_8$ is the Euler characteristic of $M_8$. If the manifold $M_8$ has a nowhere-vanishing spinor, $E_8(M_8)$ and $X_8(M_8)$ are related in the sense that their integrals over $M_8$ are proportional to the Euler characteristic of $M_8$. The details of this correspondence are provided in section \ref{03-x8-computation}. The variation of $E_8(M_8)$ with respect to the internal metric can be derived using the definition \eqref{03-generalized-euler} or much easier from \eqref{03-e8-euler} to be
\begin{equation}
\label{03-e8-variation}
\frac{\delta E_8(M_8)}{\delta g_{mn}} =  - \frac{1}{2} \, g^{mn} \, E_8(M_8) \,,
\end{equation}
therefore the trace of the variation is
\begin{equation}
\label{03-e8-trace}
g_{mn} \frac{\delta E_8(M_8)}{\delta g_{mn}} = - 4 E_8(M_8) \,.
\end{equation}
We want to note that the variation of $E_8$ given in \eqref{03-e8-variation} is of order $t^{-5}$ whereas its trace \eqref{03-e8-trace} is of order $t^{-4}$. Finally, for further reference, we provide the perturbative expansion for $E_8(M_8)$ and $E_6(M_8)$
\begin{subequations}
\begin{align}
E_8(M_8) &= t^{-4} \, {E_8}^{(4)}(M_8) + \ldots \,, \label{03-e8-expansion} \\
E_6(M_8) &= t^{-3} \, {E_6}^{(3)}(M_8) + \ldots \,. \label{03-e6-expansion}
\end{align}
\end{subequations}
%


\subsection[Properties of the $J_0$ Polynomial]{Properties of the $\mathbf{J_0}$ Polynomial}
\label{03-j0-computation}


In this subsection we look closely at the properties of the quartic polynomial $J_0$ defined in \eqref{03-pppJ_0}. We particularize the background to be Spin(7) holonomy compact manifold and we compute the value of $J_0$ integral on such a background. We will also calculate the first variation of $J_0$ with respect to the internal metric and the trace of its first variation.

As we will show, for a Spin(7) holonomy manifold the integral of the quartic polynomial $J_0$ vanishes. Below we provide the detailed proof of this statement. The essential fact that constitutes the basis of the demonstration is the existence of the covariantly constant spinor on a compact manifold which has Spin(7) holonomy.

The quartic polynomial $J_0$ can be expressed as a sum of an internal and an external polynomial \cite{Haack:2001jz}. Furthermore, these polynomials can be written only in terms of the internal and external Weyl tensors \cite{Banks:1998nr, Gubser:1998nz}. Since the Weyl tensor vanishes in three dimensions we are left only with the contribution from the internal polynomial
\begin{equation}
\label{03-j0-integral-11d}
\int_{M_{11}} J_0(M_{11}) \, \sqrt{-g} \, d^{11}x = \int_{M_8} J_0(M_{8}) \, \sqrt{g} \, d^8y \,.
\end{equation}
Because the internal manifold has a nowhere-vanishing spinor, the integral of the remaining internal part can be replaced by the kinematic factor which appears in the four-point scattering amplitude for gravitons, as explained in \cite{Isham:1988jb}
\begin{equation}
\int_{M_8} J_0(M_8) \, \sqrt{g} \, d^8y = \int_{M_8} Y \, \sqrt{g} \, d^8y\,,
\end{equation}
where we have denoted the kinematic factor by $Y$. As a matter of fact $J_0$ represents the covariant generalization of $Y$ and the modifications of the equations of motion are given in terms of $Y$ and its variation with respect to the internal metric. This kinematic factor was written in \cite{Gross:1986iv} as an integral over $SO(8)$ chiral spinors\footnote{The eight-rank tensor $``t''$ that appears in \cite{Gross:1986iv} is different from our convention.}
\begin{equation}
\label{03-so8-integral}
Y = \int d \psi_L\, d \psi_R \exp(R_{abmn}\, \bar\psi_L\, \Gamma^{ab}\, \psi_L\, \bar\psi_R\, \Gamma^{mn}\, \psi_R)\,,
\end{equation}
where \eqref{03-so8-integral} is evaluated using the rules of Berezin integration. As argued in \cite{Gross:1986iv}, $Y$ is zero for Ricci-flat and K\"{a}hler manifolds, but for general Ricci-flat manifolds it does not necessarily have to vanish. In our case, the ${\bf 8}_s$ of $SO(8)$ decomposes under Spin(7) as ${\bf 7} \oplus {\bf 1}$. The singlet in this decomposition corresponds to the Killing spinor $\eta$ of the Spin(7) manifold. If the holonomy group of the eight-di\-men\-sio\-nal manifold is Spin(7) and not some proper subgroup, then the covariantly constant spinor $\eta$ is the only zero mode of the Dirac operator, as proved in \cite{Joyce}. Moreover, the parallel spinor obeys the integrability condition (e.g. see \cite{Gibbons:1990er})
\begin{equation}
R_{abmn} \Gamma^{mn} \eta = 0 \,,
\end{equation}
therefore the integrand of \eqref{03-so8-integral} does not depend on the Killing spinor $\eta$ and implies the vanishing of $Y$ for $M_8$ with Spin(7) holonomy
\begin{equation}
Y = 0 \quad \mathrm{for} \quad \mathrm{Hol}[g(M_8)] = Spin(7) \,,
\end{equation}
which implies the vanishing of the integral \eqref{03-j0-integral-11d}. It has been shown in \cite{Lu:2003ze} that $Y$ vanishes in the $G_2$ holonomy case as well. The Calabi-Yau case is another example where the polynomial $Y$ vanishes \cite{Gross:1986iv}. The fact that the manifold is Ricci-flat and K\"{a}hler ensures the existence of the covariantly constant spinors, which is sufficient to imply $Y=0$ as explained in \cite{Freeman:1986br}. We conclude that the integral of $J_0$ vanishes if the internal manifold admits at least one covariantly constant spinor, in particular it vanishes for an internal manifold which has Spin(7) holonomy.

In what follows we will derive the first variation of $Y$ with respect to the internal metric. One can use \eqref{03-so8-integral} to compute the variation of $Y$ and the following result is obtained \cite{Lu:2003ze}
\begin{equation}
\label{03-Y-variation-1}
\begin{split}
\delta Y &= 4 \, \varepsilon^{\alpha_1 \ldots \alpha_8} \, \varepsilon^{\beta_1 \ldots \beta_8} \, (\Gamma^{i_1 i_2})_{\alpha_1 \alpha_2} \ldots (\Gamma^{i_7 i_8})_{\alpha_7 \alpha_8} \, (\Gamma^{j_1 j_2})_{\beta_1 \beta_2} \ldots \, \\
& \quad \cdot (\Gamma^{j_7 j_8})_{\beta_7 \beta_8} \, R_{i_1 i_2 j_1 j_2} \, R_{i_3 i_4 j_3 j_4} \, R_{i_5 i_6 j_5 j_6} \, \delta R_{i_7 i_8 j_7 j_8} \,.
\end{split}
\end{equation}
Because the internal manifold has a nowhere vanishing spinor we can transform from the spinorial representation to the vector representation ${\bf 8_v}$ of $SO(8)$. From \cite{Gibbons:1990er} we have the following relation between these representations
\begin{equation}
V^a = -i (\overline{\eta} \Gamma^a)_\alpha \psi^\alpha \,,
\end{equation}
where $\eta$ is the unit Killing spinor. After performing the change of representation in  \eqref{03-Y-variation-1} and using the identity \eqref{03-id3} and relation \eqref{03-Riemann-variation} we obtain
\begin{equation}
\delta Y = - 2^{15} \, z^{k_7 k_8 m_7 m_8} \, {{\overline{\Omega}}^{\, i_7 i_8}}_{k_7 k_8} \, {\overline{\Omega}^{\, j_7 j_8}}_{m_7 m_8} \, \nabla_{i_7} \nabla_{j_7} \delta g_{i_8 j_8} \,,
\end{equation}
where we have introduced
\begin{equation}
\label{03-omega-bar}
{\overline{\Omega}^{\,ab}}_{mn} = {\Omega^{ab}}_{mn} + \delta^{ab}_{mn} \,,
\end{equation}
$\Omega$ being the Cayley calibration of the Spin(7) holonomy manifold $M_8$. To provide a perturbative expansion for $\Omega$ we have to remember that the volume ${\cal V}_{M_8}$ of the internal manifold $M_8$ can be expressed in terms of the Cayley calibration
\begin{equation}
\int \Omega \wedge \star \Omega = 14 {\cal V}_{M_8} \,,
\end{equation}
hence the Cayley calibration perturbative expansion is
\begin{equation}
\label{03-Cayley-expansion}
\Omega_{mnpr} = t^2 \, \Omega_{mnpr}^{(2)} + t \, \Omega_{mnpr}^{(1)} + \ldots \,.
\end{equation}
The polynomial $z^{k_7 k_8 m_7 m_8}$ is cubic in the eight-di\-men\-sio\-nal Riemann tensor and it is defined by
\begin{equation}
\label{03-z-tensor}
z^{k_7 k_8 m_7 m_8} = |g|^{-1} \, \varepsilon^{a_1 \cdots a_6 k_7 k_8}\, \varepsilon^{b_1 \cdots b_6 m_7 m_8} \, R_{a_1 a_2 b_1 b_2} \, R_{a_3 a_4 b_3 b_4} \, R_{a_5 a_6 b_5 b_6} \,.
\end{equation}
It is obvious that the perturbative expansion of $z^{mnpr}$ has the following form
\begin{equation}
z^{mnpr} = t^{-5} \, {[z^{(5)}]}^{mnpr} + \ldots \,.
\end{equation}

Finally we determine the expression of the first variation of $Y$ with respect to the internal metric
\begin{equation}
\label{03-Y-variation}
\frac{\delta Y}{\delta g_{i_8 j_8}} = - 2^{15} \, {{\overline{\Omega}}^{\,i_7 i_8}}_{k_7 k_8} \, {\overline{\Omega}^{\,j_7 j_8}}_{m_7 m_8} \nabla_{i_7} \nabla_{j_7} z^{k_7 k_8 m_7 m_8} \,,
\end{equation}
which contributes to the internal Einstein equation. It is obvious that the leading order of \eqref{03-Y-variation} is $t^{-5}$, i.e., the leading order of $z^{mnpr}$. However the term $\delta Y / \delta g^{ij}$ which appears in the equation of motion \eqref{03-eom-g} is of order $t^{-3}$. In other words, $t^{-3}$ is the order at which the equations of motion receive contributions from the quantum correction terms. As we have explained in section \ref{03-eq-motion}, it is natural to suppose that the warping effects are visible to the same order in the perturbation theory and this is why we have considered the ansatz \eqref{03-warp-exp}.

In addition we also need the trace of \eqref{03-Y-variation} with respect to the internal metric. We provide in what follows the main steps of the derivation. We begin the computation by using the definition \eqref{03-omega-bar} for ${\overline{\Omega}}$ and we obtain that
\begin{equation}
\label{03-first-step}
\begin{split}
g_{i_8 j_8} \frac{\delta Y}{\delta g_{i_8 j_8}} &= - 2^{15} \, \Big[ \, g_{i_8 j_8} {\Omega^{i_7 i_8}}_{k_7 k_8} \, {\Omega^{j_7 j_8}}_{m_7 m_8} \nabla_{i_7} \nabla_{j_7} z^{k_7 k_8 m_7 m_8} \\
& \quad + g_{i_8 j_8} \, {\Omega^{i_7 i_8}}_{k_7 k_8} \, \delta^{j_7 j_8}_{m_7 m_8} \nabla_{i_7} \nabla_{j_7} z^{k_7 k_8 m_7 m_8} \\
& \quad + g_{i_8 j_8} \, \delta^{i_7 i_8}_{k_7 k_8} \, {\Omega^{j_7 j_8}}_{m_7 m_8} \nabla_{i_7} \nabla_{j_7} z^{k_7 k_8 m_7 m_8} \\
& \quad + 4 \nabla^a \nabla_b \left( {z_{am}}^{bm} \right) \, \Big] \,.
\end{split}
\end{equation}
We denote the first, the second and the third terms in the square parentheses of \eqref{03-first-step} with $T_1$, $T_2$ and $T_3$, respectively. Using \eqref{03-omega-contraction-1}, $T_1$ can be rewritten as
\begin{equation}
\begin{split}
T_1 &= 2 \triangle \left({z_{mn}}^{mn} \right) + \triangle \left( \Omega \cdot z \right) \\
& \quad - 2  \left( \nabla_a \nabla^b +\nabla^b \nabla_a \right) \left( \Omega \cdot z \right) - 4 \nabla^a \nabla_b \left( {z_{am}}^{bm} \right) \,,
\end{split}
\end{equation}
where $\triangle = \nabla_a \nabla^a$ is the Laplacian and $\Omega \cdot z$ is a short notation for the full contraction between the Cayley calibration $\Omega$ and the $z$ polynomial. The sum of the second and the third terms in \eqref{03-first-step} can be rewritten as
\begin{equation}
T_2 + T_3 = 2  \left( \nabla_a \nabla^b +\nabla^b \nabla_a \right) \left( \Omega \cdot z \right) \,.
\end{equation}
It was noted in \cite{Lu:2004ng} that
\begin{equation}
\Omega \cdot z = 0 \,,
\end{equation}
therefore we obtain an elegant and compact expression for the trace of \eqref{03-Y-variation}
\begin{equation}
\label{03-trace-j0}
g_{mn} \frac{\delta Y}{\delta g_{mn}} = - 2^{16} \, \triangle {z_{mn}}^{mn} \,.
\end{equation}
With the observation that
\begin{equation}
{z_{mn}}^{mn} = 2 \, E_6(M_8) \,,
\end{equation}
the result \eqref{03-trace-j0} can be expressed as
\begin{equation}
\label{03-Y-trace}
g_{mn} \frac{\delta Y}{\delta g_{mn}} =  - 2^{17} \, \triangle \, E_6(M_8) \,,
\end{equation}
where $E_6(M_8)$ is given by \eqref{03-generalized-euler} for $n=6$ and $D=8$. It is very interesting to note the similarity of formula \eqref{03-Y-trace} with the corresponding one for Calabi-Yau manifolds \cite{Becker:2001pm}. We also want to emphasize that the shift of the Cayley calibration toward $\overline{\Omega}$ is exactly what is needed in order to obtain the simple form of the trace given in \eqref{03-Y-trace}. A simple analysis of formula \eqref{03-Y-trace} reveals that the trace of the first variation of $Y$ is of order $t^{-4}$.
\begin{equation}
g_{mn} \frac{\delta Y}{\delta g_{mn}} =  - 2^{17} \, \triangle^{(1)} \, E_6^{(3)}(M_8) \, t^{-4} + \ldots \,,
\end{equation}
where $E_6^{(3)}(M_8)$ was introduced in equation \eqref{03-e6-expansion} and $\triangle^{(1)}$ was defined in \eqref{03-laplace-p}.


\subsection[Properties of the $X_8$ Polynomial]{Properties of the $\mathbf{X_8}$ Polynomial}
\label{03-x8-computation}


The integral of $X_8(M_8)$ over an eight-di\-men\-sio\-nal manifold $M_8$ is related to the Euler characteristic $\chi_8$ of the manifold if $M_8$ admits at least one nowhere vanishing spinor
\begin{equation}
\label{03-x8-euler}
\int_{M_8} X_8(M_8) = - \, \frac{\chi_8}{24} \,.
\end{equation}
In our calculation $M_8$ has Spin(7) holonomy, so there is a Killing spinor on $M_8$ and therefore we can use the above property in our derivations.

In what follows we will justify the relation \eqref{03-x8-euler}. The eight-form $X_8$ is defined by relation \eqref{03-x8} and can be expressed in terms of the first two Pontryagin forms $P_1$ and $P_2$
\begin{subequations}
\begin{align}
P_1 &= - \frac{1}{8 \pi^2} {\rm Tr} \,{\cal R}^2 \,, \\
P_2 &=  \frac{1}{128 \pi^4} \, [({\rm Tr} \, {\cal R}^2)^2 - 2 {\rm Tr} \, {\cal R}^4] \,,
\end{align}
\end{subequations}
as follows
\begin{equation}
X_8 = \frac{1}{192} \, [P_1^2-4P_2]\,,
\end{equation}
where ${\cal R}$ is the curvature two-form. The existence of a covariantly constant spinor on a Spin(7) holonomy manifold means that we have a nowhere vanishing spinor field on the eight-di\-men\-sio\-nal manifold. It has been shown in \cite{Isham:1988jb} that under these circumstances there is a necessary and sufficient condition which relates the Euler class and the first two Pontryagin classes of the manifold
\begin{equation}
e - \frac{1}{2} \, P_2 + \frac{1}{8} \, P_1^2 = 0 \,,
\end{equation}
where $e$ is the Euler integrand of $M_8$. Hence, the eight-form $X_8$ is proportional to the Euler integrand of $M_8$
\begin{equation}
X_8(M_8) = - \frac{1}{24} \, e(M_8) \,,
\end{equation}
and from here the relation in \eqref{03-x8-euler} follows immediately.



\chapter[Compactification of M-theory on Spin(7) Holonomy Manifolds]{COMPACTIFICATION OF M-THEORY ON SPIN(7) HOLONOMY MANIFOLDS}
\label{chapter-compactification}

\setcounter{equation}{0}
\renewcommand{\theequation}{\thechapter.\arabic{equation}}


In this section, we perform the compactification of the bosonic part of M-theory action on a Spin(7) holonomy manifold $M_8$. Since Spin(7) holonomy manifolds admit only one covariantly constant spinor, we will obtain a theory with ${\cal N}=1$ supersymmetry in three dimensions. We use the following assumptions and conventions. The eight-di\-men\-sio\-nal manifold $M_8$ is taken to be compact and smooth. As seen before in chapter \ref{chapter-vacua} we shall assume the large volume limit in which case the size of the internal eight-ma\-ni\-fold $l_{M_8} = ({\cal V}_{M_8})^{1/8}$ is much bigger than the e\-le\-ven-di\-men\-sio\-nal Planck length $l_{11}$. Here ${\cal V}_{M_8}$ denotes the volume of the internal manifold.

It was shown in \cite{Becker:1996gj, Sethi:1996es, Becker:2000jc} that compactifications of M-theory on both conformally Calabi-Yau four-folds and Spin(7) holonomy manifolds should obey the tadpole cancelation condition
\begin{equation}
\label{02-anomaly-condition}
\frac{1}{4\kappa^2_{11}} \,
\int_{M_8}\hat{F}_2 \wedge \hat{F}_2 +N_2= T_2 \, \frac{\chi_8}{24}\,,
\end{equation}
where $\hat{F}_2$ is the internal part of the background flux, $\chi_8$ is the Euler characteristic of the internal manifold and $N_2$ represents the number of space-time filling membranes. We have slightly changed the notation in the sense that a symbol with a ``hat'' above it denotes the corresponding background value. $\kappa_{11}$ is the e\-le\-ven-di\-men\-sio\-nal gravitational coupling constant, which is related to the membrane tension $T_2$ by
\begin{equation}
T_2 = \left( \frac{2 \pi^2}{\kappa_{11}^2} \right)^{1/3} \,.
\end{equation}
Equation \eqref{02-anomaly-condition} is important because it restricts the topology of the internal manifold as the Euler characteristic is expressed in terms of the internal fluxes. In our computation, we consider the case $N_2=0$, in other words the Euler characteristic of the internal manifold depends only on the internal flux and we have no membranes in our analysis. Under this assumption, in the case when the background fluxes are zero, i.e. $\hat{F}_2=0$, the tadpole cancelation condition \eqref{02-anomaly-condition} restricts the class of internal manifolds to those which have zero Euler characteristic. In this case there is no need for a warped geometry and the target space is simply the direct product $M_3 \otimes M_8$. In section \ref{02-zero-flux}, we consider this particular case and we show that no scalar potential for the moduli fields arises under these circumstances. To relax the constraint and allow for manifolds with non-vanishing Euler characteristic we have to consider a non-zero value for the internal background flux $\hat{F}_2$. Consequently, we will have to use a warped metric ansatz as we did in the analysis from chapter \ref{chapter-vacua} and we will impose the requirement \eqref{warp-factors-condition} for the warp factors. In section \ref{02-non-zero-flux}, we show that the appearance of background fluxes generates a scalar potential for some of the moduli fields appearing in the three-di\-men\-sio\-nal low energy effective action. Later on in section \ref{02-moduli} we will see that the anti-self-dual part of the four-form $\hat{F}$ is the one that generates the scalar potential.


\section{Compactification with Zero Background Flux}
\label{02-zero-flux}

\setcounter{equation}{0}
\renewcommand{\theequation}{\thesection.\arabic{equation}}


We want to compactify the action \eqref{03-start-point} on a compact and smooth Spin(7) ho\-lo\-no\-my manifold whose Euler characteristic is zero. Because of this property and because of the exceptional holonomy of the internal manifold, the quantum correction terms \eqref{03-pppS_1} vanish upon an integration over the internal manifold. Therefore, the only contribution to the three-di\-men\-sio\-nal effective action will come from \eqref{03-zero}, i.e. from the bosonic truncation of the e\-le\-ven-di\-men\-sio\-nal supergravity. In order to achieve our goal, we make the spontaneous compactification ansatz for the e\-le\-ven-di\-men\-sio\-nal metric $g_{MN}(x,y)$, which respects the maximal symmetry of the external space which is described by the metric $\eta_{\mu \nu}(x)$
\begin{equation}
\label{02-g-ansatz}
ds^2 = g_{MN} \, dX^M \, dX^N = \eta_{\mu \nu }(x) \, dx^{\mu} dx^{\nu} + g_{mn}(x,y) \, dy^m dy^n \,,
\end{equation}
where $g_{mn}(x,y)$ is the internal metric. Here $x$ represents the external coordinates labeled by $\mu=0,1,2$, while $y$ represents the internal coordinates labeled by $m=3, \ldots ,10$, and $M, ~N$ run over the complete e\-le\-ven-di\-men\-sio\-nal coordinates. In addition, ${g}_{mn}(x,y)$ depends on a set of parameters which characterize the possible deformations of the internal metric. These parameters, called moduli, appear after compactification as massless scalar fields in the three-di\-men\-sio\-nal effective action. In other words, an arbitrary vacuum state is characterized by the vacuum expectation values of these moduli fields. In the compactification process we choose an arbitrary vacuum state or equivalently an arbitrary point in moduli space and consider infinitesimal displacements around this point. Consequently, the metric will have the following form
\begin{equation}
\label{02-g-var}
{g}_{mn}(x,y) = \hat{g}_{mn}(y) +\delta g_{mn}(x,y)\,,
\end{equation}
where $\hat{g}_{mn}$ is the background metric and $\delta {g}_{mn}$ is its deformation. The deformations of the metric are expanded in terms of the transverse traceless zero modes of the Lichnerowicz operator
\begin{equation}
\label{Laplacian-Lichnerowicz}
\triangle_L \, e_{ab} = - \Box \, e_{ab} - 2 R_{abmn} e^{mn} + 2 {R_{(a}}^m e_{b)m} \,,
\end{equation}
where $e_{ab}$ is some symmetric second rank tensor. The transverse traceless zero modes of $\triangle_L$ describe variations of the internal metric leaving the Ricci tensor invariant to linear order. Furthermore, it was shown in \cite{Gibbons:1990er}, that for a Spin(7) holonomy manifold, the zero modes of the Lichnerowicz operator $e_A$ are in one to one correspondence with the anti-self-dual harmonic four-forms $\xi_A$ of the internal manifold
\begin{subequations}
\begin{align}
e_{A\,mn}(y) &= \tfrac{1}{6}\,\xi_{A\,mabc}(y) \, {\Omega_{n}}^{abc}(y)\,, \label{02-e-tensor} \\
\xi_{A\,abcd}(y) &= - \,{e_{A\,[a}}^m(y)\Omega_{bcd]m}(y)\,, \label{02-e-tensor-inv}
\end{align}
\end{subequations}
where $A=1, \ldots b_4^-$ and $\Omega$ is the Cayley calibration of the internal manifold, which in our convention is self-dual. The tensor $e^I_{mn}$ is symmetric and traceless (see \cite{Gibbons:1990er}). $b_4^-$ is the Betti number that counts the number of anti-self-dual harmonic four-forms of the internal space.

Besides the zero modes of the Lichnerowicz operator there is an additional volume changing modulus, which corresponds to an overall rescaling of the background metric. So the metric deformations take the following form
\begin{equation}
\label{02-g-expan}
\delta g_{mn}(x,y)= \phi(x) \, \hat{g}_{mn}(y) + \sum_{A=1}^{b_4^-} \phi^A(x) \, e_{A\,mn}(y) \,,
\end{equation}
where $ \phi$ is the radial modulus fluctuation and $\phi^A$ are the scalar field fluctuations that characterize the deformations of the metric along the directions $e_A$. Therefore the internal metric has the following expression
\begin{equation}
\label{02-internal-metric}
g_{mn}(x,y)= \hat{g}_{mn}(y) + \phi(x) \,
\hat{g}_{mn}(y) + \sum_{A=1}^{b_4^-}\phi^A(x) \, e_{A\,mn}(y) \,.
\end{equation}

The three-form potential and the corresponding field strength have fluctuations around their backgrounds $\hat C(y)$ and $\hat F(y)$, respectively, which in this section are considered to be zero. The fluctuations of the three-form potential are decomposed in terms of the zero modes of the Laplace operator. Taking into account that for Spin(7) holonomy manifolds there are no harmonic one-forms (see \eqref{02-cohomology} ) the decomposition of the three-form potential has two pieces
\begin{equation}
\label{02-dc-expand}
\begin{split}
\delta C(x,y) &= \delta C_1(x,y) + \delta C_2(x,y)\\
&= \sum_{I=1}^{b_2} A^I(x) \wedge \omega_I(y) + \sum_{J=1}^{b_3} \rho^J(x) \, \zeta_J(y)\,,
\end{split}
\end{equation}
where $\omega_I$ are harmonic two-forms and $\zeta_J$ are harmonic three-forms. The set of $b_2$ vector fields $A^I(x)$ and the set of $b_3$ scalar fields $\rho^J(x)$ are infinitesimal quantities that characterize the fluctuation of the three-form potential around its background value. The fluctuations of the field strength $F$ are then
\begin{equation}
\label{02-df-expand}
\begin{split}
\delta F(x,y) &= \delta F_1(x,y) + \delta F_2(x,y)\\
&= \sum_{I=1}^{b_2} dA^I(x) \wedge \omega_I(y) + \sum_{J=1}^{b_3} d\rho^J(x) \wedge \zeta_J(y)\,.
\end{split}
\end{equation}

Substituting \eqref{02-internal-metric}, \eqref{02-dc-expand} and \eqref{02-df-expand} into $S$ and considering the lowest order contribution in moduli fields we obtain
\begin{equation}
\label{02-zero-flux-action}
\begin{split}
S_{3D}&= \frac{1}{2\kappa_3^2} \, \int_{M_3}d^3x \, \sqrt{-\eta} \, \Big\{ R(M_3) - 18 (\partial_\alpha\phi)(\partial^\alpha\phi) \\
& \quad - \sum_{I,J=1}^{b_2} {\cal K}_{IJ} ~ f_{\alpha\beta}^I f^{J \alpha \beta} - \sum_{I,J=1}^{b_3} {\cal L}_{IJ} (\partial_\alpha\rho^I) (\partial^\alpha\rho^J) \\
& \quad - \sum_{A,B=1}^{b_4^-} {\cal G}_{AB} (\partial_\alpha \phi^A) (\partial^\alpha \phi^B) \Big\} ~+~ \ldots \,,
\end{split}
\end{equation}
where $\eta=det(\eta_{\mu\nu})$ and the ellipsis denotes higher order terms in moduli fluctuations. $\kappa_3$ is the three-di\-men\-sio\-nal gravitational coupling constant
\begin{equation}
\kappa_3^2 = {\cal V}^{-1}_{M_8} \, \kappa_{11}^2 \,,
\end{equation}
and ${\cal V}_{M_8}$ is the volume of the internal manifold
\begin{equation}
\label{02-volume}
{\cal V}_{M_8} = \int_{M_8} d^8y \, \sqrt{\hat{g}} \,,
\end{equation}
where $\hat{g}=det(\hat{g}_{mn})$. The details of the dimensional reduction of the Einstein-Hilbert term can be found in appendix \ref{02-appendix-dimensional}. The other quantities appearing in \eqref{02-zero-flux-action} are the field strength $f^I$ of the $b_2$ $U(1)$ gauge fields $A^I$
\begin{equation}
\label{02-field-strength}
f^I_{\alpha \beta} = \partial_{[\alpha}A^I_{\beta]} = \tfrac{1}{2}
(\partial_\alpha A^I_\beta-\partial_\beta A^I_\alpha) \,,
\end{equation}
and the metric coefficients for the kinetic terms
\begin{subequations}
\label{02-moduli-metrics}
\begin{align}
{\cal K}_{IJ} &= \frac{3}{2 {\cal V}_{M_8}} \, \int_{M_8} \omega_I \wedge \star \, \omega_J \,, \label{02-moduli-metric-k} \\
 {\cal L}_{IJ} &= \frac{2}{{\cal V}_{M_8}} \, \int_{M_8} \zeta_I \wedge \star \, \zeta_J \,, \label{02-moduli-metric-m}\\
{\cal G}_{AB} &= \frac{1}{4{\cal V}_{M_8}} \int_{M_8} d^8y\,\sqrt{\hat{g}} \, e_{A \, am} \, e_{B\,bn} \, \hat{g}^{ab}\, \hat{g}^{mn}\,. \label{02-moduli-metric-l}
\end{align}
\end{subequations}
With the help of \eqref{02-e-tensor} and \eqref{02-e-tensor-inv} we can rewrite \eqref{02-moduli-metric-l} as follows
\begin{equation}
\label{02-moduli-metric-l2}
{\cal G}_{AB} = \frac{1}{{\cal V}_{M_8}} \, \int_{M_8} \xi_A \wedge \star \, \xi_B \,.
\end{equation}
Note that the Hodge $\star$ operator used in the previous relations is defined with respect to the background metric. As we can see in the zero flux case, the action contains only the gravitational part plus kinetic terms of the massless moduli fields and no scalar potential. Therefore, if the flux is zero we have no constraint on the dynamics of the moduli fields and the vacuum of the three-di\-men\-sio\-nal theory remains arbitrary.


\section{Compactification with Non-Zero Background Flux}
\label{02-non-zero-flux}

\setcounter{equation}{0}
\renewcommand{\theequation}{\thesection.\arabic{equation}}


In this section we relax the topological constraint imposed on the internal manifold and allow for a non-vanishing background value for the field strength of M-theory, i.e. we consider manifolds with non-vanishing Euler characteristic. Because of this assumption we will have nonzero contributions in the three-di\-men\-sio\-nal action which come from the quantum correction terms \eqref{03-pppS_1}. We start with the warped ansatz \eqref{03-full-metric} for the metric
\begin{equation}
\label{02-warped-metric}
\begin{split}
ds^2 &= \widetilde{g}_{MN} \, dX^M \, dX^N \\
&= e^{2A(y)} \, \eta_{\mu \nu }(x) \, dx^{\mu} dx^{\nu} + e^{-A(y)} \, g_{mn}(x,y) \, dy^m dy^n \,,
\end{split}
\end{equation}
where we have imposed the condition \eqref{warp-factors-condition} on the warp factors. In equation \eqref{02-warped-metric} $A(y)$ represents the scalar warp factor, $\eta_{\mu \nu}(x)$ is the metric for the maximally symmetric external space, i.e. Minkowski, and ${g}_{mn}(x,y)$ has Spin(7) holonomy. As we did in the previous section we will decompose the field fluctuations in terms of harmonic forms. The metric fluctuations will have the same decomposition as in \eqref{02-g-expan} and also the field strength and its associated potential will have the decompositions \eqref{02-df-expand} and \eqref{02-dc-expand}, respectively. Maximal symmetry of the external space restricts the form of the background flux to
\begin{subequations}
\label{02-bf-expand}
\begin{align}
&\hat{F}(y) = \hat{F}_1(y)+\hat{F}_2(y)\,, \\
&\hat{F}_1(y) = \tfrac{1}{3!} \, \varepsilon_{\alpha \beta\gamma} \, \partial_m f(y) \: dx^\alpha \wedge dx^\beta \wedge dx^\gamma \wedge dy^m\,, \\
&\hat{F}_2(y) = \tfrac{1}{4!} \, F_{mnpq}(y) \: dy^m \wedge dy^n \wedge dy^p \wedge dy^q\,,
\end{align}
\end{subequations}
therefore, $C$ has the following background
\begin{subequations}
\label{02-bc-expand}
\begin{align}
&\hat{C}(y) = \hat{C}_1(y)+\hat{C}_2(y)\,, \\
&\hat{C}_1(y) = -\, \tfrac{1}{3!} \, \varepsilon_{\alpha \beta\gamma} \, f(y) \: dx^\alpha \wedge dx^\beta \wedge dx^\gamma\,, \\
&\hat{C}_2(y) = \tfrac{1}{3!} \, C_{mnp}(y) \: dy^m \wedge dy^n \wedge dy^p\,.
\end{align}
\end{subequations}

Next we consider the compactification of the e\-le\-ven-di\-men\-sio\-nal action. We start with the Einstein-Hilbert term which becomes
\begin{equation}
\label{02-warped Einstein-Hilbert}
\begin{split}
\frac{1}{2\kappa_{11}^2} \,&\int_{M_{11}} \, d^{11}x \, \sqrt{-\widetilde{g}_{11}} \, \widetilde{R}(M_{11}) \, = \, \frac{1}{2\kappa_3^2} \,\int_{M_3} d^3x \,\sqrt{-\eta}\, \Big\{ R(M_3) \\
&- 18(\partial_\alpha \phi) \, (\partial^\alpha \phi) - \, \sum_{I,J=1}^{b_4^-} \, {\cal G}_{IJ} \,(\partial_\alpha \phi^I) \, (\partial^\alpha \phi^J)\Big\} ~+~\ldots \,,
\end{split}
\end{equation}
where $\widetilde{g}_{11}=det(\widetilde{g}_{MN})$ and ${\cal G}_{IJ}$ is given in \eqref{02-moduli-metric-l2}. The details of the dimensional reduction can be found in appendix \ref{02-appendix-dimensional}. Let us take a closer look at the second term in \eqref{03-pppS_1}. We have showed in section \ref{03-j0-computation} that the exceptional holonomy implies the vanishing of integral of the quartic polynomial $J_0$. Also, regarding the term of $S_1$ which involves the quartic polynomial $E_8$, we can use properties \eqref{03-e8-property} and \eqref{03-e8-euler} to obtain
\begin{equation}
\label{02-relation}
b_1 T_2 \int_{M_{11}} {d^{11}x \,\sqrt{-g_{11}}\, \left( J_0 - \frac{1}{2} E_8
\right)} = \int_{M_3} d^3x \,\sqrt{-\eta}\, T_2 \,
\frac{\chi_8}{24} \,.
\end{equation}
We want to emphasize that the metric used in computing \eqref{02-relation} is the un-warped one. In other words, in the language of chapter \ref{chapter-vacua}, we are considering only the leading order contribution in a perturbative series in the ``t'' parameter and therefore we can neglect in the first approximation the contribution which comes from warping.

The remaining terms in $S$ consist of the kinetic term for $C$, the Chern-Simons term, and the tadpole anomaly term, i.e. the term proportional to $X_8$. The expressions \eqref{02-df-expand} and \eqref{02-bf-expand} of the field strength $F$ imply that
\begin{equation}
\begin{split}
\int_{M_{11}} F \wedge \star \,F &= \int_{M_{11}} \hat{F}_1 \wedge \star \, \hat{F}_1 + \int_{M_{11}} \hat{F}_2 \wedge \star \, \hat{F}_2 \\
& \quad + \int_{M_{11}} \delta F_1 \wedge \star \, \delta F_1 + \int_{M_{11}} \delta F_2 \wedge \star \, \delta F_2\,,
\end{split}
\end{equation}
where the first term is subleading and will be neglected. To leading order, the last two terms in the above sum can be expressed as
\begin{equation}
\begin{split}
\frac{1}{4\kappa_{11}^2} \,&\int_{M_{11}}[\delta F_1 \wedge \star \, \delta F_1 + \delta F_2 \wedge \star \, \delta F_2] \\
& = \frac{1}{2\kappa_3^2} \,\int_{M_3}d^3x \, \sqrt{-\eta} \, \Big\{ \sum_{I,J=1}^{b_2} {\cal K}_{IJ} \, f_{\alpha\beta}^I\: f^{J \alpha \beta}\\
& \quad + \sum_{I,J=1}^{b_3} {\cal L}_{IJ} \, (\partial_\alpha\rho^I) (\partial^\alpha\rho^J) \Big\} \,,
\end{split}
\end{equation}
where $f^I$, ${\cal K}_{IJ}$ and ${\cal L}_{IJ}$ were defined in \eqref{02-field-strength} and \eqref{02-moduli-metrics}. Due to the specific structure of $C(x,y)$ and $F(x,y)$, which are given in equations \eqref{02-dc-expand}, \eqref{02-df-expand}, \eqref{02-bf-expand} and \eqref{02-bc-expand}, the Chern-Simons term will have the following form to leading order in moduli field fluctuations
\begin{equation}
\label{02-chern-simons}
\begin{split}
\int_{M_{11}}C \wedge F \wedge F &= 3 \int_{M_{11}} \hat{C}_1 \wedge \hat{F}_2 \wedge \hat{F}_2\\
& \quad + 2 \int_{M_{11}} \delta C_2 \wedge \delta F_2 \wedge \hat{F}_2 ~+~\ldots \,.
\end{split}
\end{equation}
Since the first term in \eqref{02-chern-simons} cancels the tadpole anomaly term, we obtain the following result
\begin{equation}
\label{02-chern-simons-final}
\begin{split}
\frac{1}{12 \kappa_{11}^2} \, \int_{M_{11}}C \wedge F \wedge F + T_2 &\int_{M_{11}} {C \wedge X_8}\\
&= \frac{1}{6 \kappa_{11}^2} \, \int_{M_{11}} \delta C_2 \wedge\delta F_2 \wedge \hat{F}_2 ~+~\ldots \,.
\end{split}
\end{equation}
Using the harmonic decomposition for the field fluctuations \eqref{02-dc-expand} and \eqref{02-df-expand} we derive
\begin{equation}
\frac{1}{6 \kappa_{11}^2} \, \int_{M_{11}} \delta C_2
\wedge\delta F_2 \wedge \hat{F}_2 = \frac{1}{2\kappa_3^2} \,\sum_{I,J=1}^{b_2}
{\cal E}_{IJ}\int_{M_3} A^I \wedge dA^J\,, \end{equation}
where we have defined
\begin{equation}
\label{02-cs-coef}
{\cal E}_{IJ}= \frac{1}{3 {\cal V}_{M_8}} \, \int_{M_8}\omega^I \wedge \omega^J \wedge
\hat{F}_2\,.
\end{equation}
The coefficient \eqref{02-cs-coef} is proportional to the internal flux and this is the reason why we did not obtain a Chern-Simons term in section \ref{02-zero-flux}. This completes the compactification of M-theory action on Spin(7) holonomy manifolds. Using the above results we obtain to leading order in moduli fields the following expression for the low energy effective action
\begin{align}
\label{02-non-zero-flux-action}
\begin{split}
S_{3D} &= \frac{1}{2\kappa_3^2} \, \int_{M_3}d^3x \, \sqrt{-\eta} \, \Big\{ R(M_3) - 18 (\partial_\alpha\phi)(\partial^\alpha\phi)\\
& \quad - \sum_{I,J=1}^{b_3} {\cal L}_{IJ} \, (\partial_\alpha\rho^I) (\partial^\alpha\rho^J) - \sum_{I,J=1}^{b_4^-} {\cal G}_{IJ} (\partial_\alpha \phi^I) (\partial^\alpha \phi^J) \\
& \quad - \sum_{I,J=1}^{b_2} \left[ \, {\cal K}_{IJ} \, f_{\alpha\beta}^I \: f^{J \alpha \beta} + {\cal E}_{IJ} ~ \varepsilon^{\mu \nu \sigma} A_\mu^I ~ f_{\nu \sigma}^J \, \right] - V \Big\} + \ldots \,,
\end{split}
\end{align}
where we have denoted by $V$ the scalar potential
\begin{equation}
\label{02-potential}
V = \, \frac{1}{2{\cal V}_{M_8}} \, \int_{M_8}\hat{F}_2\wedge
\star \hat{F}_2 - 2 \kappa_3^2~T_2 \, \frac{\chi_8}{24} \,.
\end{equation}
We can see clary now that besides the Einstein-Hilbert term, the kinetic terms and the Chern-Simons term we have an additional piece in the action because we choose to have some non-vanishing value for the background of the field strength. The relation \eqref{02-potential}, which defines the scalar potential, is very similar with the tadpole anomaly cancelation condition \eqref{02-anomaly-condition} and we will see in chapter \ref{chapter-superpotential} that this property will determine that $V$ depends only on the anti-self-dual part of $\hat{F}_2$ whereas the self-dual part of $\hat{F}_2$ is dynamical in nature and under special conditions can break the supersymmetry of the theory.


\chapter[Minimal Three-dimensional Supergravity Coupled to Matter]{MINIMAL THREE-DIMENSIONAL SUPERGRAVITY COUPLED TO MATTER}
\label{chapter-3dsugra}

\setcounter{equation}{0}
\renewcommand{\theequation}{\thechapter.\arabic{equation}}


Some of the vacua obtained after compactification are supersymmetric and they correspond to a minimal supergravity theory in three dimensions. The analysis of the properties of these vacua requires the knowledge of the supergravity action. Hence, this chapter is dedicated to the derivation of the most general off-shell three-di\-men\-sio\-nal ${\cal N}=1$ supergravity action coupled to an arbitrary number of scalars and $U(1)$ gauge fields. Component formulations of supergravity in various dimensions with extended supersymmetry have been known for a long time \cite{Salam:1989fm}. In general, the extended supergravities can be obtained by dimensional reduction and truncation of higher dimensional supergravities. For example, a four-di\-men\-sio\-nal supergravity with ${\cal N} = 1 $ supersymmetry leads to a three-di\-men\-sio\-nal supergravity with ${\cal N}=2$ supersymmetry after compactification. For this reason the component form of three-di\-men\-sio\-nal ${\cal N}=2$ supergravity is known. Although there has been much activity in three dimensions \cite{Rocek:1986bk, vanNieuwenhuizen:1985cx, Karlhede:1987qd, Karlhede:1987qf, Park:1999cw, Ivanov:2000tz, Zupnik:1999tf, Zupnik:1997nn, Alexandre:2003qb}, there is no general {\it off-shell} component or superspace formulation of three-di\-men\-sio\-nal ${\cal N}=1$ supergravity in the literature. There are, however, on-shell realizations with ${\cal N} \geq 1$ given  in \cite{deWit:1992up, deWit:2003ja}. The ${\cal N}=1$ theory cannot be obtained by dimensional reduction from a four-di\-men\-sio\-nal theory and requires a formal analysis.

Although the off-shell formulation of ${\cal N}=1$ three-di\-men\-sio\-nal supergravity has been around since 1979 \cite{Brown:1979ma}, there has been little work done on understanding this theory with the same precision and detail of the minimal supergravity in four dimensions. The spectrum of the ${\cal N}=1$ three-di\-men\-sio\-nal supergravity theory consists of a dreibein, a Majorana gravitino and a single real auxiliary scalar field. Since our formal analysis yields an off-shell formulation, we can freely add distinct super invariants to the action. The resulting theory corresponds to a non-linear sigma model and copies of $U(1)$ gauge theories coupled to supergravity. We will present the complete superspace formulation in the hope that the presentation will familiarize the reader with the techniques required to reach our goals. In this section we use the Ectoplasmic Integration theorem to derive the component action for the general form of supergravity coupled to matter. The matter sector includes $U(1)$ gauge fields and a non-linear sigma model.


\section{Supergeometry}
\label{02-supergeometry}

\setcounter{equation}{0}
\renewcommand{\theequation}{\thesection.\arabic{equation}}


Calculating component actions from manifestly supersymmetric supergravity descriptions is a complicated process. However, knowing the supergravity density projector simplifies dramatically this procedure. The density projector arises from the following observation. Every supergravity theory that is known to possess an off-shell formulation for a superspace with space-time dimension $D$, and fermionic dimension $\cal N$ can be be shown to obey an equation of the form
\begin{equation}
\int d^{D} x \, d^{\cal N} \theta \, \, {\rm E}{}^{-1}{\cal L}
= \int d^{D}x ~ {\rm e}{}^{-1}({\cal D}^{\cal N} {\cal L}|) \,.
\end{equation}
E${}^{-1}$ is the super determinant of the super frame fields E${}_{A}{}^{M}$, ${\cal D}^{\cal N}$ is a differential operator called the supergravity density projector, and the symbol $|$ denotes taking the anti-commuting coordinate to zero. This relation has been dubbed the Ectoplasmic Integration Theorem and shows us that knowing the form of the density projector allows us to evaluate the component structure of any Lagrangian just by evaluating $({\cal D}^{\cal N} {\cal L}|)$. Thus, the problem of finding components for supergravity is relegated to computing the density projector.

Two well defined methods for calculating the density projector exist in the literature. The first method is based on super $p$-forms and the Ethereal Conjecture. This conjecture states that in all supergravity theories, the topology of the superspace is determined {\it {solely}} by its bosonic submanifold. The second method is called the ectoplasmic normal coordinate expansion \cite{Grisaru:1997ub, Gates:1997ag}, and explicitly calculates the density projector. The normal coordinate expansion provides a proof of the ectoplasmic integration theorem. Both of these techniques rely heavily on the algebra of superspace supergravity covariant derivatives. The covariant derivative algebra for three dimensional supergravity was first given in \cite{Brown:1979ma}. In this paper, we have modified the original algebra by coupling it to $n ~ U(1)$, gauge fields:\footnote{We do not consider non-abelian gauged supergravity because the compactifications of M-theory on Spin(7) manifolds that we consider lead to abelian gauged supergravities.}
\vspace*{-2ex}
\begin{subequations}
\label{02-algebra}
\begin{align}
[ \nabla_{\a} ,~ \nabla_{\b} \} &= (\g^c)_{\a \b} ~ \nabla_c - (\g^c)_{\a \b}R \, {\cal M}_c \,, \\
\begin{split}
[ \nabla_\a ,~ \nabla_b \} &= \tfrac{1}{2} \, {(\g_b)_{\a}}^{\d} R \nabla_{\d} + (\nabla_{\a} R ) {\cal M}_b + \tfrac{1}{3} \, (\g_b)_\a^{~\b}W_\b^I t_I\\
& \quad - \big[2(\g_b)_{\a}{}^{\d} \S_{\d} {}^d + \tfrac{2}{3} \, (\g_b\g^d)_{\a}{}^{\e} ( \nabla_{\e} R ) \big] {\cal M}_d \,,
%
\end{split}\\
\begin{split}
[ \nabla_a ,~ \nabla_b \} &= -2 \e_{abc} \big[ \,  \S^{ \a c} + \tfrac{1}{3} \, (\g^c)^{\a \b} (\nabla_{\b} R) \, \big] \nabla_{\a} + \e_{abc} \big[~ {\widehat {\cal R}}{}^{cd}\\
& \quad  - \tfrac{2}{3} \, \eta^{cd} (2\nabla^2 R + \tfrac{3}{2}\, R^2 )~ \big] {\cal M}_d + \tfrac{1}{3} \, \e_{abc}(\g^{c})_\b^{~\d}\nabla^\b W_\d^I t_I\,,
%
\end{split}
\end{align}
\end{subequations}
where
\begin{subequations}
\label{02-secondfs}
\begin{align}
&\nabla^\d W_\d^I = 0 \,, \\
&{\widehat {\cal R}}{}^{ab} - {\widehat {\cal R}}{}^{ba} = \eta_{ab}{\widehat {\cal R}}{}^{ab} = (\g_d)^{\a \b} \S_{\b}{}^d = 0 \,, \\
\intertext{and}
&\nabla_\a \S_\b^{~f} = -\tfrac{1}{4} \,(\g^e)_{\a\b} \, \widehat {\cal R}_e^{~f} +\tfrac{1}{6} \, \big[ \, C_{\a\b}\h^{fd} + \tfrac{1}{2} \, \e^{fde}\, (\g_e)_{\a\b} \, \big]\nabla_d R\,.
\end{align}
\end{subequations}
The superfields $R, ~\S_{\a}{}^b$ and $\widehat{\cal R}^{ab}$ are the supergravity field strengths, and $W_\a^I$ are the $U(1)$ super Yang-Mills fields strengths. $t_I$ are the $U(1)$ generators with $I=1\dots n$. ${\cal M}_a$ is the 3D Lorentz generator. Our convention for the action of ${\cal M}_a$ is given in appendix \ref{02-3d-sugra-conventions}. An explicit verification of the algebra \eqref{02-algebra} is performed in appendix \ref{02-3D-algebra}, where it is shown that the algebra closes off-shell.


\section{Closed Irreducible Super Three-forms}
\label{02-super-3forms}

\setcounter{equation}{0}
\renewcommand{\theequation}{\thesection.\arabic{equation}}


Indices of topological significance in a D-di\-men\-sio\-nal space-time manifold can be calculated from the integral of closed but not exact D-forms. The Ethereal Conjecture suggests that this reasoning should hold for superspace. Thus, in order to use the Ethereal Conjecture \cite{Gates:1998hy}, we must first have the field strength description of a super three-form. In this section, we derive the super three-form associated with the covariant derivative algebra \eqref{02-algebra}.

We start with the general formulas for the super two-form potential and super three-form field strength. A super two-form $\G_2$ has the following gauge transformations
\begin{equation}
\label{02-FBianchi1}
\d\G_{AB} = \nabla_{[A}K_{B)}-\tfrac{1}{2} \, T_{[AB)}^{~~~E} K_{E} \,,
\end{equation}
which expresses the fact that the gauge variation of the super two-form is the super exterior derivative of a super one-form $K_1$.  The field strength $G_3$ is the super exterior derivative of $\G_2$
\begin{equation}
G_{ABC} = \tfrac{1}{2} \, \nabla_{[A}\G_{BC)}-\tfrac{1}{2} \, T_{[AB| }^{~~~~E}\G_{E|C)} \,.
\end{equation}
We have a few comments about the notation in these expressions. First, upper case roman indices are super vector indices which take values over both the spinor and vector indices. Also, letters from the beginning(middle) of the alphabet refer to flat(curved) indices. Finally, the symmetrization symbol $[\,)$ is a graded symmetrization. A point worth noting here is the that the superspace torsion appears explicitly in these equations. This means that the super form is intimately related to the type of supergravity that we are using. The appearance of the torsion in these expressions is {\it {not}} peculiar to supersymmetry. Whenever forms are referred to using a non-holonomic basis this phenomenon occurs.

A super form is a highly reducible representation of supersymmetry. Therefore, we must impose certain constraints on the field strength to make it an irreducible representation of supersymmetry. In general, there are many types of constraints that we can set. Different constraints have specific consequences. A conventional constraint implies that one piece of the potential is related to another. In this case if we set the conventional constraint
\begin{equation}
G_{\a\b\g} = \tfrac{1}{2} \, \nabla_{(\a}\G_{\b\g)}-\tfrac{1}{2} \, (\g^c
)_{(\a\b|}\G_{c|\g)}=0 \,,
\end{equation}
we see that the potential $\G_{c\a}$ is now related to the spinorial derivative of the potential $\G_{\a\b}$. This constraint eliminates six superfield degrees of freedom.

Since $G_3$ is the exterior derivative of a super two-form it must be closed, i.e. its exterior derivative $F_4$ must vanish. This constitutes a set of Bianchi identities
\begin{equation}
\label{02-FBianchi}
F_{ABCD} = \tfrac{1}{3!} \, \nabla_{[A}G_{BCD)}- \tfrac{1}{4} \, T_{[AB|}^{~~~~E} G_{E|CD)} = 0 \,.
\end{equation}
Once a constraint has been set, these Bianchi identities are no longer identities. In fact, the consistency of the Bianchi identities after a constraint has been imposed implies an entire set of constraints. By solving the Bianchi identities with respect to the conventional constraint, we can completely determine the irreducible super three-form field strength. Since we have set $G_{\a\b\g}=0$, it is easiest to solve $F_{\a\b\g\d}=0$ first
\begin{equation}
F_{\a\b\g\d} = \tfrac{1}{6} \, \nabla_{(\a}G_{\b\g\d)}- \tfrac{1}{4} \, T_{(\a\b|}^{~~~E} G_{E|\g\d)} = -\tfrac{1}{4} \, (\g^e)_{(\a\b|} \,G_{e|\g\d)} \,.
\end{equation}
To solve this equation, we must write out the Lorentz irreducible parts of $G_{a\b\g}$. We first convert the last two spinor indices on $G_{e\g\d}$ to a vector index by contracting with the gamma matrix: $G_{e\g\d}=(\g^f)_{\g\d}G_{ef}$. Further, $G_{a\b\g}=G_{\b\g a}$ implies that $G_{ab}$ is a symmetric tensor, so we make the following decomposition: $G_{ab} = \overline G_{ab} + \frac{1}{3} \h_{ab}G^d_{~d}$, where the bar on $\overline G_{ab}$ denotes tracelessness. With this decomposition, the Bianchi identity now reads
\begin{equation}
F_{\a\b\g\d}=- \tfrac{1}{4} \,(\g^e)_{(\a\b}(\g^f)_{\g\d)}\overline G_{ef}=0 \,,
\end{equation}
where the term containing $G^d{}_d$ vanishes exactly. The symmetric traceless part of this gamma matrix structure does not vanish, so we are forced to set $\overline G_{ab}=0$. Thus, our conventional constraint implies the further constraint $G_{a\b\g}=(\g_a)_{\b\g}G$. The next Bianchi identity reads
\begin{equation}
\begin{split}
F_{\a\b\g d} &= \tfrac{1}{2} \, \nabla_{(\a} \, G_{\b\g)d} -\tfrac{1}{3!} \, \nabla_d \, G_{(\a\b\g)} -\tfrac{1}{2} \, {T_{(\a\b|}}^E \, G_{E|\g)d} \\
& \quad + \tfrac{1}{2} \, {T_{d(\a|}}^E \, G_{E|\b\g)} \,.
\end{split}
\end{equation}
Using our newest constraint and substituting the torsions we have
\begin{equation}
\label{02-Fbianchi2}
\begin{split}
F_{\a\b\g d} &= \tfrac{1}{2} \, (\g_d)_{(\b\g} \nabla_{\a)} \, G + \tfrac{1}{2} \, (\g^e)_{(\a\b|} \, G_{\g)ed} \\
&= \tfrac{1}{2} \, (\g_d)_{(\b\g}\nabla_{\a)} \, G + \tfrac{1}{2} \,  (\g^e)_{(\a\b|} \, {\e_{ed}}^a \Big[ {(\g_a)_{\g)}}^\d \, G_{\d}+ \widehat{G}_{\g)a} \Big] \,,
\end{split}
\end{equation}
here we have replaced the antisymmetric vector indices with a Levi-Civita tensor via; $G_{\g ed}= {\e_{ed}}^a \, G_{\g a}$, and further decomposed $G_{\g a}$ into spinor and gamma traceless parts; $G_{\g a}=(\g_a)_\g{}^\b \, G_\b+\hat G_{\g a}$, respectively. Contracting \eqref{02-Fbianchi2} with $\e_{cbe} \, (\g^e)^{\a\b} \, {\d_\s}^\g$ implies $G_\s = \nabla_\s G$. Substituting this result back into \eqref{02-Fbianchi2} implies that $\hat G_\g{}^a=0$. Thus, we have derived another constraint on the field strength
\begin{equation}
G_{\a bc} = \e_{bc}^{~~a}(\g_a)_\a^{~\s}\nabla_\s G \,.
\end{equation}
The third Bianchi identity will completely determine the super three-form
\begin{equation}
\begin{split}
F_{\a\b cd} &= \nabla_{(\a}G_{\b)cd} + \nabla_{[c}G_{d] \a\b} - T_{\a\b}^{~~E}G_{Ecd} - T_{cd}^{~~E}G_{E\a\b} - {T_{(\a|[c|}}^E \, G_{E|d]\b)} \\
&= {\e_{cd}}^e (\g_e)_{(\a}^{~~\s} \nabla_{\b)} \nabla_\s G + (\g_{[d})_{\a\b} \nabla_{c]} G -(\g^e)_{\a\b} G_{ecd} + (\g_{[c} \g_{d]})_{\a\b} \, RG \,.
\end{split}
\end{equation}
Note that $G_{\a b \g}= -(\g_b)_{\a\g}G$. Contracting with $(\g_b)^{\a\b}$ yields the following equation for the vector three-form
\begin{equation}
G_{bcd}=2\e_{bcd} \big[ \nabla^2 G + RG \big] \,.
\end{equation}
The final two Bianchi identities are consistency checks and vanish identically
\begin{subequations}
\begin{align}
&F_{\a bcd} = \tfrac{1}{3!} \, \nabla_\a \, G_{[bcd]} - \tfrac{1}{2} \nabla_{[b} \, G_{cd]\a} -\tfrac{1}{2} \, {T_{\a [b|}}^E \, G_{E|cd]} + \tfrac{1}{2} \, {T_{[bc|}}^E \, G_{E|d]\a} = 0 \,,\\
&F_{abcd} = \tfrac{1}{3!}\nabla_{[a}G_{bcd]} - \tfrac{1}{4} T_{[ab|}^{~~~E}G_{E|cd]}=0 \,.
\end{align}
\end{subequations}
We have shown that the super three-form field strength related to the supergravity covariant derivative algebra \eqref{02-algebra} is completely determined in terms of a scalar superfield $G$. In 3D, a scalar superfield is an irreducible representation of supersymmetry, and therefore the one conventional constraint was enough to completely reduce the super three-form.


\section{Ectoplasmic Integration}
\label{02-ectoplasmic}

\setcounter{equation}{0}
\renewcommand{\theequation}{\thesection.\arabic{equation}}


In order to use the Ethereal Conjecture, we must integrate a three-form over the bosonic sub-ma\-ni\-fold. The super three-form derived in the previous section is
\begin{subequations}
\begin{align}
&G_{\a\b\g} = 0 \,, \\
&G_{\a \b c} = (\g_c)_{\a\b} \, G \,, \\
&G_{\a bc} = \e_{bcd} \, (\g^d)_{\a}^{~\s} \, \nabla_\s \, G \,, \\
&G_{abc} = 2\e_{abc} \big[ \, \nabla^2 \, G + RG \big] \,.
\end{align}
\end{subequations}
The only problem with this super three-form is that it has flat indices. We worked in the tangent space so that we could set supersymmetric constraints on the super three-form. Now we require the curved super three-form to find the generally covariant component three-form. In general, the super three-form with flat indices is related to the super three-form with curved indices via
\begin{equation}
\label{02-flattocurved}
{\cal G}_{MNO}=(-1)^{[3/2]}\,{\rm E}{}_M^{~A}{\rm E}{}_N^{~B} {\rm E}{}_O^{~C}G_{CBA} \,,
\end{equation}
where we have used a different symbol for the curved super three-form just to avoid any possible confusion. As it turns out, the component three-form is the lowest component of the curved super three-form $g_{mno}={\cal G}_{mno}|$. Using the usual component definitions for the super frame fields; E${}_m^{~a}|=e_m^{~a} ,~ {\rm E}_m^{~\a}|=-\psi_{m}^{~\a}$, we can write the lowest component of the vector three-form part of \eqref{02-flattocurved}
\begin{equation}
g_{mno} = -G_{onm}| -\tfrac{1}{2} \, {\psi_{[m}}^a \, G_{no]\a}| - \tfrac{1}{2} \, {\psi_{[m}}^\a \, {\psi_{n}}^\b \, G_{o]\a\b}| + {\psi_m}^\a \, {\psi_n}^\b \, {\psi_o}^\g \, G_{\a\b\g}| \,.
%
\end{equation}
Since this is a $\q$ independent equation, we can convert all of the curved indices to flat ones using e${}_a^{~m}$
\begin{equation}
\begin{split}
g_{abc} &= G_{abc}| -\tfrac{1}{2} \, {\psi_{[a}}^\a \, G_{bc]\a}| - \tfrac{1}{2} \, {\psi_{[a}}^\a \, {\psi_b}^\b \, G_{c]\a\b}| + {\psi_{a}}^\a \, {\psi_{b}}^\b \, {\psi_{c}}^\g \, G_{\a\b\g}| \\
&= \Big\{ \, 2 \, \e_{abc} \, \big[\, \nabla^2 + R| \, \big] - \tfrac{1}{2} \, \psi_{[a}^{~\a} \, \e_{bc]d} \, (\g^d)_{\a}^{~\s} \, \nabla_\s - \tfrac{1}{2} \, \psi_{[a}^{~\a} \, \psi_b^{~\b} \, (\g_{c]})_{\a\b} \, \Big\} \, G| \,.
\end{split}
\end{equation}
We note in passing that this equation is of the form ${\cal D}^2G|$. Since $g_{abc} $ is part of a closed super three-form, it is also closed in the ordinary sense. Thus, any volume three-form $\o^{a b c}$ = $\o$ $\e^{a b c }$ may be integrated against $g_{abc} $ and will yield an index of the 3D theory if $g_{abc} $ is not exact. We are led to define an index $\Delta$ by
\begin{equation}
\Delta = ~ \int \, \o ~ \e^{a b c } \, g_{a b c } \,.
\end{equation}
If we define $\frac{1}{6} \e^{abc} g_{abc}  = {\cal D}^2G|$ we can read off the density projector\pagebreak
\begin{equation}
{\cal D}^2 = -2\nabla^2 + \psi_d^{~\a} \, (\g^d)_\a^{~\s} \, \nabla_\s - \tfrac{1}{2} \, \psi_a^{~\a} \, \psi_b^{~\b} \, \e^{abc} \, (\g_c)_{\a\b} - 2R \,.
\end{equation}
The Ethereal Conjecture asserts that for all superspace Lagrangians $\cal L$ the local integration theory for 3D, $\cal N$ = 1 superspace supergravity takes the form
\begin{equation}
\int d^3 x d^2 \q {\rm E}{}^{-1}{\cal L}=\int d^3 x \, {\rm e}{}^{-1}({\cal D}^2 {\cal L}|) \,.
\end{equation}


\section{Obtaining Component Formulations}
\label{02-component-formalism}

\setcounter{equation}{0}
\renewcommand{\theequation}{\thesection.\arabic{equation}}


We are interested in describing at the level of component fields the following general gauge invariant Lagrangian containing two derivatives for 3D, $\cal N = 1$ gravity coupled to matter
\begin{equation}
\begin{split}
{\cal L} &= \k^{-2} \, K(\F) \, R + g^{-2} \, h(\F)_{IJ} \, W^{\a I} \, W_\a^J + g(\F)_{ij} \, \nabla^\a \F^i \, \nabla_\a \F^j \\
& \quad + Q_{IJ} \, \G^I_\b \, W^{J \b} + W(\F) \,.
\end{split}
\end{equation}
This action encompasses all possible terms which can arise from the compactification of M-theory which we are considering. The first term is exactly 3D supergravity when $K(\F)=1$. The second term is the kinetic term for the gauge fields. The third term is the kinetic part of the sigma model for the scalar matter fields $\F^i$. The fourth term represents the Chern-Simons term for the gauge fields. Finally, $W(\F)$ is the superpotential.

In order to obtain the usual gravity fields we must know how to define the components of the various field strengths and curvatures. This is done in a similar manner as before when we determined the three-form component field of the super three-form. In this case, we go to a Wess-Zumino gauge to write all of the torsions, curvatures and field strengths at $\q=0$
\begin{subequations}
\label{02-FSX}
\begin{align}
&T_{ab}^{~~\g}| = t_{ab}^{~~\g} + \psi_{[a}^{~\d} ~T_{\d b]}^{~~\g} | + \psi_{[a}^{~\d}\psi_{b]}^{~\g}T_{\d\g}^{~~\g} | \,, \\
&T_{ab}^{~~c}| = t_{ab}^{~~c} + \psi_{[a}^{~\d} ~T_{\d b]}^{~~c} | + \psi_{[a}^{~\d}\psi_{b]}^{~\g}T_{\d\g}^{~~c} | \,, \\
&R_{ab}^{~~c}| = r_{ab}^{~~c}+ \psi_{[a}^{~\d}~R_{\d b]}^{~~c}| + \psi_{[a}^{~\d}~\psi_{b]}^{~\g}~R_{\d\g}^{~~c} | \,, \\
&{\cal F}_{ab}^{~~I} | = f_{ab}^{~~I} + \psi_{[a}^{~\d}{\cal F}_{\d b]}^{~~I}| + \psi_{[a}^{~\d}\psi_{b]}^{~\g}{\cal F}_{\d\g}^{~~I} | \,.
\end{align}
\end{subequations}
The leading terms in each of these equations, i.e. $t_{ab}^{~~\g} $, $t_{ab}^{ ~c} $, $r_{ab}^{~~c} $ and $ f_{ab}^{~~I} $ correspond, respectively, to the exterior derivatives of $\psi {}_a{}^\g$, e${}_a{}^m$, $\o {}_a{}^c$ and $A{}_a{}^I$, using the bosonic truncation of the definition of the exterior derivative of a super two-form given in \eqref{02-FBianchi1}. By definition the super covariantized curl of the gravitino is the lowest component of the torsion $T_{ab}{}^\g$. Substituting from \eqref{02-algebra} we have
\begin{equation}
f_{ab}^{~~\g}:= T_{ab}^{~~\g}|= -2\e_{abc} \big[~ \S^{\g c}| +
\tfrac{1}{3} \, (\g^c)^{\g \b} (\nabla_{\b} R)| ~\big] \,.
\end{equation}
This equation implies the following:
\begin{subequations}
\label{02-curlcomp}
\begin{align}
\nabla_\a R| &= -\tfrac{1}{4} \, (\g_d)_{\a\g} \, \e^{abd} \, f_{ab}^\g \,, \\
\S^{\g d}| &= \tfrac{1}{6} \, \big[ \, \e^{abd} \, {f_{ab}}^\g - {(\g_b)_\d}^\g \, f^{bd\d} \, \big] \,.
\end{align}
\end{subequations}
The lowest component of $\S^{\d d}$ is indeed gamma traceless. The other torsion yields information about the component torsion
\begin{equation}
T_{ab}^{~~c}|=0=t_{ab}^{~~c}+(\g^c)_{\a\b}\,\psi_{[a}^{~~\a} \,\psi_{b]}^{~~\b} \,,
\end{equation}
which can be solved in the usual manner to express the spin connection in terms of the anholonomy and gravitino. The super curvature leads us to the component definition of the super covariantized curvature tensor
\begin{equation}
\begin{split}
R_{ab}^{~~c}| &= r_{ab}^{~~c} -\psi_{[a}^{~\d}\psi_{b]}^{~\g}(
\g^c)_{\d\g}R| + \psi_{[a}^{~\d} \, \big[ \, -2 \, (\g_{b]})_\d^{~\a}\S_\a^{~c}| - \tfrac{2}{3} \, (\g_{b]}\g^c)_\d^{~\a}\nabla_\a R|\\
& \quad + \nabla_\d \, {R|\d_{b]}}^c \big] = \e_{abd} \, \big[ \, \widehat{\cal R}^{dc}| + \tfrac{2}{3} \, \h^{cd} \, \big( -2 \nabla^2R|-\tfrac{3}{2} R^2| \big) \, \big] \,.
\end{split}
\end{equation}
Contracting this equation with $\e^{ab}{}_c$ and using the component definitions \eqref{02-curlcomp} leads to the component definition
\begin{equation}
\nabla^2R| = -\tfrac{3}{4} \, R^2| -\tfrac{1}{4} \, \e^{abc} \, \psi_a^{~\a} \, \psi_b^{~\b} \, (\g_c)_{\a\b} \, R| + \tfrac{1}{8} \, \e^{abc} \, r_{abc} + \tfrac{1}{4} \, \psi^{a\b} \, (\g^b)_\b^{~\g} \, f_{ab\g} \,.
\end{equation}
The super field strength satisfies
\begin{equation}
{\cal F}_{ab}^{~~I}| =f_{ab}^{~~I}+\tfrac{1}{3} \psi_{[a}^{~\d} (\g_{b]})_\d^{~\a}W_\a^I | = \tfrac{1}{3} \e_{abc}(\g^c)_\b^{~\d}\nabla^\b W_\d^I | \,,
\end{equation}
which implies
\begin{equation}
\nabla_\a W_\b^I| = -\tfrac{3}{4} \e^{abc}(\g_c)_{\a\b}f_{ab}^I -
\tfrac{1}{2} \e^{abc}(\g_c)_{\a\b}(\g_b)_\d^{~\g}\psi_a^{~\d}W_\g^I| \,.
\end{equation}
From \eqref{02-secondfs} we have $\nabla_{\a} \nabla^{\b} W_\b^I$ = 0 so we can derive
\begin{equation}
\nabla^2W_\a^I|=\tfrac{1}{2} (\g^c)_\a^{~\b}\nabla_cW_\b^I| -\tfrac{3}{4} \, R|W_\a^I| \,.
\end{equation}
We now have complete component definitions for $R$ and $W_\a$ and enough of the components of $\S^{a\b}$ and $\hat R_{ab}$ to perform the ectoplasmic integration. Since the gauge potential $\G^I_\a$ for the $U(1)$ fields appears in our Lagrangian we must also make component definitions for it. $\G^I_\a$ has the gauge transformation
\begin{equation}
\d\G^I_\a=\nabla_\a K^I \,,
\end{equation}
so we can choose the Wess-Zumino gauge
\begin{equation}
\G^I_\a |= \nabla^\a\G^I_\a |= 0 \,.
\end{equation}

We are now in a position to derive the full component action. We introduce the following definitions for the component fields
\begin{equation}
\label{02-component-fields}
\begin{split}
&R| = B \,, \quad W_\a^I| = \l_\a^I \,, \\
&\F^i| =\f^i \,, \quad \nabla_\a \, \F^i| = \chi_\a^i \,, \quad \nabla^2\F^i|=F^i \,, \\
&\nabla_\a \, \G^I_\b| = \tfrac{1}{2} \, (\g^c)_{\a\b}A^I_c \,, \quad \nabla^\b \, \nabla_\b \, \G^I_\a | = \tfrac{2}{3} \, \l^I_\a \,,
\end{split}
\end{equation}
\vspace*{-2.5ex}
\hbox{\ }\\
in addition to the curl of the gravitino defined in \eqref{02-FSX}. Using these component definitions the terms in the action become
\vspace*{1ex}
\begin{align}
\begin{split}
&\int d^3x d^2\q {\rm E}^{-1} K (\F) R = \, \int d^3x \, {\rm e}^{-1} ({\cal D}^2K(\F)R)| \\
&= \, \int d^3x \, {\rm e}{}^{-1}\Big\{-2B\nabla^2K| +\nabla_\a K|\big[-\tfrac{1}{2} \,(\g_a)_\b^{~\a}\e^{abc}f_{bc}^{~~\b} \\
& \quad -\psi_d^{~\b} (\g^d)_\b^{~\a}B \, \big] + K| \big[-\tfrac{1}{2} \,B^2 -\tfrac{1}{4} \, \e^{abc} \, r_{abc} + \tfrac{1}{4}\, \psi_a^{~\b} \, \e^{abc} \, f_{bc\b} \, \big] \Big\} \,,
\end{split}\label{02-compact1}\\
\noalign{\vskip4ex}
\begin{split}
&\int d^3xd^2\q {\rm E}{}^{-1}h_{IJ}W^{\a I}W_{\a}^{~J} = \int d^3x \, {\rm e}{}^{-1}({\cal D}^2h_{IJ}W^{\a I}W_{\a}^{~J})| \\
&=\int d^3 x \, {\rm e}{}^{-1} \Big\{ +\nabla_\a h_{IJ}|\big[-3 \e^{abc} \, (\g_c)^{\a\b} \, f_{ab}^{~~I} \, \l_\b^{~J}  \\
& \quad + (\g^a)_\d^{~\a} \, \psi_a^{~\d} \, \l^{\b I} \, \l_\b^{~J} \, \big] + h_{IJ}|\big[-2(\g^c)^{\a\b}(\nabla_c\l_\a^{~J})\l_\b^{~J} \\
& \quad - \tfrac{1}{2} \, \psi^{a\a} \, \psi_{a\a} \, \l^{\b I} \, \l_\b^{~J} + 3(\g_e)_\s^{~\r} \, f^{deI} \, \psi_d^{~\s} \, \l_\r^{~J} + B\l^{\b I}\l_\b^{~J} \\
& \quad + \tfrac{9}{2} \, f^{abI}f_{ab}^{~~J} - \tfrac{3}{2} \, \e^{abc} \, \psi_c^{~\g} \, \l_\g^{~J} \, f_{ab}^{~~I} \, \big] -2\nabla^2 \, h_{IJ}| \, \l^{\b I} \, \l_\b^{~J} \, \Big\} \,,
\end{split}\label{02-compact2}\\
\noalign{\vskip4ex}
\begin{split}
&\int d^3 x \, {\rm e}{}^{-1}Q_{IJ}\G^I_\b W^{J \b}=\int d^3 x\, {\rm e}{}^{-1} ( {\cal D}^2 Q_{IJ} \G^I_\b W^{J \b} )| \\
&=\int d^3 x \, {\rm e}{}^{-1}\Big\{ \tfrac{2}{3} \, Q_{IJ}|\l^I_\b\l^{\b J} - \nabla^\a Q_{IJ}| (\g^c)_{\a\b} A^I_c \l^{\b J} \\
& \quad - \tfrac{1}{2} \, Q_{IJ}|A^{aI}\psi^{~\a}_a\l^I_\a - \tfrac{1}{2} \, Q_{IJ}|\e^{abc}(\g_a)^{~\b}_\a A^I_b \psi^{~\a}_c\l^J_\b - \tfrac{3}{2} \, Q_{IJ}|\e^{abc} A^I_a f^J_{bc}\Big\} \,,
\end{split}\\
\noalign{\vskip3ex}
\begin{split}
&\int d^3x d^2\q {\rm E}{}^{-1}g_{ij}\nabla^\a\F^i\nabla_\a\F^j = \int d^3 x \, {\rm e}{}^{-1}({\cal D}^2g_{ij}\nabla^\a\F^i\nabla_\a \, \F^j)| \\
&=\int d^3 x \, {\rm e}{}^{-1}\Big\{ \, 4g_{ij}|\big[ \tfrac{1}{2} \, (\g^c)_{\a\b} \, \nabla_c \chi^{\b i} -\tfrac{1}{4} \, B \chi_\a^{~i}\big] \chi^{\a j} \\
& \quad + 2 \, g_{ij}|\big[-\tfrac{1}{2} \nabla^c \, \f^i \, \nabla_c \, \f^j + 2 \, F^i \, F^j \, \big] \\
& \quad +g_{ij}|\big[\psi_d^{~\a}\nabla^d \, \f^i \chi_\a^{~j} + \e^{abc} (\g_c)_{\b}^{\a} \psi_a^{~\b} \chi_\a^i \nabla_b\f^j + 2(\g^a)_{\a \b} \psi_a^{\b} F^j \chi^{\a i} \big] \\
& \quad-2\nabla^2g_{ij}| \chi^{\a i} \chi_\a^{~j} - \big[ 2B + \tfrac{1}{2} \, \psi_a^{~\a} \psi_b^{~\b} \e^{abc} (\g_c)_{\a\b} \big] g_{ij}|\, \chi^{\a i} \chi_\a^{~j} \\
& \quad +\nabla_\b g_{ij}| \big[ 2 (\g^c)^{\a\b} \nabla_c\f^i \chi_\a^{~j} - 4F^i \chi^{\b j} - \psi_d^{~\a} (\g^d)_\a^{~\b} \chi^{\a i} \chi_\a^{~j} \big] \Big\} \,,
\end{split}\\
\noalign{\vskip3ex}
\begin{split}
&\int d^3 x \, {\rm e}{}^{-1}W(\F)=\int d^3 \, {\rm e}{}^{-1}({\cal D}^2W)| = \int d^3 x \, {\rm e}{}^{-1}\Big\{ -2\nabla^2W|\\
& \quad + \psi_d^{~\a}(\g_d)_\a^{~\b}\nabla_\b W| - \big[ 2B + \tfrac{1}{2} \, \psi_a^{~\a}\psi_b^{~\b} \e^{abc}(\g_c)_{\a\b}\big]W| \Big\} \,.
\end{split}
\end{align}
This component action is completely off-shell supersymmetric. We now put it on-shell by integrating out $B$ and $F^i$. The equation of motion for $F^i$ leads to
\begin{equation}
\begin{split}
F^i &= \frac{1}{4} \, g^{ij}|\Big\{\frac{\d W}{\d\F^j} \Big| + \frac{\d g_{kl}}{\d\F^j} \Big| \chi^{\a k} \chi_\a^{~l} + g^{-2} \frac{\d h_{IJ}}{\d\F^j} \Big|\l^{\a I}\l_\a^{~J} \\
& \quad + 2\nabla_\a g_{jl}| \chi^{\a l} + \k^{-2} \frac{\d K}{\d\F^j} \Big|B - g_{jl} (\g^a)_{\a\b} \psi_a^\b \chi^{\a l} \Big\} \,,
\end{split}
\end{equation}
and the equation of motion for $B$ yields
\begin{equation}
\label{02-deeznuts}
\begin{split}
B &=\k^2K|^{-1}\Big\{g^{-2}h_{IJ}|\l^{\a I}\l_\a^{~J}-2W| - g_{ij}| \chi^{\a i} \chi_\a^{~j} \\
& \quad -\k^{-2}(\nabla_\a K|\psi_d^{~\b}(\g^d)_\b^{~\a} + 2\nabla^2K|) \, \Big\} \,.
\end{split}
\end{equation}
To be completely general we assume that the coupling functions depend on some combination of matter fields, ${\cal F}^a$, thus:
\begin{equation}
\begin{split}
\nabla^2K| &= \frac{1}{2} \, \sum_a \sum_b \frac{\d^2 K}{\d{\cal F}^b\d{\cal F}^a} \Big| \nabla^\a {\cal F}^b| \nabla_\a {\cal F}^a| + \sum_a \frac{\d K}{\d{\cal F}^a} \Big|\nabla^2{\cal F}^a| \\
&\equiv \widetilde{\nabla}^2 \, K|+ \frac{\d K}{\d \F^i} \, \Big| \, F^i \,.
\end{split}
\end{equation}
With this definition, we can substitute for $F^i$ in \eqref{02-deeznuts}, leading to
\begin{equation}
\begin{split}
B&=\k^2K|^{-1}\Big[1+\frac{1}{2} \,\k^{-2}K|^{-1}g^{ij}| \frac{\d K}{\d \F^i} \Big| \frac{\d K}{\d \F^j} \Big| \Big]^{-1}\Big\{-2W| \\
& \quad - \frac{1}{2} \, \k^{-2}g^{ij}| \frac{\d W}{\d \F^i} \Big| \frac{\d K}{\d \F^j} \Big| - 2 \k^{-2} \widetilde{\nabla}^2K| - \k^2 g^{ij}| \frac{\d K}{\d \F^i} \Big|\nabla_\a g_{jl}|\chi^{\a l} \\
& \quad + \Big[-g_{kl}| - \frac{1}{2} \, \k^{-2} g^{ij}| \frac{\d K}{\d \F^i} \Big| \frac{\d g_{kl}}{\d \F^j} \Big| \, \Big] \, \chi^{\a k}\chi_\a^{~l} \\
& \quad + \Big[ \, g^{-2} \, h_{IJ}| - \frac{1}{2} \, \k^{-2}g^{-2}g^{ij}| \frac{\d K}{\d \F^i} \Big| \frac{\d h_{IJ}}{\d \F^j} \Big| \Big]\l^{\a I} \l_\a^{~J} \Big\} \,.
\end{split}
\end{equation}
\vspace*{-2.5ex}
\hbox{\ }\\
This equation for the scalar field $B$ is what is required to obtain the on-shell supersymmetry variation of the gravitino. To see this we begin with the off-shell supersymmetry variation of the gravitino
\begin{equation}
\begin{split}
\d_Q \psi_a^{~\b} &= D_a\e^\b-\e^\a(T_{\a a}^{~~\b}|+T_{\a a}^{~~b}|\psi_b^{~\b}) - \e^\a\psi_a^{~\g}(T_{\a\g}^{~~\b}| + T_{\a\g}^{~~e}|\psi_e^{~\b}) \\
&=D_a\e^\b-\tfrac{1}{2} \, \e^\a(\g_a)_\a^{~\b}B - \e^\a\psi_a^{~\g} (\g^e)_{\a\g}\psi_e^{~\b} \,.
\end{split}
\end{equation}
By converting to curved indices and keeping in mind the variation of e${}_m^{~a}$
\begin{equation}
\d_Q {\rm e}{}_{a}{}^{m} = -[\e^\b T_{\b a}^{~~d}|+\e^\b\psi_a^{~\g} T_{\b\g}^{~~d}|] {\rm e}{}_d^{~m} = -\e^\b\psi_a^{~\g}(\g^d)_{\b\g} {\rm e}{}_d^{~m} \,,
\end{equation}
the supersymmetry variation of the gravitino can be put into a more canonical form
\vspace*{-5ex}
\begin{equation}
\d_Q \psi_m^{~\b} = D_m\e^\b-\tfrac{1}{2} \, \e^\a(\g_m)_\a^{~\b}B \,.
\end{equation}
The other fields have the following supersymmetry transformations
\begin{subequations}
\begin{align}
&\d_Q {\rm e}{}_m{}^a = \e^\b\phi_m^\g(\g^a)_{\b\g} \,, \\
&\d_Q B = \tfrac{1}{4} \, \e^\a(\g_a)_{\a\g}\e^{abc}f_{bc}^\g \,, \\
&\d_Q A_c^I = -\tfrac{1}{3} \, \e^\g(\g_c)_\g^\b\l_\b \,, \\
&\d_Q \l_\a^I = \e^\b\e^{abc}(\g_c)_{\a\b}(\tfrac{3}{4}\, f_{ab}^I+\tfrac{1}{2}\, (\g_b)_\d^\g\psi_a^\d \l_\g^I \,, \\
&\d_Q\f^i = -\e^\a\chi_\a^i \,, \\
&\d_Q \chi_\a^i = -\tfrac{1}{2} \,\e^\b(\g^c)_{\a\b}\nabla_c\f^i + \e_\a F^i \,, \\
&\d_Q F^i = -\e^\a(\tfrac{1}{2} \, (\g^c)_\a^\b\nabla_c\chi_\b^i + \tfrac{1}{4} \, B\chi_\a^i) \,.
\end{align}
\end{subequations}
The purely bosonic part of the lagrangian is
\begin{equation}
\label{02-compbosact}
\begin{split}
{\cal S}_B &= \int d^3 x \, {\rm e}{}^{-1}\, \big[ \, - \tfrac{1}{4} \, \k^{-2}\e^{abc} \, r_{abc} + \tfrac{9}{2} \, g^{-2} h_{IJ}|f^{ab I} \, f_{ab}^J - g_{ij}| \nabla^c \f^i \nabla_c \f^j \\
%
%
& \quad -\tfrac{3}{2} \, Q_{IJ}|\e^{abc}A_a^I f_{bc}^J -\tfrac{1}{2} \, \k^{-2}B^2 - 2 \partial_i W|F^i -2BW| +4g_{ij}F^i F^j \,\big] \,.
\end{split}
\end{equation}
The equations of motion for B and $F^i$ with $K(\F)$ = 1 and fermions set to zero are
\begin{subequations}
\label{02-ppp-beom}
\begin{align}
&B = -2 \, \k^2 \, W| \,, \\
&F^i = \tfrac{1}{4} \, g^{ij} \, \partial_j W| \,.
\end{align}
\end{subequations}
Substituting these back into the bosonic Lagrangian we have
\begin{equation}
\label{02-bosons}
\begin{split}
{\cal S}_B &= \int d^3 x \, {\rm e}{}^{-1}\big[ -\tfrac{1}{4} \,\k^{-2}
\e^{abc}r_{abc} +\tfrac{9}{2} \,g^{-2} h_{IJ}|f^{ab I}f_{ab}^J -
g_{ij}|\nabla^c\f^i\nabla_c\f^j \\
& \quad - \tfrac{3}{2} \,Q_{IJ}| \e^{abc}A_a^I f_{bc}^J - \, (\, \tfrac{1}{4} \, g^{ij}|\partial_i W| \partial_j W| -2\k^2 W|^2 \, ) ~ \big] \,.
\end{split}
\end{equation}
The scalar potential for this theory can be read off from above and is given by
\begin{equation}
\label{02-ppp-sugra-potential}
V(\phi) = \tfrac{1}{4} \, g^{ij}\partial_i W| \partial_j W| -2\k^2 W|^2 \,,
\end{equation}
and the on-shell supersymmetry variation of the gravitino takes the form
\begin{equation}
\label{02-gavidelta}
\d_Q \psi_m^{~\b} = D_m\e^\b - \k^2 \, \e^\a (\g_m)_\a^{~\b} \, W| \,.
\end{equation}
We want to note that the superpotential $W$ determines the scalar potential in the action and also it appears in the gravitino transformation law. These properties give us the possibility to determine the explicit form of the superpotential, as we will see in section \ref{01-gravitino}, and to subsequently show in section \ref{02-moduli} that the three dimensional action obtained from compactification of M-theory on a Spin(7) holonomy manifold is a particular form of \eqref{02-bosons}.

From the form of \eqref{02-algebra} and the discussion of the above section, it is clear the issue of an $AdS_3$ background is described in the usual manner known to superspace practitioners. In the limit
\begin{equation}
\label{02-limit}
\begin{split}
&R=\sqrt{\lambda} \,, \quad {\Sigma_\alpha}^b=0 \,, \\
&{W_\alpha}^J=0 \,, \quad \widehat{\cal R}^{ab}=0 \,,
\end{split}
\end{equation}
the commutator algebra in \eqref{02-algebra} remains consistent in the form
\begin{subequations}
\label{02-algebraAdS}
\begin{align}
&[ \nabla_{\a} ,~ \nabla_{\b} \} = (\g^c)_{\a \b}  \nabla_c - \sqrt \l \, (\g^c)_{\a \b}{}\, {\cal M}_c \,, \\
&[ \nabla_\a ,~ \nabla_b \} =  \tfrac{1}{2} \, \sqrt \l \, (\g_b)_{\a} {}^{\d}\, \nabla_{\d} \,, \\
&[ \nabla_a ,~ \nabla_b \} = -  \l \, \e_{abc} \, {\cal M}^c \,,
\end{align}
\end{subequations}
and clearly the last of these shows that the curvature tensor is given by $R_{a \, b}{}^c = - \, \l \, \e_{ab}{}^c $. This in turn implies that the curvature scalar is
\begin{equation}
\e^{abc} \, R_{abc} = - 6 \lambda \,.
\end{equation}
Through the equation of motion for $B$ in \eqref{02-ppp-beom} we see that
\begin{equation}
\label{02-w}
{\sqrt \l} = - 2 \, \k^2 \,W | \,,
\end{equation}
where we have used the definition \eqref{02-component-fields} for $B$ and the imposed limit \eqref{02-limit} on $R$. Thus, there is a supersymmetry preserving $AdS_3$ background whenever the condition
\begin{equation}
W | \,< 0
\end{equation}
is satisfied, i.e. the space has a constant negative curvature. On the other hand, supersymmetry is broken whenever
\begin{equation}
W | \,> 0 \,.
\end{equation}
It is obvious from the condition \eqref{02-w} that $W|=0$ corresponds to a supersymmetric Minkowski background. We will see in section \ref{03-susy-solutions} that the proportionality between $W|$ and the curvature scalar plays an important role in selecting the possible supersymmetric solutions obtained after performing the compactification.



\chapter[The Superpotential Conjecture]{THE SUPERPOTENTIAL CONJECTURE}
\label{chapter-superpotential}

\setcounter{equation}{0}
\renewcommand{\theequation}{\thechapter.\arabic{equation}}

In the previous chapter we have derived the most general off-shell three-di\-men\-sio\-nal ${\cal N}=1$ supergravity action coupled to an arbitrary number of scalars and $U(1)$ gauge fields and we have also identified the scalar potential and the gravitino supersymmetry transformation. We have now all the information necessary to analyze the properties of the theory derived in chapter \ref{chapter-compactification}. We know that the three-di\-men\-sio\-nal action obtained from compactification includes a superpotential, whose concrete form has been conjectured in \cite{Gukov:1999gr}
\begin{equation}
\label{01-axi}
W=\int_{M} F \wedge \Omega \,,
\end{equation}
where $M$ denotes the internal manifold. In section \ref{01-gravitino} we will check directly that the superpotential is indeed given by \eqref{01-axi}, by performing a Kaluza-Klein reduction of the gravitino supersymmetry transformation law. We will repeat the procedure for the heterotic theory on Calabi-Yau three-folds and we will check the above conjectured form for $W$ for this case as well.

An important question is the relationship between the three dimensional action obtained from the compactification of M-theory on a Spin(7) holonomy manifold and the general form of the action shown in \eqref{02-bosons}. This aspect is analyzed in section \ref{02-moduli}. We check that the form of the scalar potential and subsequently the complete action is a particular case of the more general class of models discussed in chapter \ref{chapter-3dsugra} as expected.

The constraints imposed by supersymmetry on these compactifications were derived in \cite{Becker:2000jc} and \cite{Hawking:1998bg}. In \cite{Acharya:2002vs} it was shown that these constraints can be derived from the superpotential \eqref{01-axi}. In section \ref{03-susy-solutions} we identify the subset of supersymmetric solutions by analyzing these supersymmetry conditions and we derive conditions that the internal flux has to satisfy for a supersymmetric background.


\section{The Superpotential}
\label{01-gravitino}

\setcounter{equation}{0}
\renewcommand{\theequation}{\thesection.\arabic{equation}}


In the next two sub-sections we consider the compactification of the gravitino supersymmetry transformation law for M-theory on Spin(7) manifolds and Heterotic theory on Calabi-Yau three folds. We determine by direct comparison the conjectured form \eqref{01-axi} for $W$ in both cases.


\subsection{M-theory on Spin(7) Holonomy Manifolds}
\label{01-spin}


Let us start with M-theory on Spin(7) manifolds. The eleven-di\-men\-sio\-nal supersymmetry transformation of the gravitino $\Psi_M$ takes the form
\begin{equation}
\label{01-axiii}
\delta \Psi_M = \nabla_M \zeta - \tfrac{1}{288} \, ({\Gamma_M}^{PQRS}-8 \delta_M^P \Gamma^{QRS} ) ~ \zeta \, F_{PQRS}  \,,
\end{equation}
where capital letters denote e\-le\-ven-di\-men\-sio\-nal indices and $\zeta$ is an anticommuting Majorana spinor. In order to compactify this theory on a Spin(7) holonomy manifold, we will make the ansatz \eqref{02-warped-metric} for the metric. The e\-le\-ven-di\-men\-sio\-nal spinor $\zeta$ is decomposed as
\begin{equation}
\label{01-axv}
\zeta = \varepsilon \otimes \xi \,,
\end{equation}
where $\varepsilon$ is a three-di\-men\-sio\-nal anticommuting Majorana spinor and $\xi$ is an eight-di\-men\-sio\-nal Majorana-Weyl spinor. Furthermore, we will make the following decomposition of the gamma matrices
\begin{subequations}
\label{01-axvi}
\begin{align}
&\Gamma_{\mu} =\gamma_{\mu} \otimes \gamma_9 \,, \\
&\Gamma_m = 1 \otimes \gamma_m \,,
\end{align}
\end{subequations}
where $\gamma_{\mu}$ and $\gamma_m$ are the gamma matrices of the external and internal space, respectively. We choose the matrices $\gamma_m$ to be real and antisymmetric. $\gamma_9$ is the eight-di\-men\-sio\-nal chirality operator, which anti-commutes with all the $\gamma_m$'s. In compactifications with maximally symmetric three-di\-men\-sio\-nal space-time the non-vanishing components of the four-form field strength $F$ have the form given in \eqref{03-flux-form}. Using the particular form for $F$ and the decomposition \eqref{01-axvi} for the $\gamma$ matrices, we obtain the external component of the gravitino supersymmetry transformation
\begin{equation}
\label{01-axviii}
\begin{split}
\delta \Psi_\mu &= \nabla_\mu \zeta - \tfrac{1}{288} \, e^{-3A} ( \gamma_{\mu} \otimes \gamma^{mnpq}) F_{mnpq} \zeta \\
& \quad + \tfrac{1}{6} \, e^{-3A} (\gamma_\mu \otimes \gamma^m) f_m \zeta - \tfrac{1}{2} \, \partial_n A (\gamma_{\mu} \otimes \gamma^n) \zeta \,,
\end{split}
\end{equation}
where we have used a positive chirality eigenstate $\gamma_9 \xi=\xi$. Considering a negative chirality spinor corresponds to an eight-ma\-ni\-fold with a reversed orientation \cite{Isham:1988jb, Isham:1987qe}. We decompose the e\-le\-ven-di\-men\-sio\-nal gravitino as we did with $\zeta$ in \eqref{01-axv}
\begin{equation}
\label{01-axviib}
\Psi_\mu = \psi_\mu \otimes \xi \,,
\end{equation}
where $\psi_{\mu}$ is the three-di\-men\-sio\-nal gravitino. After inserting \eqref{01-axv} and \eqref{01-axviib} in \eqref{01-axviii}, we multiply both sides of this equation from the left with the transposed spinor $\xi^T$. To evaluate the resulting expression we notice that on these eight-ma\-ni\-folds it is possible to construct different types of $p$-forms in terms of the eight-di\-men\-sio\-nal spinor $\xi$ as
\begin{equation}
\label{01-ppp-aii}
\omega_{a_1 \ldots a_p} = \xi^T \gamma_{a_1\ldots a_p} \xi \,.
\end{equation}
Since ${\xi}$ is Majorana-Weyl, \eqref{01-ppp-aii} is non-zero only for $p=0,4$ or 8 (see \cite{Gibbons:1990er}). By this argument we notice that the expectation values of the last two terms appearing in \eqref{01-axviii} vanish, as they contain only one internal gamma matrix. The Spin(7) calibration, also called the Cayley calibration, is given by the closed self-dual four-form \cite{Becker:2000jc, Hawking:1998bg}
\begin{equation}
\label{01-ppp-aiii}
\Omega_{mnpq} = \xi^T \gamma_{mnpq} \xi \,.
\end{equation}
Neglecting the contribution from the warp factor we obtain from \eqref{01-axviii}
\begin{equation}
\label{01-bxi}
\delta \psi_\mu = \nabla_\mu \varepsilon - \gamma_\mu \varepsilon ~ \int_{M_8} F \wedge \Omega \,,
\end{equation}
where we have again dropped a multiplicative constant in front of the second term on the right hand side. By comparison with formula \eqref{02-gavidelta} we can then read off the form of the superpotential
\begin{equation}
\label{01-axi-ppp}
W = \int_{M_8} F \wedge \Omega \,,
\end{equation}
which is what we wanted to show. In the next section we perform the same analysis for the case of heterotic theory on Calabi-Yau three-folds. We note that we had to rescale the superpotential, i.e. $\kappa^2 W| \rightarrow W|$, in order to have agreement. We will see in section \ref{02-moduli} that the same rescaling of the superpotential lowest component $W|$ is needed in order to have agreement between the compactified action and the supergravity result.


\subsection{The Heterotic String on Calabi-Yau Three-folds}
\label{01-heterotic}


In order to derive the superpotential for the four-di\-men\-sio\-nal heterotic string, we will consider the compactification of the gravitino supersymmetry transformation law, as we did in the previous section. Recall that the most general gauge invariant ${\cal N}=1$, $D=4$ supergravity action can be described in terms of three functions (see e.g. \cite{Gates:1983nr, Wess:1992cp}). These are the superpotential $W$, the K\"ahler potential $K$, and a holomorphic function $H_{ab}$, which plays the role of the gauge coupling. In the following we will take $H_{ab}={\delta}_{ab}$. The theory is formulated in terms of massless chiral multiplets, containing a complex scalar $\phi$ and a Weyl spinor $\psi$ and massless vector multiplets, containing the field $A^a_{\mu}$ with field strength $F^a_{\mu\nu}$ and a Weyl spinor ${\lambda}^a$. We shall be adding a real auxiliary field $D^a$ to the vector multiplets. The bosonic part of the Lagrangian takes the following form\footnote{We will be following the conventions of \cite{Wess:1992cp}.}
\begin{equation}
\label{01-aixbb}
{\cal G} = -  \tfrac{1}{2}R - K_{{\bar i}j} D_{\mu}{\phi}^{i*}D^{\mu}{\phi}^j - \tfrac{1}{4} F^a_{\mu\nu} F^{a\mu\nu} - V(\phi,{\phi}^*) \,.
\end{equation}
Here $V(\phi,\phi^*)$ describes the scalar potential given by
\begin{equation}
\label{01-aix}
V(\phi,\phi^*) = exp(K) (K^{{\bar i}j} W^*_i W_j - 3W^* W) + \tfrac{1}{2} D^a D_a.
\end{equation}
In this formula $K^{{\bar i}j}$ is the inverse matrix to $K_{{\bar i}j}=\partial_{\bar{i}} \partial_j K(\phi, \phi^*)$ where the partial derivatives are with respect to the scalar fields $\phi$,
and $W_i=\partial_i W + \partial_i K ~ W$. The complete Lagrangian is invariant under ${\cal N}=1$ supersymmetry. The relevant part of the supersymmetry transformations takes the form
\begin{subequations}
\label{01-ax}
\begin{align}
\delta \lambda^a &= F^a_{\mu \nu}~{\sigma^{\mu \nu}} \varepsilon -i D^a~{\varepsilon} \,, \\
\delta \psi_\mu &= 2 \nabla_{\mu}\varepsilon +i~e^{K/2}~{\gamma}_{\mu}{\varepsilon^*}W \,.
\end{align}
\end{subequations}
Here ${\lambda}^a$ and ${\psi}_{\mu}$ are positive chirality Weyl spinors, describing the gluino and gravitino, respectively, $\varepsilon$ is a four-di\-men\-sio\-nal Weyl spinor of positive chirality, while ${\varepsilon^*}$ is the complex conjugate spinor with negative chirality. If the space-time is flat, the complete supersymmetry transformations tell us that supersymmetry demands (see \cite{Wess:1992cp})
\begin{equation}
\label{01-bixa}
W_i = D^a = W = 0 \,.
\end{equation}
In what follows we will use the above supersymmetry transformations to determine the superpotential and $D$-term for the heterotic string compactified on a Calabi-Yau three-fold.

It has been known for a long time that gluino condensation triggers spontaneous supersymmetry breaking in the heterotic string compactified on a Calabi-Yau three-fold $Y_3$ (with no warp factors) without producing a vacuum energy \cite{Dine:1985rz}. In this process the Neveu-Schwarz three-form $H$ of the heterotic string acquires a vacuum expectation value proportional to the holomorphic three-form ${\Omega}$ of the Calabi-Yau three-fold. It was shown in \cite{Dine:1985rz} that this generates a superpotential, which will break the supersymmetry completely. In a more recent context\footnote{For an earlier discussion of the form of the superpotential see \cite{Dine:1985rz}.} it was argued in \cite{Behrndt:2000zh} that the superpotential which is induced by such a non-vanishing $H$-field extends the conjecture \eqref{01-axi} to superpotentials with non-vanishing fluxes of Neveu-Schwarz type, i.e.
\begin{equation}
\label{01-axba}
W = \int_{Y_3} H \wedge \Omega \,.
\end{equation}
The argument, which motivated the above formula, was based on the identification of $BPS$ domain walls with branes wrapped over supersymmetric cycles. More concretely, the $BPS$ domain wall of the ${\cal N}=1$, $D=4$ theory originates from the heterotic five-brane wrapping a special Lagrangian submanifold of $Y_3$. This is because the five-brane is a source for the Neveu-Schwarz three-form field strength $H$. Here we would like to compute the form of the superpotential and the form of the $D$-term appearing in \eqref{01-ax} in this particular model by a direct Kaluza-Klein reduction of the gravitino and gluino supersymmetry transformation, respectively. Recall that the ten-di\-men\-sio\-nal ${\cal N}=1$ supergravity multiplet contains a metric $g_{MN}$, a spin-$\frac{3}{2}$ field ${\Psi}_M$, a two-form potential $B_{MN}$, a spin-$\frac{1}{2}$ field $\lambda$ and a scalar field $\phi$. The super Yang-Mills multiplet contains the Yang-Mills field $F^a_{MN}$ and a spin-$\frac{1}{2}$ field ${\chi}^a$, the so-called gluino. The relevant part of the ${\cal N}=1$ supersymmetry transformations in the ten-di\-men\-sio\-nal string frame takes the form
\begin{subequations}
\label{01-axiiib}
\begin{align}
&\delta \Psi_{\mu} = \nabla_{\mu}\zeta + \tfrac{1}{48} ({\gamma}_{\mu} {\gamma}_5 ~{\otimes}~ \gamma^{abc} H_{abc})~ \zeta \,, \\
&\delta {\chi}^{\alpha} = - \tfrac{1}{4} F^{\alpha}_{ab} {\gamma}^{ab} \zeta \,.
\end{align}
\end{subequations}
Here $\mu$ describes the coordinates of the four-di\-men\-sio\-nal Minkowski space, and $a,b, \ldots$ describe the six-di\-men\-sio\-nal internal indices, while $\alpha$ describes the gauge index.

We consider a Majorana representation for ten-di\-men\-sio\-nal Dirac matrices with $\Gamma_M$ real and hermitian, apart from $\Gamma_0$ which is real and antihermitian. The matrices $\Gamma_M$ can be represented as tensor products of $\gamma_{\mu}$, the matrices of the external space, with $\gamma_m$, the matrices of the internal space
\begin{subequations}
\label{01-pingoi}
\begin{align}
&\Gamma_{\mu} = \gamma_{\mu} \otimes 1 \,, \\
&\Gamma_m = \gamma_5 \otimes \gamma_m \,,
\end{align}
\end{subequations}
with
\begin{equation}
\label{01-pingoii}
\gamma_5 = \tfrac{i}{4!} \, \varepsilon_{\mu \nu \rho \sigma} \, \gamma^{\mu \nu \rho \sigma} \,.
\end{equation}
We can also introduce the matrix
\begin{equation}
\label{01-pingoiii}
\gamma = \tfrac{i}{6!} \sqrt{g_{(6)}} \, \varepsilon_{mnpqrs} \, \gamma^{mnpqrs} \,,
\end{equation}
which determines the chirality in the internal space. Here $g_{(6)}$ represents the determinant of the internal metric. Thus $\gamma_{\mu}$ are real and hermitian, apart from $\gamma_0$ which is real and antihermitian, and $\gamma_m$ are imaginary and hermitian as are $\gamma_5$ and $\gamma$. The relation between $\Gamma$, the matrix which determines the chirality in ten-dimensions, $\gamma_5$ and $\gamma$ is
\begin{equation}
\label{01-pingoiv}
\Gamma = - \, \gamma_5 \otimes \gamma \,.
\end{equation}
Consider $\zeta$ a ten-di\-men\-sio\-nal Majorana-Weyl spinor of positive chirality. In order to compactify transformations \eqref{01-axiiib} to four dimensions, we decompose this ten-di\-men\-sio\-nal spinor in terms of the covariantly constant spinors of the internal manifold:
\begin{equation}
\label{01-eiviv}
\zeta=\varepsilon^* ~{\otimes}~\xi_+ + \varepsilon ~{\otimes}~\xi_- \,,
\end{equation}
where $\xi_+$ and $\xi_- = (\xi_+)^*$ are six-di\-men\-sio\-nal Weyl spinors with positive and negative chirality, respectively, and $\varepsilon$ is a four-di\-men\-sio\-nal Weyl spinor of positive chirality, whose complex conjugate is $\varepsilon^*$. Similarly, we decompose the ten-di\-men\-sio\-nal gravitino as:
\begin{equation}
\label{01-eivvi}
\Psi_{\mu} = \psi^*_\mu ~ \otimes ~ \xi_+ + \psi_\mu ~ \otimes ~ \xi_- \,,
\end{equation}
where $\psi_{\mu}$ is a four-di\-men\-sio\-nal Weyl spinor of positive chirality, that represents the four-di\-men\-sio\-nal gravitino.

In complex coordinates the gravitino supersymmetry transformation takes the form
\begin{equation}
\label{01-eivvii}
\begin{split}
\delta \Psi_{\mu} & = \nabla_{\mu}\zeta + \tfrac{1}{48} \bigl[ {\gamma}_{\mu}{\gamma}_5 ~{\otimes}~ ( \gamma_{mnp} H^{mnp}+ \gamma_{\bar m \bar n \bar p} H^{\bar m \bar n \bar p}) \bigr] ~ \zeta \\
& \quad + \tfrac{1}{48} \bigl[ {\gamma}_{\mu} {\gamma}_5 ~{\otimes}~ (\gamma_{mn{\bar p}} H^{mn{\bar p}}+\gamma_{m{\bar n}{\bar p}} H^{m{\bar n}{\bar p}}) \bigr] ~ \zeta \,.
\end{split}
\end{equation}
To evaluate the resulting expressions we use the identities (see e.g. \cite{Becker:1995kb} or \cite{Marino:1999af})
\begin{subequations}
\label{01-eivviii}
\begin{align}
&\gamma_{\bar{m}} \,\xi_+ = 0 \,, \label{01-eivv}\\
&\gamma_{ m n p} \,\xi_+  = \|\xi_+\|^{-2} ~\Omega_{ m n p} \,\xi_- \,, \\
&\gamma_{mn{\bar p}} \,\xi_+  = 2i ~\gamma_{[m}J_{n] {\bar p}}\,\xi_+ \,, \\
&\gamma_{m{\bar n}{\bar p}} \,\xi_+  = \gamma_{{\bar m}{\bar n}{\bar p}} \,\xi_+ = 0 \,.
\end{align}
\end{subequations}
We now decompose our ten-di\-men\-sio\-nal spinors as in \eqref{01-eiviv} and \eqref{01-eivvi} and make use of formulas \eqref{01-eivv} and \eqref{01-eivviii}. Multiplying the resulting expression from the left with $\xi_-^{\dagger}=\xi^T_+$, we obtain the transformation:
\begin{equation}
\label{01-eivxab-ppp}
\delta \psi_\mu = \nabla_\mu \varepsilon - \tfrac{1}{48} \gamma_\mu \varepsilon^* ~ \|\xi_+\|^{-2} ~ H_{{\bar m}{\bar n}{\bar p}} ~ \Omega^{{\bar m}{\bar n}{\bar p}} \,.
\end{equation}
After integration over the internal manifold we obtain:
\begin{equation}
\label{01-eivxab}
\delta \psi_\mu = \nabla_\mu \varepsilon + i \gamma_\mu \varepsilon^* \, \|\xi_+\|^{-2}\, e^{K_2} ~\int_{Y_3} H \wedge \Omega \,,
\end{equation}
where we have used that:
\begin{equation}
\label{01-eivxp}
V = \tfrac{1}{48} ~ \int_{Y_3} J \wedge J \wedge J = \tfrac{1}{64} \, e^{-K_2} \,,
\end{equation}
with $V$ being the volume of the internal Calabi-Yau manifold. If we choose
\begin{equation}
\label{01-eivxabo}
\|\xi_+\|^{-2} = e^{K/2-K_2} \,,
\end{equation}
and rescale the fields:
\begin{subequations}
\label{01-eivxabp}
\begin{align}
&\psi \rightarrow \tfrac{1}{2} \,\psi \,, \\
&H \rightarrow \tfrac{1}{2}\, H\,,
\end{align}
\end{subequations}
then we obtain the four-di\-men\-sio\-nal supersymmetry transformation for gravitino,
\begin{equation}
\label{01-eivxabs}
\delta \psi_\mu = 2 \nabla_\mu \varepsilon + i \gamma_\mu \varepsilon^* ~ e^{K/2} ~ \int_{Y_3} H \wedge \Omega \,.
\end{equation}
In the above formulas $K=K_1+K_2$ is the total K\"ahler potential, where $K_1$ is the K\"ahler potential for complex structure deformations
\begin{equation}
\label{01-eivix}
K_1 = -\log \left( i~\int_{Y_3} \Omega \wedge {\bar \Omega} \right) \,,
\end{equation}
and $K_2$ is the K\"ahler potential for the K\"ahler deformations
\begin{equation}
\label{01-eivx}
K_2 = -\log \left( \frac{4}{3} ~\int_{Y_3} J \wedge J \wedge J \right) \,.
\end{equation}
Comparing this result with \eqref{01-ax} we find the superpotential
\begin{equation}
\label{01-eivxiv}
W = \int_{Y_3} H \wedge \Omega \,,
\end{equation}
as promised.

Let us now consider the gluino supersymmetry transformations in \eqref{01-axiiib}. If we again decompose the gluino as in \eqref{01-eivvi} and the spinor $\zeta$ as in \eqref{01-eiviv}, we obtain after comparing with \eqref{01-ax} the form of the four-di\-men\-sio\-nal $D$-term up to a multiplicative constant
\begin{equation}
\label{01-eab}
D^a = F^a_{m \bar n} J^{m \bar n} \,.
\end{equation}
Here we have used
\begin{equation}
\label{01-eivxvi}
J_{m \bar n} = - i \xi_+^{\dagger} \gamma_{m \bar n} \xi_+ \,,
\end{equation}
while the expectation value for the other index contractions appearing in the four-di\-men\-sio\-nal gluino supersymmetry transformation vanish. As we have mentioned in the previous section, supersymmetry demands $D^{(a)}=0$, which in this case gives the well known Donaldson-Uhlenbeck-Yau equation
\begin{equation}
\label{01-eivxviii}
J^{m \bar n} F_{m \bar n}^a=0 \,.
\end{equation}
The fact that the Donaldson-Uhlenbeck-Yau equation originates from a $D$-term constraint was first discussed in \cite{Witten:1985bz}. Furthermore, supersymmetry demands
\begin{equation}
\label{01-eivxix}
W_i=0 \,,
\end{equation}
where we are using again the notation $W_i = \partial_i W + \partial_i K_1 ~ W$, where $K_1$ is the K\"ahler potential for complex structure deformations and is given in \eqref{01-eivix}. It is straightforward to evaluate this constraint to obtain $W_i = \int_{Y_3} {\phi}_i \wedge \Omega = 0$, where ${\phi}_i$ is a complete set of $(2,1)$ forms \cite{Candelas:1990pi}. This implies that $H$ is of type $(0,3)$. However in this case
\begin{equation}
\label{01-eivxx}
W \neq 0 \,,
\end{equation}
and we therefore see that no supersymmetric solutions can be found. It is expected that this situation changes if we consider instead a ``warped'' compactification of the heterotic string \cite{Strominger:1986uh, deWit:1986xg}. The resulting background is in this case a complex manifold with non-vanishing torsion. Manifolds with non-vanishing torsion have also been discussed some time ago in e.g. \cite{Gates:1983py}. As opposed to the previous reference, the manifolds we shall be interested in have a torsion that is not closed. It is expected that supersymmetric ground states can be found in this case. In \cite{Becker:2003yv} it was computed the form of this superpotential and checked, that it takes the same form as \eqref{01-eivxiv}.


\section[The Scalar Potential and the Effective Action]{The Scalar Potential and the Effective Action}
\label{02-moduli}

\setcounter{equation}{0}
\renewcommand{\theequation}{\thesection.\arabic{equation}}


In this section we show that the action \eqref{02-non-zero-flux-action} is a particular case of the more general construction presented in section \eqref{02-component-formalism}. Let us elaborate this in detail. First, we will show that the scalar potential \eqref{02-potential} can be written in terms of the superpotential
\begin{equation}
\label{02-ppp-superpotential} W=\int_{M_8}\hat{F}_2 \wedge
\Omega \,,
\end{equation}
and subsequently we will perform a series of field redefinitions to obtain agreement between the relevant terms in both actions. The form of the superpotential $W$ was conjectured in \cite{Gukov:1999gr} and we have checked this conjecture in the previous section. More explicitly we have used the supersymmetry transformation for the gravitino \eqref{02-gavidelta} to identify the form of the superpotential. Using the anomaly cancelation condition \eqref{02-anomaly-condition}, the scalar potential becomes
\begin{equation}
\label{scalar-pot-int-flux}
V = \, \frac{1}{{\cal V}_{M_8}} \, \int_{M_8} \hat{F}_{2 \,-} \wedge \star \, \hat{F}_{2 \,-} \,,
\end{equation}
where
\begin{equation}
\hat{F}_{2\,-} = \tfrac{1}{2} \, \big[ \hat{F}_2 - \star \, \hat{F}_2 \big]
\end{equation}
is the anti-self-dual part of the internal flux $\hat{F}_2$. Using the definition \eqref{02-moduli-metric-l2} for ${\cal G}_{AB}$ we can obtain the functional dependence of the scalar potential $V$ in terms of the superpotential \eqref{02-ppp-superpotential}
\begin{equation}
\label{02-scalar-potential}
V[W] = \sum_{A,B=1}^{b_4^-} {\cal G}^{AB} \, D_AW \, D_BW \,,
\end{equation}
where ${\cal G}^{AB}$ is the inverse matrix of ${\cal G}_{AB}$ and we have introduced the operator
\begin{equation}
\label{02-D-operator}
D_A \Omega = \partial_A \Omega + K_A \Omega \,.
\end{equation}
From equation \eqref{02-partail-a-omega} we can see that the action of $D_A$ on the Cayley calibration produces an anti-self-dual harmonic four-form and this motivates the appearance of ${\cal G}^{AB}$ in formula \eqref{02-scalar-potential}.

What we observe from \eqref{02-scalar-potential} is that the external space is restricted to three-di\-men\-sio\-nal Minkowski because the scalar potential is a perfect square, in agreement with \cite{Acharya:2002vs}. Furthermore, when $D_A W=0$ the scalar potential vanishes. This relation provides a set of $b_4^-$ equations for $b_4^- + 1$ fields, which lives us with the radial modulus unfixed at this level. Its rather possible that non-perturbative effects will lead to a stabilization of this field, as in \cite{Kachru:2003aw}.

A few remarks are in order before we can compare the compactified action to the supergravity action. For a consistent analysis, we must take into account all of the kinetic terms for the metric moduli. Furthermore, the scalar potential \eqref{02-scalar-potential} does not seem to be a special case of \eqref{02-ppp-sugra-potential}. The discrepancy arises for two reasons. First, in the general case the superpotential may depend on all of the scalar fields existing in the theory and the summation in \eqref{02-ppp-sugra-potential} is taking into account all of these scalars, whereas in the compactified version the superpotential depends only on the metric moduli
\begin{equation}
\partial_i W \neq 0 \quad i=0,~1, \ldots b_4^- \,,
\end{equation}
where $``0"$ labels the radial modulus. Second, because of \eqref{02-partial-0-omega} the superpotential has a very special radial modulus dependence in the sense that
\begin{equation}
\partial_0 W = 2W \,,
\end{equation}
and this is the reason why the summation in \eqref{02-scalar-potential} does not include the radial modulus. Keeping these remarks in mind we proceed to show that the result coming from compactification is a particular case of the general supergravity analysis.

We begin by rescaling some of the fields in the supergravity action \eqref{02-bosons}
\begin{subequations}
\begin{align}
&2 \kappa_3^2~g_{ij}| = L_{ij} \,, \\
&\frac{1}{2 \kappa_3^2}~g^{ij}| = L^{ij} \,, \\
&\kappa_3^2~ W| = \widetilde{W} \,,
\end{align}
\end{subequations}
therefore the relevant terms in the supergravity action, which we denote by $S^{\kern.1em\prime}_B$, can be written as
\begin{equation}
\label{02-qqq-sugra}
\begin{split}
S^{\kern.1em\prime}_B &= \frac{1}{2 \kappa_3^2}~\int_{M_3}d^3x ~ \sqrt{-g^{(3)}} ~ \Big\{ - L_{ij} ~ \partial_\alpha \bar{\f}^i ~ \partial^\alpha \bar{\f}^j \\
& \quad  - \big[\, L^{ij} ~ \bar{\partial}_i \widetilde{W} ~ \bar{\partial}_j \widetilde{W} - 4 ~ \widetilde{W}^2 \, \big] ~ \Big\} \,.
\end{split}
\end{equation}
In the above equation the indices $i,~j = 0,~ 1, \ldots b_4^-$. In what follows we will drop the label ``0'' from the radion $\phi_0=\phi$ and the derivative with respect to it $\partial_0=\partial$ and we will denote by $A,~B \ldots$ the remaining set of indices, i.e. $A,~B=1,\ldots,b_4^-$. In \eqref{02-qqq-sugra} we have placed bars on the scalar fields and the derivative operators which involve them in order to avoid confusion since we require one more field redefinition.

The relevant terms in the compactified action \eqref{02-non-zero-flux-action}, which we denote by $S^{\kern.1em\prime}_{3D}$, have the following form
\begin{equation}
\label{02-qqq-compact}
\begin{split}
S^{\kern.1em\prime}_{3D} &= \frac{1}{2\kappa_3^2} \, \int_{M_3}d^3x \sqrt{-g^{(3)}} \Big\{ -18 (\partial \phi)^2 -{\cal G}_{AB} (\partial_\alpha\phi^A)(\partial^\alpha\phi^B) \\
& \quad - {\cal G}^{AB} D_A W D_B W ~\Big\} \\
&= \frac{1}{2\kappa_3^2} \, \int_{M_3}d^3x \, \sqrt{-g^{(3)}} \, \Big\{ -18 (\partial_{\alpha} \phi) (\partial^{\alpha}\phi) -{\cal G}_{AB} (\partial_\alpha\phi^A)(\partial^\alpha\phi^B) \\
& \quad - \big[ \, {\cal G} ^{AB} (\partial_A W) (\partial_B W) + 4 {\cal G}^A (\partial_A W)W + {\cal G} W^{2} \, \big] ~\Big\}.
\end{split}
\end{equation}
In the above equation we have used expression \eqref{02-D-operator} for $D_A$ and we have introduced ${\cal G}^{A}={\cal G}^{AB}K_B$ and ${\cal G}={\cal G}^{AB} K_A K_B$.

In order to make the comparison between \eqref{02-qqq-sugra} and \eqref{02-qqq-compact}, we have to redefine the fields in \eqref{02-qqq-sugra} in the following manner
\begin{subequations}
\begin{align}
&\phi = L_{00} \bar{\phi} + L_{0A} \bar{\phi}^A \,, \\
&\phi^A = \bar{\phi}^A \,.
\end{align}
\end{subequations}
Keeping track that $\phi$ is the radial modulus, we obtain the following form for $S^{\kern.1em\prime}_B$
\begin{equation}
\begin{split}
S^{\kern.1em\prime}_B &= \frac{1}{2\kappa_3^2} \, \int_{M_3}d^3x \sqrt{-g^{(3)}} \Big\{ - \frac{1}{L_{00}} \, (\partial_\alpha \phi) (\partial^\alpha \phi)\\
& \quad - \Big( L_{AB}- \frac{L_{0A} L_{0B}}{L_{00}} \Big) \, (\partial_\alpha \phi^A)(\partial^\alpha \phi^B) - \big[ \, L^{AB} (\partial_A \widetilde{W})(\partial_B \widetilde{W})\\
& \quad  + 4L^{0A} L_{00} \widetilde{W} (\partial_A \widetilde{W}) + 4(L^{00}L_{00}^2+3L_{00}+1) \widetilde{W}^2 \, \big]\Big \} \,.
%
\end{split}
\end{equation}
Surprisingly, we have that
\begin{equation}
\Big( \, L_{AB}- \frac{L_{0A}L_{0B}}{L_{00}} \, \Big) \, L^{BC}={\delta_A}^C \,,
\end{equation}
and as a consequence we can perform the following identifications
\begin{subequations}
\begin{align}
&L_{00} = \frac{1}{18} \,, \\
&L_{AB}- \frac{L_{0A}L_{0B}}{L_{00}} = {\cal G}_{AB} \,, \\
&L^{AB} = {\cal G}^{AB} \,, \\
&L^{0A}L_{00} = {\cal G}^{AB}K_B \,, \\
&4(L^{00}L_{00}^2+3 L_{00}+1) = {\cal G}^{AB}K_A K_B \,.
\end{align}
\end{subequations}
With these identifications, both actions are seen to coincide. The remaining kinetic terms and the Chern-Simons terms that were left in the actions \eqref{02-bosons} and \eqref{02-non-zero-flux-action} can be easily identified and we conclude that the compactified action is in perfect agreement with the general supergravity action. Therefore we can conclude that M-theory compactified on manifolds with Spin(7) holonomy produces a low energy effective action that corresponds to a particular case of the minimal three dimensional supergravity coupled with matter. Hence under particular conditions we can obtain supersymmetric solutions. In the next section we address this issue.


\section{The Internal Flux and Supersymmetry Breaking}
\label{03-susy-solutions}

\setcounter{equation}{0}
\renewcommand{\theequation}{\thesection.\arabic{equation}}


We have seen that the internal flux is responsible for generating a scalar potential for some of the moduli fields. Equation \eqref{scalar-pot-int-flux} shows clearly that only the anti-self-dual part of $\hat{F}_2$ is responsible for the emergence of $V$. The self-dual part is dynamic in nature and we have seen in section \ref{03-eq-motion} that this component of the internal flux is the solution of the equations of motion. Let us investigate under what conditions the solution is supersymmetric.

Since the expressions in which the internal flux appears are of order $t^{-4}$ or higher our analysis will be performed on an undeformed and unwarped background as we have previously considered in section \ref{03-quartic-polynomials}. In other words to $t^{-4}$ order in perturbation theory we do not have equations which involve both the internal flux and the corrections due to the warping or due to the deformation of the internal manifold. As a consequence in this section the manifold $M_8$ is considered to be a Spin(7) holonomy manifold. Under these circumstances it is very easy to select the supersymmetric solutions. We want to emphasize that our discussion involves the leading order perturbative expansion coefficients of the Cayley calibration $\Omega^{(2)}$ and the one for the internal flux $F^{(0)}$. However, the discussion is not influenced by this fact because the perturbative expansion of the fields does not affect their representation, i.e., all the expansion coefficients will belong to the same representation as the original field. Our analysis is based only on the fields representations therefore in what follows we drop the upper index which denote the order in perturbation theory and the discussion will make no distinction between the full quantities $F$ and $\Omega$ and their leading contributions $F^{(0)}$ and $\Omega^{(2)}$, respectively.

The constraints imposed to the internal flux in order to obtain a supersymmetric vacua in three dimensions have been determined in \cite{Becker:2000jc}. Later on in \cite{Acharya:2002vs} these constraints have been derived from certain conditions imposed to the superpotential $W$ whose expression was conjectured in \cite{Gukov:1999gr} to be
\begin{equation}
\label{03-superpotential}
W=\int_{M_8} F \wedge \Omega \,.
\end{equation}
We have shown in section \ref{01-gravitino} that \eqref{03-superpotential} is indeed the superpotential of the theory and we have proved in section \ref{02-moduli} that the scalar potential of the effective three-di\-men\-sio\-nal theory is
\begin{equation}
\label{03-scalar-potential}
V[W] = \sum_{A,B=1}^{b^4_{{\bf 35}^-}} {\cal G}^{AB} \, D_AW \, D_BW \,,
\end{equation}
where ${\cal G}^{AB}$ is the inverse matrix of ${\cal G}_{AB}$, which was defined in \eqref{02-moduli-metric-l2}. We have denoted by $b^4_{{\bf 35}^-}$ the refined Betti number which represents the number of anti-self-dual harmonic four-forms $\xi_A$. The covariant derivative $D_A$ was defined in \eqref{02-D-operator} through its action on the Cayley calibration
\begin{equation}
D_A \, \Omega =\xi_A \,.
\end{equation}
As we have discussed in section \ref{02-moduli} the minimum of the scalar potential \eqref{03-scalar-potential} is zero due to its quadratic expression. Therefore an $AdS_3$ solution is excluded and a three-di\-men\-sio\-nal supersymmetric effective theory is obtained only when the scalar potential vanishes, i.e., for a Minkowski background. In other words supersymmetry requires
\begin{equation}
\label{03-susy-condition-1}
D_A W=0 \qquad A=1,\ldots, b^4_{{\bf 35}^-} \,.
\end{equation}
Unlike for the four-di\-men\-sio\-nal minimal supergravity case we do not have to impose $W=0$ as well in order to obtain $V=0$\footnote{The scalar potential in the four-di\-men\-sio\-nal case contains a term proportional to $W^2$.}. However having set the Minkowski background in the three-di\-men\-sio\-nal theory we are left with no choice and we must impose $W=0$. This is because the three-di\-men\-sio\-nal scalar curvature is proportional to the superpotential\footnote{See the discussion at the end of the section \ref{02-component-formalism} for more details.}. To summarize, the conditions that $W$ has to satisfy in order to have a supersymmetric vacua are
\begin{equation}
\label{03-susy-conditions}
W = 0 \quad \mathrm{and} \quad {D_A W}=0 \,,
\end{equation}
where $D_A W$ indicates the covariant derivative of $W$ with respect to the moduli fields which correspond to the metric deformations of the Spin(7) holonomy manifold. Therefore if we want to break the supersymmetry of the effective three-di\-men\-sio\-nal theory, all we have to do is to impose
\begin{equation}
\label{03-non-susy-conditions}
W \neq 0 \quad \mathrm{or} \quad D_A W \neq 0 \,,
\end{equation}
and as we will see below the only way we can break the supersymmetry is through the first condition in \eqref{03-non-susy-conditions} because the second condition in \eqref{03-susy-conditions} is always valid. It is natural to have $D_A W = 0$, i.e. a Minkowski background, because this is the only solution found in section \ref{03-eq-motion}. Since the general solution that emerges from our analysis invalidates the second condition in \eqref{03-non-susy-conditions} from the beginning, the only way we can break the supersymmetry is to have a non-vanishing value for the vacuum expectation value of the superpotential $W$. It is obvious that once the supersymmetry is broken the relation that exists between $W$ and $D_A W$ is no longer valid and we can have, for example, $D_A W = 0$ and $W \ne 0$ without generating any inconsistencies. In other words the superpotential is no longer proportional to the three-di\-men\-sio\-nal scalar curvature for a non-supersymmetric theory. Having said that let us see what constraints we should impose on the internal background flux in order to obtain a supersymmetric solution.

All the fields on $M_8$ form representations of the Riemannian holonomy group Spin(7). In particular, the space of differential forms on $M_8$ can be decomposed into irreducible representations of Spin(7) and because the Laplace operator preserves this decomposition the de Rham cohomology groups have a similar decomposition into smaller pieces. Here we are interested in the four-form internal flux of the field strength of M-theory, therefore we need the decomposition of the fourth cohomology group of $M_8$
\begin{equation}
\label{03-general-decomp}
H^4(M_8, \IR)=H_{{\bf 1}^+}^4(M_8, \IR) \, {\oplus} \, H^4_{{\bf 7}^+}(M_8, \IR) \, {\oplus} \, H_{{\bf 27}^+}^4(M_8, \IR) \, {\oplus} \, H_{{\bf 35}^-}^4(M_8, \IR) \,.
\end{equation}
In the above expression the numerical sub-index represents the dimensionality of the representation and the $\pm$ stands for a self-dual or anti-self-dual representation. We denote by $b^n_{\bf r}$ the refined Betti number which represents the dimension of $H_{{\bf r}}^n(M_8, \IR)$. It is shown in \cite{Joyce} that for a compact manifolds which has Spin(7) holonomy $b^4_{{\bf 7}^+} =0$, hence the decomposition \eqref{03-general-decomp} becomes
\begin{equation}
H^4(M_8, \IR)=H_{{\bf 1}^+}^4(M_8, \IR) \, {\oplus} \, H_{{\bf 27}^+}^4(M_8, \IR) \, {\oplus} \, H_{{\bf 35}^-}^4(M_8, \IR) \,.
\end{equation}
Therefore on a compact Spin(7) holonomy manifold the internal flux can have three pieces
\begin{equation}
\label{03-F-general}
\hat{F}_2=F_{{\bf 1}^+} \oplus F_{{\bf 27}^+} \oplus F_{{\bf 35}^-} \,.
\end{equation}
However we have showed in section \ref{03-eq-motion} that the most general solution which emerges from the equations of motion has to be self-dual, whereas the anti-self-dual piece generates the scalar potential as shown in section \ref{02-moduli}. Therefore the dynamical component of the internal flux is $F_{{\bf 1}^+} \oplus F_{{\bf 27}^+}$. The non-dynamical component vanishes, i.e. $F_{{\bf 35}^-} = 0$, because the scalar potential vanishes. Although not obvious, the vanishing of the $F_{{\bf 35}^-}$ piece is related to the second set of equations in \eqref{03-susy-conditions}. To see this let us rewrite $D_A W$ using the definition \eqref{03-superpotential}
\begin{equation}
D_A W = \int_{M_8} D_A \Omega \wedge \hat{F}_2 \,.
\end{equation}
The variation of the Cayley calibration with respect to the metric moduli belongs to $H_{{\bf 35}^-}^4(M_8, \IR)$ and therefore
\begin{equation}
D_A W = \int_{M_8} \xi_A \wedge F_{{\bf 35}^-} \,,
\end{equation}
hence $D_A W$ vanishes when $F_{{\bf 35}^-}$ vanishes
\begin{equation}
F_{{\bf 35}^-} = 0 \quad \Rightarrow \quad D_A W = 0 \,.
\end{equation}
In other words the general solution for the internal flux precludes a non-vanishing cosmological constant in the effective three-di\-men\-sio\-nal theory.

Regarding the first condition in \eqref{03-susy-conditions} we can easily see that it is satisfied as long as $F_{{\bf 1}^+}=0$ because the Cayley calibration belongs to $H_{{\bf 1}^+}^4(M_8, \IR)$ and therefore we have that
\begin{equation}
W=\int_{M_8} \Omega \wedge \hat{F}_2 = \int_{M_8} \Omega \wedge F_{{\bf 1}^+} \,,
\end{equation}
hence
\begin{equation}
F_{{\bf 1}^+} = 0 \quad \Rightarrow \quad W = 0 \,.
\end{equation}
This result means that the only piece from the internal flux that accommodates a supersymmetric vacuum is
\begin{equation}
\label{03-susy-flux}
\hat{F}_2 = F_{{\bf 27}^+} \,.
\end{equation}
Due to the fact that we obtain from the equations of motion that the internal flux has to be self-dual and the external space is Minkowski, i.e. the $F_{{\bf 35}^-}$ piece is identically zero, the only way we can break supersymmetry is by turning $F_{{\bf 1}^+}$ on in such a way that we obtain a non-vanishing value for the superpotential. It is interesting to note that breaking the supersymmetry in this way does not affect the value of the cosmological constant which remains zero. Such an interesting scenario with a vanishing cosmological constant and broken supersymmetry has already appeared in a number of different contexts \cite{Gukov:2002iq, Haack:2001jz, Giddings:2001yu, Becker:2001pm} and \cite{Becker:2002nn}. However, in contrast to the superpotentials appearing in the previous references, it is expected that the superpotential \eqref{01-axi} receives perturbative and non-perturbative quantum corrections. For an analysis of some aspects of these corrections see \cite{Acharya:2002vs}. This completes our discussion about M-theory compactifications on Spin(7) holonomy manifolds.



\chapter[Conclusions]{CONCLUSIONS}
\label{chapter-conclusions}

\setcounter{equation}{0}
\renewcommand{\theequation}{\thechapter.\arabic{equation}}


In this thesis we have analyzed the properties of the most general warped vacua which emerge from a compactification with flux of M-theory on Spin(7) holonomy manifolds. More specifically we have looked at the conjecture made in \cite{Gukov:1999gr} for the superpotential which arises in such compactifications. We have also computed the scalar potential generated by the internal flux and we have determined the conditions imposed on the flux by a supersymmetric solution.

The existence of the quantum corrections terms in the low energy effective action of M-theory forced us to perform in chapter \ref{chapter-vacua} a perturbative analysis of the problem. The perturbative parameter ``$t$'' was defined in \eqref{03-t-def}. We have determined that a consistent solution of the equations of motion requires the self-duality of the leading order term of the internal flux. We have also shown that the internal manifold remains Ricci flat to the $t^{-2}$ order in the perturbation theory, as shown in \eqref{03-Ricci-t2}. By analyzing \eqref{03-i-t3} we have shown that the Ricci flatness of the internal manifold is in general lost to order $t^{-3}$. However, imposing the restrictive condition \eqref{03-R-flat-condition} the internal manifold remains Ricci flat to this order as well. As a matter of fact this is the order in the perturbation theory where the influence of the quantum corrections terms is felt in the equations of motion and it is natural to expect deformations of the internal manifold to occur at this order. We have also derived a relation between the warp factor ``$A$'' and the external flux given by \eqref{03-warp-eflux}. We have collected some of the properties related to the quartic polynomials in section \ref{03-quartic-polynomials}. In particular in section \ref{03-j0-computation} we have shown that $J_0$ vanishes on a Spin(7) background and we have computed its first variation on a Spin(7) holonomy background. We have also determined a nice formula \eqref{03-Y-trace} for the trace of the first variation of $J_0$.

The results obtained in chapter \ref{chapter-vacua} helped us to understand what approximations and what kind of ansatz have to be employed in the compactification procedure for the background metric. In chapter \ref{chapter-compactification} we have performed the Kaluza-Klein compactification of M-theory on a Spin(7) holonomy manifold with and without fluxes. When fluxes are included, we generate a scalar potential for moduli fields. The main results from chapter \ref{chapter-compactification} are the formula that gives the three-dimensional low energy effective action \ref{02-non-zero-flux-action} and the expression \ref{02-potential} of the scalar potential generated by the background flux of the field strength.

The analysis performed in chapter \ref{chapter-superpotential} uses information about the form and the structure of minimal supergravity in three dimensions. Therefore in chapter \ref{chapter-3dsugra} we have derived the general form of 3D, ${\cal N}=1$ supergravity coupled to matter. The off-shell component action is the sum of \eqref{02-compact1} and \eqref{02-compact2}. In addition the on-shell bosonic action is given in \eqref{02-bosons}. The supersymmetry variation of the gravitino, \eqref{02-gavidelta}, was shown to be proportional to the superpotential. The latter statement was an important ingredient in order to check the form of the superpotential for compactifications of M-theory on Spin(7) holonomy manifolds conjectured in \cite{Gukov:1999gr}.

Chapter \ref{chapter-superpotential} contains various analyses related to the compactified theory. In section \ref{01-gravitino} we have checked the conjecture made in \cite{Gukov:1999gr} regarding the form of the superpotential, which is induced when non-trivial fluxes are turned on in different types of string theory and M-theory compactifications. We have accomplished this task by performing a Kaluza-Klein reduction of the gravitino supersymmetry transformation. In section \ref{01-spin} we have considered warped compactifications of M-theory on Spin(7) holonomy manifolds. Since the gravitino supersymmetry transformation contains a term proportional to $W$, we were able to verify the conjecture of \cite{Gukov:1999gr} by a direct calculation of the superpotential. As it is well known from \cite{Dine:1985rz}, a compactification of the heterotic string on a Calabi-Yau three-fold leads to a superpotential, which breaks the supersymmetry completely. We have checked that this superpotential can be written in the form \eqref{01-eivxiv}, which extends the conjecture made in \cite{Gukov:1999gr} to fluxes of Neveu-Schwarz type \cite{Behrndt:2000zh}. In section \ref{02-moduli} we have showed that the scalar potential can be expressed in terms of the superpotential. Interestingly, formula \eqref{02-scalar-potential} shows that the potential is a perfect square, hence only compactifications to three-di\-men\-sio\-nal Minkowski space can be obtained in agreement with \cite{Acharya:2002vs}. It is plausible that non-perturbative effects will modify this result to three-di\-men\-sio\-nal de-Sitter space along the lines of \cite{Kachru:2003aw}. This will be an interesting question for the future. In section \ref{03-susy-solutions} we have determined the condition \eqref{03-susy-flux} which has to be satisfied by the internal flux in order to obtain a supersymmetric solution. The analysis was based on the set of conditions \eqref{03-susy-conditions} that were imposed to the superpotential. We have shown the existence of solutions to the three-di\-men\-sio\-nal equations of motion, which break supersymmetry and have a vanishing three-di\-men\-sio\-nal cosmological constant. Such an interesting scenario has recently appeared many times in the literature.

Contrary to the superpotential appearing in compactifications of M-theory on Calabi-Yau four-folds, it is known that this ${\cal N}=1$ superpotential receives perturbative and non-perturbative quantum corrections \cite{Acharya:2002vs}. The computation of these corrections along the lines of \cite{Becker:2002nn} represents an interesting open question.



\appendix


\newsavebox{\symmtii}
\savebox{\symmtii}{
\begin{picture}(16,8)
\put(0,0){\line(1,0){16}}
\put(0,8){\line(1,0){16}}
\put(0,0){\line(0,1){8}}
\put(8,0){\line(0,1){8}}
\put(16,0){\line(0,1){8}}
\end{picture}}


\newsavebox{\symmtiii}
\savebox{\symmtiii}{
\begin{picture}(24,8)
\put(0,0){\line(1,0){24}}
\put(0,8){\line(1,0){24}}
\put(0,0){\line(0,1){8}}
\put(8,0){\line(0,1){8}}
\put(16,0){\line(0,1){8}}
\put(24,0){\line(0,1){8}}
\end{picture}}


\newsavebox{\symmtiv}
\savebox{\symmtiv}{
\begin{picture}(16,16)
\put(0,0){\line(1,0){16}}
\put(0,8){\line(1,0){16}}
\put(0,16){\line(1,0){16}}
\put(0,0){\line(0,1){16}}
\put(8,0){\line(0,1){16}}
\put(16,0){\line(0,1){16}}
\end{picture}}


\newsavebox{\corner}
\savebox{\corner}{
\begin{picture}(16,16)
\put(0,0){\line(1,0){8}}
\put(0,8){\line(1,0){16}}
\put(0,16){\line(1,0){16}}
\put(0,0){\line(0,1){16}}
\put(8,0){\line(0,1){16}}
\put(16,8){\line(0,1){8}}
\end{picture}}


\newsavebox{\gun}
\savebox{\gun}{
\begin{picture}(24,16)
\put(0,0){\line(1,0){8}}
\put(0,8){\line(1,0){24}}
\put(0,16){\line(1,0){24}}
\put(0,0){\line(0,1){16}}
\put(8,0){\line(0,1){16}}
\put(16,8){\line(0,1){8}}
\put(24,8){\line(0,1){8}}
\end{picture}}



\chapter[Conventions, Identities and Derivations]{CONVENTIONS, IDENTITIES AND DERIVATIONS}
\label{app-compact}

\setcounter{equation}{0}
\renewcommand{\theequation}{\thechapter.\arabic{equation}}


\section{Conventions and Useful Identities}
\label{03-conventions}

\setcounter{equation}{0}
\renewcommand{\theequation}{\thesection.\arabic{equation}}


This appendix contains the conventions and the main formulas used in the computations related to the Kaluza-Klein procedure. In what follows we present some conventions related to the Levi-Civita tensor density, some algebraic identities which involve generalized Kronecker delta and Levi-Civita symbols, and a few useful $\Gamma$-matrix identities. We choose to have
\begin{equation}
\varepsilon^{1 \ldots n} = \, 1 \,,
\end{equation}
and because the covariant tensor density $\varepsilon_{b_1 \ldots b_n}$ is obtained from $\varepsilon^{a_1 \ldots a_n}$ by lowering the indices with the help of the metric coefficients $g_{ab}$, we will have that
\begin{equation}
\varepsilon_{1 \ldots n} = \, g \,,
\end{equation}
where $g=det(g_{mn})\,$. Therefore the product of two Levi-Civita symbols can be reexpressed as
\begin{equation}
\label{03-e-e}
\varepsilon^{a_1 \ldots a_n} \, \varepsilon_{b_1 \ldots b_n} = g \, \delta^{a_1 \ldots a_n}_{b_1 \ldots b_n} \,.
\end{equation}
The generalized Kronecker delta symbol which has appeared in \eqref{03-e-e} is defined as
\begin{equation}
\delta^{a_1 \ldots a_n}_{b_1 \ldots b_n} = n! \delta^{a_1}_{[b_1} \ldots \delta^{a_n}_{b_n]} \,,
\end{equation}
where the antisymmetrization implies a $1/n!$ pre-factor, e.g.,
\begin{equation}
\delta^a_{[m} \delta^b_{n]} = \, \frac{1}{2!} \, ( \, \delta^a_m \delta^b_n - \delta^a_n \delta^b_m \, ) \,.
\end{equation}
For a single contraction of a $(p+1)$-delta symbol in an $n$-di\-men\-sio\-nal space we have
\begin{equation}
\delta^{a_1 \ldots a_p m}_{b_1 \ldots b_p m} = (n-p) \, \delta^{a_1 \ldots a_p}_{b_1 \ldots b_p} \,,
\end{equation}
therefore in an $n$-di\-men\-sio\-nal space a $p$-delta symbol is related to an $n$-delta symbol as follows
\begin{equation}
\delta^{a_1 \ldots a_p m_1 \ldots m_{n-p}}_{b_1 \ldots b_p m_1 \ldots m_{n-p}} = (n-p)! \, \delta^{a_1 \ldots a_p}_{b_1 \ldots b_p} \,.
\end{equation}

As above, the antisymmetrization of two or more gamma matrices implies a  $1/n!$ pre-factor, e.g.,
\begin{equation}
\Gamma_{mn} = \Gamma_{[m} \Gamma_{n]} = \frac{1}{2!} \, ( \, \Gamma_m\Gamma_n - \Gamma_n\Gamma_m \, ) \,.
\end{equation}
Using the fundamental relation
\begin{equation}
\{\Gamma_m, \Gamma^n\} = 2 \delta_m^n \,,
\end{equation}
one can deduce the following gamma matrix identities
\begin{subequations}
\label{03-gamma-id1}
\begin{align}
&[\Gamma_m, \Gamma^r] =2{\Gamma_m}^r \,, \\
&\{\Gamma_{mn}, \Gamma^r \} = {2\Gamma_{mn}}^r \,, \\
&[\Gamma_{mnp}, \Gamma^r ] =2{ \Gamma_{mnp}}^r \,,
\end{align}
\end{subequations}
and
\begin{subequations}
\label{03-gamma-id2}
\begin{align}
&\{ \Gamma_m, \Gamma^r\} = 2{\delta_m}^r \,, \\
&[\Gamma_{mn} ,\Gamma^r ] = -4{\delta^r}_{[m} \Gamma_{n]} \,, \\
&\{ \Gamma_{mnp}, \Gamma^r\} = 6 {\delta^r}_{[m} \Gamma_{np]} \,.
\end{align}
\end{subequations}

In what follows we present our conventions regarding differential forms. If $\alpha_p$ is a differential form of order $p$, or an $p$-form, then its expansion in components is given by
\begin{equation}
\alpha_p = \frac{1}{p!}\: \alpha_{m_1,\ldots,m_p} \: dx^{m_1}\wedge
\ldots \wedge dx^{m_p}\,.
\end{equation}
Let us consider the wedge product between a $p$-form $\alpha_p$ and a $q$-form $\beta_q$. $\alpha_p \wedge \beta_q$ is a $(p+q)$-form, so
\begin{equation}
\alpha_p \wedge \beta_q = \frac{1}{\left(p+q\right)!} \: (\alpha_p
\wedge \beta_q)_{m_1,\ldots,m_{p+q}} \: dx^{m_1}\wedge \ldots
\wedge dx^{m_{p+q}}\,.
\end{equation}
On the other hand, by definition
\begin{equation}
\alpha_p \wedge \beta_q = \frac{1}{p!\, q!} \:
\alpha_{[m_1,\ldots,m_p} \, \beta_{m_{p+1},\ldots,m_{p+q}]} \:
dx^{m_1}\wedge \ldots \wedge dx^{m_{p+q}}\,,
\end{equation}
therefore
\begin{equation}
\label{02-3.7} \left( \alpha_p \wedge \beta_q
\right)_{m_1,\ldots,m_{p+q}} = \frac{\left(p+q\right)!}{p! \, q!}
\: \alpha_{[m_1,\ldots,m_p} \, \beta_{m_{p+1},\ldots,m_{p+q}]}\,.
\end{equation}
The definition for the exterior derivation is
\begin{equation}
d\alpha_p= \frac{1}{p!} \: \partial_{[m_1} \,
\alpha_{m_2,\ldots,m_{p+1}]} \: dx^{m_1}\wedge \ldots \wedge
dx^{m_{p+1}}\,.
\end{equation}
Since $d\alpha_p$ is a $(p+1)$-form
\begin{equation}
d\alpha_p= \frac{1}{\left(p+1\right)!} \:
(d\alpha_p)_{m_1,\ldots,m_{p+1}} \: dx^{m_1}\wedge \ldots \wedge
dx^{m_p}\,,
\end{equation}
we have that
\begin{equation}
\label{02-3.13} \left( d\alpha_p \right)_{m_1,\ldots,m_{p+1}}= (p+1)
\:\partial_{[m_1}\, \alpha_{m_2,\ldots,m_{p+1}]}\,.
\end{equation}
The Hodge $\star$ operator of some $p\,$-form on a real $n\,$-di\-men\-sio\-nal manifold is defined as
\begin{equation}
\begin{split}
\star \, \alpha_p &= \frac{\sqrt{g}}{p!(n-p)!} \, \alpha_{k_1 \ldots k_p} \, g^{k_1 m_1} \ldots g^{k_p m_p} \, \varepsilon_{m_1 \ldots m_p m_{p+1} \ldots m_n} \\
& \quad \cdot dx^{m_{p+1}} \wedge \ldots \wedge dx^{m_n}\,,
\end{split}
\end{equation}
where
\begin{equation}
\varepsilon_{1 \ldots n} \,=\, +\,1\,.
\end{equation}
Regarding the integration of some $p\,$-form $\alpha_p$ on a $p\,$-cycle ${\cal{C}}_p$ we have that
\begin{equation}
\int_{{\cal{C}}_p} \alpha_p = \frac{1}{p!} \int_{{\cal{C}}_p}
\alpha_{m_1 \ldots m_p} dx^{m_1} \wedge \ldots \wedge dx^{m_p}\,.
\end{equation}
We can also introduce an inner product on the space of real $p\,$-forms defined on a $n\,$-di\-men\-sio\-nal manifold $\cal{M}$
\begin{equation}
\label{02-inner-product} \langle \alpha_p,\,\beta_p\rangle =
\int_{\cal{M}}\alpha_p \wedge \star \beta_p = \frac{1}{p!} \,
\int_{\cal{M}} \alpha_{m_1 \ldots m_p} \, \beta^{m_1 \ldots m_p}
\, \sqrt{g} \, dx^1 \wedge \ldots \wedge dx^n \,.
\end{equation}

We end this appendix with the derivations of some important formulas used in the computations performed in section \ref{03-j0-computation}. We also list some other useful identities providing the appropriate references for detailed explanations.

In general one has
\begin{equation}
[\nabla_m, \nabla_n]\eta=\frac{1}{4} R_{mnpq} \Gamma^{pq} \eta \,,
\end{equation}
therefore if $\eta$ is a Killing spinor then $\nabla_m \eta =0$ and we obtain the integrability condition
\begin{equation}
\label{03-integrability-condition}
R_{abmn} \Gamma^{mn} \eta= 0 \,.
\end{equation}
If we multiply \eqref{03-integrability-condition} from the left with $\overline{\eta}\Gamma_{cd}$ we obtain
\begin{equation}
\label{03-id0}
R_{abmn} \overline{\eta}\Gamma_{cd}\Gamma^{mn} \eta= 0 \,.
\end{equation}
Using the identities \eqref{03-gamma-id1} and \eqref{03-gamma-id2}, we can show that
\begin{equation}
\label{03-id1}
\Gamma^{a} \Gamma_{mn} = {\Gamma^{a}}_{mn} + \delta^a_m \Gamma_n - \delta^a_n \Gamma_m \,,
\end{equation}
and
\begin{equation}
\label{03-id2}
\Gamma^{ab} \Gamma_{mn} = {\Gamma^{ab}}_{mn} - \delta^{ab}_{mn} - 4 \delta^{[a}_{[m} {\Gamma^{b]}}_{n]} \,.
\end{equation}
If we sandwich the relation \eqref{03-id2} between $\overline{\eta}$ and $\eta$ we obtain that
\begin{equation}
\label{03-id3}
\overline{\eta} \Gamma^{ab} \Gamma_{mn} \eta= {\Omega^{ab}}_{mn} - \delta^{ab}_{mn} \,,
\end{equation}
where
\begin{equation}
\Omega_{abmn} = \overline{\eta} \Gamma_{abmn} \eta
\end{equation}
is the Cayley calibration of the Spin(7) holonomy manifold and the Killing spinor is normalized to unity, i.e., $\overline{\eta} \eta= 1$. We remind the reader that for a Spin(7) holonomy manifold, terms like $\overline{\eta} \Gamma_{m_1 \ldots m_p} \eta$ are not zero only when $p=0,4$ or $8$. For details see reference \cite{Gibbons:1990er}. This is the reason why we have no contribution from the last term in \eqref{03-id2}. Using \eqref{03-id3} we can recast \eqref{03-id0} as
\begin{equation}
\label{03-id4}
R_{abmn} {\Omega^{mn}}_{cd} = 2 R_{abcd} \,.
\end{equation}

We have the following one index contraction between two Cayley calibrations (see for example \cite{Dundarer:1984fe} and \cite{deWit:1984gs})
\begin{equation}
\label{03-omega-contraction-1}
\Omega^{tabc} \Omega_{tmnp} = 6 \delta^a_{[m} \delta^b_n \delta^c_{p]} + 9 \delta^{[a}_{[m} {\Omega^{bc]}}_{np]}\,.
\end{equation}

We use the following expression for the variation of the Riemann tensor in terms of the metric fluctuations
\begin{equation}
\label{03-Riemann-variation}
\delta R_{abmn} = - \nabla_{[a|} \nabla_m \delta g_{|b]n} + \nabla_{[a|} \nabla_n \delta g_{|b]m} \,.
\end{equation}
The above result can be easily derived using the relation which exists between the derivative operators associated with two conformally related metrics.


\section{The Inverse Metric and Other Derivations}
\label{03-scaling}

\setcounter{equation}{0}
\renewcommand{\theequation}{\thesection.\arabic{equation}}


In this appendix we derive the power expansion in $t$ for the inverse metric and we provide some useful relations used in the analysis of section \ref{03-eq-motion}. We start with the derivation of the inverse internal metric $g^{mn}$ followed naturally by the expansions for the Riemann tensor, the Ricci tensor and the scalar curvature that correspond to $g_{mn}$ which has Spin(7) holonomy. Once we know this expansions we can perform a conformal transformation to find the corresponding tensorial quantities for the internal manifold\footnote{We have to take into account that the full metric \eqref{03-full-metric} is warped.}. Also at the end of this appendix we provide the expressions for the external and internal e\-ner\-gy-mo\-men\-tum tensor associated with $F_1$ and $F_2$, respectively, and we list the results obtained for the term in the right hand side of \eqref{03-eom-g} for the external and the internal cases.

Let us consider two arbitrary square matrices $A$ and $B$ with real entries\footnote{The reader should not confuse these matrices with the warp factors.}. We want an expression for $(A + B)^{-1}$ in terms of $A$, $A^{-1}$, $B$ and $B^{-1}$. While there is no useful formula for $(A + B)^{-1}$, we can use a Neumann series to invert $A + B$ provided that $B$, for example, has small entries relative to $A$. This means that in magnitude we have
\begin{equation}
\lim_{n \, \rightarrow \, \infty} (A^{-1} \, B)^n = 0 \,.
\end{equation}
Under this assumption the inverse of $A+B$ matrix can be expressed as an infinite series
\begin{equation}
(A + B)^{-1} = \sum_{k=0}^\infty (-A^{-1}B)^k \, A^{-1} \,,
\end{equation}
which in a first approximation is given by
\begin{equation}
\label{03-inverse-sum}
(A + B)^{-1} = A^{-1} - A^{-1}BA^{-1} + \ldots \,.
\end{equation}
The above setup helps us to compute the inverse of the matrix $g_{mn}$ introduced in \eqref{03-metric-exp}. If we set $A_{mn}=t \, [g^{(1)}]_{mn}$ and $B_{mn}=[g^{(0)}]_{mn}$,  which is the ``small'' matrix\footnote{The perturbative parameter ``$t$'' was introduced in \eqref{03-t-def} and in the large volume limit ``t'' is much bigger than the unity.}, then the formula \eqref{03-inverse-sum} translates into
\begin{equation}
g^{mn} = t^{-1} [g^{(1)}]^{mn} + t^{-2} [g^{(2)}]^{mn} + \ldots \,,
\end{equation}
where we have defined
\begin{subequations}
\begin{align}
&[g^{(1)}]^{mn} = [g^{(1)}]^{-1}_{mn} \,, \\
&{[g^{(2)}]}^{mn} = - [g^{(1)}]^{mp} [g^{(0)}]_{pr} [g^{(1)}]^{rn} \,,
\end{align}
\end{subequations}
and as usual $g^{mn}$ represents the inverse of $g_{mn}$.

By performing the appropriate conformal transformations we obtain the internal and external components of the e\-le\-ven-di\-men\-sio\-nal Ricci tensor and the e\-le\-ven-di\-men\-sio\-nal Ricci scalar that correspond to the metric \eqref{03-full-metric}. The results are provided in terms of the un-warped quantities, denoted here without a tilde above the symbol
\begin{equation}
\label{03-Ricci-tensor-external}
\begin{split}
\widetilde{R}_{\mu \nu}(M_{11}) & =  R_{\mu \nu}(M_3) - \eta_{\mu \nu} \, e^{2(A-B)} [ \, \triangle A +3 (\nabla_m A) \, (\nabla^m A) \\
& \quad + 6 (\nabla_m A) \, (\nabla^m B) \,] \,,
\end{split}
\end{equation}
\begin{equation}
\label{03-Ricci-tensor-internal}
\begin{split}
\widetilde{R}_{mn}(M_{11}) & =  R_{mn}(M_8) - 3 \nabla_m \nabla_n A - 6 \nabla_m \nabla_n B - g_{mn} \triangle B \\
& \quad - 3 (\nabla_m A) \, (\nabla_n A) - 3 g_{mn} (\nabla_k A) \, (\nabla^k A) \\
& \quad + \, 6 (\nabla_m B) \, (\nabla_n B) - 6 g_{mn} (\nabla_k B) \, (\nabla^k B) \\
& \quad + \, 6 \, \nabla_{(m} A \, \nabla_{n)} B \,,
\end{split}
\end{equation}
\begin{equation}
\label{03-scalar-curvature}
\begin{split}
\widetilde{R}(M_{11}) & =  e^{-2A} R(M_3) + e^{-2B} R(M_8) - e^{-2B} [ \, 6 \triangle A + 14 \triangle B \\
& \quad + \, 12 (\nabla_m A) \, (\nabla^m A) + 42 (\nabla_m B) \, (\nabla^m B) \\
& \quad + 36 (\nabla_m A) \, (\nabla^m B) \,] \,,
\end{split}
\end{equation}
where $\triangle= \nabla^m \nabla_m$ is the internal Laplace operator. We mention that formulas (2.34) - (2.36) in \cite{Duff:1995an} represent a generalization of the above equations\footnote{The reader must be aware of a small typo in formula (2.35) of \cite{Duff:1995an} where the term $-d \, \partial_m A \, \partial_n B$ should read $-d \, \partial_m A \, \partial_n A$.}. Because the warp factors $A$ and $B$ are of order $t^{-3}$ we will truncate \eqref{03-Ricci-tensor-external}, \eqref{03-Ricci-tensor-internal} and \eqref{03-scalar-curvature} and we will retain only the linear contributions in $A$ and $B$. The ``linearized'' expressions are
\begin{subequations}
\begin{align}
&\widetilde{R}_{\mu \nu}(M_{11}) = R_{\mu \nu}(M_3) - \eta_{\mu \nu} \,  \triangle A + \ldots \,, \label{03-Ricci-tensor-external-lin} \\
&\widetilde{R}_{mn}(M_{11}) =  R_{mn}(M_8) - 3 \nabla_m \nabla_n A - 6 \nabla_m \nabla_n B - g_{mn} \triangle B + \ldots \,, \label{03-Ricci-tensor-internal-lin} \\
&\widetilde{R}(M_{11}) =  R(M_3) + R(M_8) -  6 \triangle A - 14 \triangle B + \ldots  \,. \label{03-scalar-curvature-lin}
\end{align}
\end{subequations}

Let us compute the external and the internal components of the e\-ner\-gy-mo\-men\-tum tensor associated with the field strength $F$. Because of the specific form \eqref{03-flux-form} of the background flux the e\-ner\-gy-mo\-men\-tum tensor defined in \eqref{03-en-mom-tensor} will have the following form
\begin{subequations}
\label{03-energy-momentum-ei}
\begin{align}
T_{\mu \nu} &= -3 \eta_{\mu \nu} (\nabla_m f ) \, (\nabla^m f ) - \tfrac{1}{8}\, \eta_{\mu \nu} F_{abmn} F^{abmn} + \ldots\,, \\
\begin{split}
T_{mn} &= - 6 (\nabla_m f ) \, (\nabla_n f ) + 3 g_{mn} (\nabla_p f ) \, (\nabla^p f ) \\
& \quad + F_{mabp} {F_n}^{abp} - \tfrac{1}{8} \,g_{mn} F_{abpr} F^{abpr} + \ldots\,,
\end{split}
\end{align}
\end{subequations}
where the ellipsis denotes higher order terms which contain warp factors. We are also interested in computing the trace of the above tensors
\begin{subequations}
\begin{align}
\begin{split}
\eta^{\mu \nu} \, T_{\mu \nu} &= -9 (\nabla_m f ) \, (\nabla^m f ) - \tfrac{3}{8} \, F_{abmn} F^{abmn} \\
&= - 9 \, [g^{(1)}]^{mn} [\nabla_m f^{(2)} ] \, [\nabla_n f^{(2)} ] \, t^{-5} \\
& \quad - \tfrac{3}{8} \, [F^{(0)}]_{abmn} [F^{(0)}]^{abmn} \, t^{-4} + \ldots \,,
\end{split}\\
g^{mn} \, T_{mn} &= 18 (\nabla_m f ) \, (\nabla^m f ) = 18 \, [g^{(1)}]^{mn} [\nabla_m f^{(2)} ] \, [\nabla_n f^{(2)} ] \, t^{-5} + \ldots \,,
\end{align}
\end{subequations}
where we have provided the leading order contribution of these terms. We want to note that trace of $T_{mn}$ vanishes to order $t^{-4}$ in the perturbation theory.

The external component of the left hand side of equation \eqref{03-eom-g} is
\begin{equation}
- \beta \frac{1}{\sqrt{-g}} \frac{\delta}{\delta \eta^{\mu \nu}} \left[ \sqrt{-g} ( J_0 - \frac{1}{2} E_8 ) \right] = \frac{\beta}{4} \, \eta_{\mu \nu} \, E_8(M_8) \,,
\end{equation}
where we have used \eqref{03-e8-property} and the fact that $J_0(M_{11})$ does not depend on the external metric and it vanishes on a Spin(7) holonomy background\footnote{See section \ref{03-j0-computation} for details.}. The internal component of the left hand side of equation \eqref{03-eom-g} is
\begin{equation}
- \beta \frac{1}{\sqrt{-g}} \frac{\delta}{\delta g^{mn}} \left[ \sqrt{-g} ( J_0 - \frac{1}{2} E_8 ) \right] =  - \beta \, \frac{\delta Y}{\delta g^{mn}}\,,
\end{equation}
where we have used equation \eqref{03-e8-variation} and $\delta Y / \delta g^{mn}$ is given in \eqref{03-Y-variation}. The trace of the above equation is
\begin{equation}
\label{03-trace-delJ-internal}
- \beta \frac{1}{\sqrt{-g}} \, g^{mn} \, \frac{\delta}{\delta g^{mn}} \left[ \sqrt{-g} ( J_0 - \frac{1}{2} E_8 ) \right] =  - 2^{17} \, \beta \, \triangle \, E_6(M_8) \,,
\end{equation}
where we have used \eqref{03-Y-trace} in its derivation. It is obvious that \eqref{03-trace-delJ-internal} is of order $t^{-4}$ in the perturbation theory.


\section{Dimensional Reduction of the Einstein-Hilbert Term}
\label{02-appendix-dimensional}

\setcounter{equation}{0}
\renewcommand{\theequation}{\thesection.\arabic{equation}}


In this appendix we present the technical details related to the compactification of the Einstein-Hilbert term. We treat first the zero flux case and then we calculate the reduction for the non-zero background flux case. As usual the Greek indices refer to the external space, the small Latin indices refer to the internal space, and finally the capital Latin indices refer to the entire eleven dimensional space.

We start with the following ansatz for the inverse metric
\begin{equation}
g^{mn}(x,y)= \hat{g}^{mn}(y) + \phi(x) \hat{g}^{mn}(y) + \sum_{A=1}^{b_4^-}\phi^A(x) ~ h^{mn}_A(x,y) + \ldots \,,
\end{equation}
where we have denoted by $g^{mn}(x,y)$ the inverse metric of $g_{mn}(x,y)$ and by $\hat{g}^{mn}(x)$ the inverse metric of $\hat{g}_{mn}(x)$
\begin{equation}
g_{mn}(x,y) \, g^{np}(x,y)={\delta_m}^p \,, \quad \hat{g}_{mn}(y) \, \hat{g}^{np}(y)={\delta_m}^p\,.
\end{equation}
Due to these facts we obtain that
\begin{equation}
h^{mn}_A(x,y)=-\hat{g}^{ma}(y) \, e_{A\,ab}(y) \, \hat{g}^{bn}(y)\,.
\end{equation}
The tracelessness of $e_{A\,ab}$ implies the tracelessness of $h^{mn}_A$. The ansatz \eqref{02-g-ansatz} implies that the only non-zero Christoffel symbols are
\begin{subequations}
\label{02-Christoffel}
\begin{align}
&\Gamma^\alpha_{\mu \nu} = \tfrac{1}{2} \, \eta^{\alpha \beta} \,(\partial_\mu \eta_{\beta \nu} + \partial_\nu \eta_{\mu \beta} - \partial_\beta \eta_{\mu \nu})\,, \\
&\Gamma^\alpha_{m \nu} = 0\,, \\
&\Gamma^\alpha_{mn} = -\, \tfrac{1}{2} \eta^{\alpha \beta} \,(\partial_\beta g_{mn})\,, \\
&\Gamma^a_{mn} = \tfrac{1}{2} g^{ab} \, (\partial_m g_{bn} + \partial_n g_{mb} - \partial_b g_{mn})\,, \\
&\Gamma^a_{\mu n} = \tfrac{1}{2} g^{ab} \, (\partial_\mu g_{bn})\,, \\
&\Gamma^a_{\mu \nu} = 0\,.
\end{align}
\end{subequations}
Using the following definition of the Ricci tensor
\begin{equation}
R_{MN}=\partial_A \Gamma^A_{MN} - \partial_N \Gamma^A_{MA}
+ \Gamma^A_{MN}\Gamma^B_{AB} - \Gamma^A_{MB}\Gamma^B_{AN}\,,
\end{equation}
we can derive the expression for the e\-le\-ven-di\-men\-sio\-nal Ricci scalar
\begin{equation}
\label{02-Ricci}
\begin{split}
R(M_{11}) &= R(M_3) + R(M_8) + g^{mn} \partial_\alpha \Gamma_{mn}^\alpha - \eta^{\mu \nu} \partial_\nu \Gamma^a_{\mu a} + \eta^{\mu \nu} \Gamma^\alpha_{\mu \nu} \Gamma^b_{\alpha b} \\
& \quad  - \left[ \eta^{\mu \nu} \Gamma_{\mu b}^a \Gamma_{a \nu}^b + g^{mn} \Gamma_{m \beta}^a \Gamma_{an}^\beta + g^{mn} \Gamma_{mb}^\alpha \Gamma_{\alpha n}^b \right] \\
& \quad + g^{mn} \Gamma^\alpha_{mn} \Gamma^\beta_{\alpha \beta} +\, g^{mn} \Gamma^\alpha_{mn} \Gamma^b_{\alpha b} \,.
\end{split}
\end{equation}
where $R(M_3)$ denotes the three-di\-men\-sio\-nal Ricci scalar and $R(M_8)$ is the eight-di\-men\-sio\-nal Ricci scalar. We can determine that
\begin{equation}
\label{02-integrate-Ricci}
\begin{split}
&\int_{M_{11}} d^{11}x \,\sqrt{-g_{11}} \, R(M_{11}) = \int_{M_{11}} d^{11}x \, \sqrt{-g_{11}} \, \Big\{ R(M_3) + \eta^{\mu \nu} \Gamma_{\mu a}^a \Gamma_{\nu b}^b \\
& \quad - (\partial_\alpha g^{mn}) \Gamma_{mn}^\alpha - \,\left[ \,\eta^{\mu \nu} \Gamma_{\mu b}^a \Gamma_{\nu a}^b +g^{mn} \Gamma_{\beta m}^a \Gamma_{an}^\beta + g^{mn} \Gamma_{mb}^\alpha \Gamma_{\alpha n}^b \,\right] \Big\} \,,
\end{split}
\end{equation}
where we have integrated by parts with respect to the internal coordinates and we have used the fact that the internal manifold is Ricci flat, i.e. $R(M_8)=0\,$. It is easy to see that we obtain the following results
\vspace*{-5ex}
\begin{subequations}
\begin{align}
\begin{split}
&\int_{M_{11}} d^{11}x \sqrt{-g_{11}} \, \eta^{\mu \nu} \, \Gamma_{\mu a}^a \, \Gamma_{\nu b}^b = 16 \, {\cal V}_{M_8} \int_{M_3} d^3x \, \sqrt{-\eta} \, (\partial_\alpha \phi) (\partial^\alpha \phi) \,,
\end{split}\\
\begin{split}
&\int_{M_{11}} d^{11}x \sqrt{-g_{11}} \, (\partial_\alpha g^{mn}) \Gamma_{mn}^\alpha = 4\, {\cal V}_{M_8} \int_{M_3} d^3 x \,\sqrt{-\eta}\, \Big\{ \,(\partial_\alpha \phi) \, (\partial^\alpha \phi) \\
& \quad + \frac{1}{2} \, \sum_{A,B=1}^{b_4^-} \, {\cal G}_{AB} \, (\partial_\alpha \phi^A) \, (\partial^\alpha \phi^B) \Big\}\,,
\end{split}\\
\begin{split}
&\int_{M_{11}} d^{11}x \sqrt{-g_{11}} \, \left[ \eta^{\mu \nu} \Gamma_{\mu b}^a \Gamma_{\nu a}^b + g^{mn} \Gamma_{\beta m}^a \Gamma_{an}^\beta + g^{mn} \Gamma_{mb}^\alpha \Gamma_{\alpha n}^b \right] \\
& = - {\cal V}_{M_8} \int_{M_3} d^3x \, \sqrt{-\eta}\,\Big\{ 2(\partial_\alpha \phi) \, (\partial^\alpha \phi) + \sum_{A,B=1}^{b_4^-} \, {\cal G}_{AB} (\partial_\alpha \phi^A) (\partial^\alpha \phi^B) \Big\} \,,
\end{split}
\end{align}
\end{subequations}
\vspace*{-4ex}
\hbox{\ }\\
where ${\cal G}_{AB}$ was defined in \eqref{02-moduli-metric-l2} and ${\cal V}_{M_8}$ represents the volume of the internal manifold and it is defined in \eqref{02-volume}.

We know that after compactification we arrive in the string frame even if we started in eleven dimensions in the Einstein frame. Therefore we have to perform a Weyl transformation for the external metric. The fact that we do not see any exponential of the radial modulus seating in front of $R(M_3)$ is because we have consistently neglected higher order contributions in moduli fields. However it is not difficult to realize that the Weyl transformation that has to be performed is
\begin{equation}
\eta_{\alpha \beta} \rightarrow e^{-8 \phi}~\eta_{\alpha \beta} \,.
\end{equation}
The only visible change in this order of approximation is the coefficient in front of the kinetic term for radion. All the other terms in the action remain unchanged. Therefore the Einstein-Hilbert term is
\begin{equation}
\label{02-Einstein-Hilbert}
\begin{split}
\frac{1}{2 \kappa_{11}^2} &\int_{M_{11}} d^{11}x \,\sqrt{-g_{11}}\, R(M_{11}) = \frac{1}{2\kappa_3^2} \, \int_{M_3} d^3x \,\sqrt{-\eta}\, \Big\{ R(M_3) \\
& \quad - 18(\partial_\alpha \phi) \, (\partial^\alpha \phi) - \sum_{A,B=1}^{b_4^-} \, {\cal G}_{AB} \,(\partial_\alpha \phi^A) \, (\partial^\alpha \phi^B)\Big\} + \ldots \,.
\end{split}
\end{equation}

Let us analyze what happens when we turn on the fluxes. It is easy to derive an expression for the Ricci scalar in the non-zero background case. For this task we rewrite the warped metric \eqref{02-warped-metric} as
\begin{equation}
\label{02-conformal-transformation} \widetilde{g}_{MN}=e^{2A(y)}
\, \bar{g}_{MN} \,,
\end{equation}
where the barred metric is given by
\begin{equation}
\label{02-conformal-metric}
\bar{g}_{MN} \, dX^M \, dX^N = \eta_{\mu \nu }(x) \, dx^{\mu} dx^{\nu} + e^{-3A(y)} \, g_{mn}(x,y) \, dy^m dy^n \,.
\end{equation}
The Christoffel symbols that correspond to the metric \eqref{02-conformal-metric} are
\begin{subequations}
\label{02-conformal-Christoffel}
\begin{align}
&\bar{\Gamma}^\alpha_{\mu \nu} = \Gamma^\alpha_{\mu \nu} \,, \\
&\bar{\Gamma}^\alpha_{m \nu} = \Gamma^\alpha_{m \nu} \,, \\
&\bar{\Gamma}^\alpha_{mn} = e^{-3A} \, \Gamma^\alpha_{mn} \,, \\
&\bar{\Gamma}^a_{mn} = \Gamma^a_{mn} - \tfrac{3}{2} \, \left[ \delta^a_m \partial_n + \delta^a_n \partial_m - g_{mn} g^{ab}\partial_b \right] \, A \,, \\
&\bar{\Gamma}^a_{\mu n} = \Gamma^a_{\mu n} \,, \\
&\bar{\Gamma}^a_{\mu \nu} = \Gamma^a_{\mu \nu} \,,
\end{align}
\end{subequations}
where the unbarred symbols are computed in \eqref{02-Christoffel}. We can repeat the computation for the Ricci scalar corresponding to the metric \eqref{02-conformal-metric} and at the end we will obtain a formula similar to \eqref{02-Ricci}. Due to the simple relations \eqref{02-conformal-Christoffel} between the Christoffel symbols, the Ricci scalar for the metric \eqref{02-conformal-metric} reduces to
\begin{equation}
\label{02-conformal-Ricci}
\bar{R}(M_{11}) = R(M_{11}) + 21 \, e^{3A} \left[ g^{ab}\nabla_a \nabla_b A - \tfrac{9}{2} \, g^{ab} \, \nabla_a A \, \nabla_b A \right] \,,
\end{equation}
where ${R}(M_{11})$ is given in \eqref{02-Ricci}.

To compute the Ricci scalar that corresponds to the metric \eqref{02-warped-metric} we have to perform the conformal transformation \eqref{02-conformal-transformation}. The result of the computation is
\begin{equation}
\label{02-warped-Ricci}
\begin{split}
\widetilde{R}(M_{11}) &= e^{-2A(y)} R(M_{11}) + e^{A(y)} \, \big[ \,  g^{ab}\nabla_a \nabla_b A(y) \\
& \quad - \tfrac{9}{2} \, g^{ab} \, \nabla_a A(y) \, \nabla_b A(y) \, \big] \,.
\end{split}
\end{equation}
Using the fact that the second term in \eqref{02-warped-Ricci} produces a total derivative term which vanishes by Stokes' theorem and the last term is subleading, we obtain that
\begin{equation}
\int_{M_{11}} \, d^{11}x \,\sqrt{-\widetilde{g}_{11}}\, \widetilde{R}(M_{11}) \, = \, \int_{M_{11}} d^{11}x \,\sqrt{-g}\, R(M_{11}) \, e^{- 3A(y)} + \ldots \,.
\end{equation}
As expected, to leading order the kinetic coefficients receive no corrections from warping. Therefore we conclude that
\begin{equation}
\label{02-Einstein-Hilbert-warped}
\begin{split}
\frac{1}{2 \kappa_{11}^2} &\int_{M_{11}} d^{11}x \,\sqrt{-\widetilde{g}_{11}}\, \widetilde{R}(M_{11}) = \frac{1}{2\kappa_3^2} \, \int_{M_3} d^3x \,\sqrt{-\eta}\, \Big\{ R(M_3) \\
& \quad - 18(\partial_\alpha \phi) \, (\partial^\alpha \phi) - \sum_{A,B=1}^{b_4^-} \, {\cal G}_{AB} \,(\partial_\alpha \phi^A) \, (\partial^\alpha \phi^B)\Big\} + \ldots \,,
\end{split}
\end{equation}
i.e. to leading order we obtain the same result as in the zero flux case.


\newpage

\chapter[Minimal Three-dimensional Supergravity Supplement]{MINIMAL THREE-DIMENSIONAL SUPERGRAVITY SUPLEMENT}
\label{app-3d}

\setcounter{equation}{0}
\renewcommand{\theequation}{\thechapter.\arabic{equation}}


This appendix contains our notations and conventions related to the derivation of the minimal three-di\-men\-sio\-nal supergravity. The conventions and notations are presented in appendix \ref{02-3d-sugra-conventions}. Appendix \ref{02-fierz} describes the derivation of the three-di\-men\-sio\-nal Fierz identities. In appendix \ref{02-3D-algebra}, we derive the closure of the three-di\-men\-sio\-nal super covariant derivative algebra.


\section{Notations and Conventions}
\label{02-3d-sugra-conventions}

\setcounter{equation}{0}
\renewcommand{\theequation}{\thesection.\arabic{equation}}


We use lower case Latin letters for three-vector indices and Greek letters for spinor indices. Supervector indices are denoted by capital Latin letters $A, M$. We further employ the early late convention: letters at the beginning of the alphabet denote tangent space indices while letters from the middle of the alphabet denote coordinate indices. The spinor metric is defined through
\begin{equation}
C_{\m \, \n} C^{\s \, \t} = \d_{\m}{}^{\s} \, \d_{\n}{}^{\t}
- \d_{\m}{}^{\t} \, \d_{\n}{}^{\s} ~ \equiv ~
\d_{\m}{}^{[ \s} \, \d_{\n}{}^{\t ]} \,,
\end{equation}
and is used to raise and lower spinor indices via
\begin{subequations}
\begin{align}
\q_{\n} &=~ \q^{\m} \, C_{\m \, \n} \,, \\
\q^{\m} &=~ C^{\m \, \n} \, \q_{\n} \,.
\end{align}
\end{subequations}
Some other conventions
\begin{subequations}
\begin{align}
&{\mathrm diag} (\, \eta_{ab} \,) =~ (-1 , 1 , 1) \,, \\
&\e_{a b c} \, \e^{d e f} =~ -\d_{[a}{}^d \d_{b}{}^e \d_{c]}{}^f  \,, \\
&\e^{0 1 2} = +1 \,.
\end{align}
\end{subequations}
The $\g$-matrices are defined through
\begin{equation}
\label{02-gamma}
(\g^a)_{\a}{}^{ \g} (\g^b)_{\g}{}^{ \b} =
\eta^{ ab} \d_{\a}{}^{\b} + \e^{ a
b c} (\g_c)_{\a}{}^{\b} \,,
\end{equation}
and satisfy the Fierz identities
\begin{subequations}
\begin{align}
&(\g^a)_{\a \b} (\g_a)^{\g \d} = - \d_{(\a}{}^\g \d_{\b)}{}^\d = - (\g^a)_{(\a}{}^\g (\g_a)_{\b)}{}^\d \,, \\
&\e^{a b c} \, (\g_b)_{\a \b} (\g_c)_\g {}^\d = C_{\a \g} (\g^a)_\b{}^\d +(\g^a)_{\a \g} \d_{\b} {}^\d \,.
\end{align}
\end{subequations}
For the Levi-Civita symbol, we have the contractions
\begin{subequations}
\begin{align}
\e^{abc}\e_{def} &=-\d_{[d}^{~~a}\d_{e}^{~b}\d_{f]}^{~c} \,, \\
\e^{abc}\e_{dec} &=-\d_{[d}^{~~a}\d_{e]}^{~b} \,, \\
\e^{abc}\e_{dbc} &=-2\d_{d}^{~a} \,.
\end{align}
\end{subequations}
The Lorentz rotation generator is realized in the following manner
\begin{equation}
\exp{\big[{-\tfrac{1}{2} \, \l_{ab}\,{\cal M}^{ab}}\big]} = \exp{\big[{ -\tfrac{1}{2} \, \e_{abc} \, \l^c \, \tfrac{1}{4} \, \g^{[a}\g^{b]}} \, \big]} = \exp{\big[{\, \tfrac{1}{2} \, \l^c \g_c} \big]} \,.
\end{equation}
Infinitesimally, the action of the Lorentz generator is
\begin{subequations}
\begin{align}
&[~ {\cal M}_a \, , \,\varphi(x) \, ] = 0 \,, \\
&[~{\cal M}_a \, , \, \r_{\a} (x) \, ] = \tfrac{1}{2} \,(\g_a) {}_{\a}{}^{\b} \, \r_{\b}(x) \,, \\
&[~ {\cal M}_a \, , \, A_b (x) \, ] = \e_b{}_a{}_c \, A^c(x) \,.
\end{align}
\end{subequations}
Some useful identities
\begin{subequations}
\begin{align}
&X_{[\a\b ]} =-C_{\a\b}X^\g_{~\g} \,, \\
&T_\g C_{\b\d}+T_\b C_{\d\g} + T_\d C_{\g\b}=0 \,.
\end{align}
\end{subequations}


\section{Derivation of Fierz Identities}
\label{02-fierz}

\setcounter{equation}{0}
\renewcommand{\theequation}{\thesection.\arabic{equation}}


Choosing the real basis
\begin{equation}
\g^0=i\s^2 \,, \quad \g^1=\s^1 \,, \quad \g^2=\s^3 \,,
\end{equation}
we can show by explicit substitution that
\begin{equation}
\label{02-fierz1}
(\g^a)_{\a\b}(\g_a)^{\g\d}=-\d_{(\a}^{~~\g}\d_{\b)}^{~~\d} \,.
\end{equation}
Basis free, we can derive that
\begin{equation}
\begin{split}
(\g^a)_{\a\b}(\g_a)^{\g\d}&= (\g^a)_\a^{~\varepsilon}(\g_a)_\eta^{~\d} C_{\varepsilon\b}C^{\g\eta}= (\g^a)_\a^{~\varepsilon} (\g_a)_\eta^{~\d}\d_{[\varepsilon}^{~\g} \d_{\b]}^{~\eta} \\
&=(\g^a)_\a^{~\g}(\g_a)_\b^{~\d}-\d_\b^{~\g}(\g^a\g_a)_\a^{~\d} = \tfrac{1}{2} \, (\g^a)_{(\a}^{~~\g}(\g_a)_{\b)}^{~~\d} - \tfrac{3}{2} \, \d_{(\a}^{~~\g} \d_{\b)}^{~~\d} \,.
\end{split}
\end{equation}
Using this last result and \eqref{02-fierz1} we also have
\begin{equation}
(\g^a)_{\a\b}(\g_a)^{\g\d}=-(\g^a)_{(\a}^{~~\g}(\g_a)_{\b)}^{~~\d} \,.
\end{equation}
The second Fierz identity can be derived directly from the defining relation \eqref{02-gamma}
\begin{equation}
\Big \{ (\g^{a})_{\g}^{~\s} (\g^{ b})_{\s}^{~\d} = \eta^{ab} \d_{\g}^{~\d} + \e^{abc} (\g_{c})_{\g}^{~\d} \Big \}(\g_b)_{\a\b} \,.
\end{equation}
Using \eqref{02-fierz1} we can simplify this relation
\begin{equation}
\begin{split}
\e^{abc} (\g_b)_{\a\b}(\g_c)_{\g}^{~\d} &= (\g^a)_{\g\s} \d_{(\a}^{~~\s} \d_{\b)}^{~~\d} - (\g^a)_{\a\b} \d_\g^{~\g} \\
&= (\g^a)_{\a\g} \d_\b^{~\d} + (\g^a)_{\b[\g} \d_{\a]}^{~~\d} = (\g^a)_{\a\g} \d_\b^{~\d} + C_{\a\g} (\g^a)_\b^{~\d} \,.
\end{split}
\end{equation}
A consequence of this identity is the following relation
\begin{equation}
\label{02-cons}
(\g_{[c})_{(\a}^{~~\d}(\g_{d]})_{\b )}^{~~\s}=-2\,\e_{acd}\,C^{\d\s}\,(\g^a)_{\a\b} \,.
\end{equation}


\section[Supergravity Covariant Derivative Algebra]{Supergravity Covariant Derivative Algebra}
\label{02-3D-algebra}

\setcounter{equation}{0}
\renewcommand{\theequation}{\thesection.\arabic{equation}}


The algebra of supergravity covariant derivatives given in the literature is not written in our conventions, and does not contain the gauge fields. To get the correct algebra we take the form given in the literature with arbitrary coefficients and add the superfield strengths ${\cal F}_{\a {b}}$ and ${\cal F}^{{c} I}$ associated with the $U(1)$ gauge theory
\begin{subequations}
\begin{align}
[ \nabla_{\a} ,~ \nabla_{\b} \} &=  (\g^{{c}})_{\a \b} ~ \nabla_{{c}}
- (\g^{{c}})_{\a \b}R \, {\cal M}_{{c}} \,, \\
\begin{split}
[ \nabla_\a ,~ \nabla_{{b}} \} &= -a \, (\g_{{b}})_{\a}{}^{\d} R \nabla_{\d} + c \, ( \nabla_{\a} R ) {\cal M}_{{b}} + {\cal F}_{\a{b}}^I\,t_I \\
& \quad + \big[ -2(\g_{{b}})_{\a}{}^{\d} \S_{\d} {}^{{d}} + b \, \tfrac{4}{3} \, (\g_{{b}}\g^{{d}})_{\a}{}^{\e} ( \nabla_{\e} R ) \, \big] {\cal M}_{{d}} \,,
\end{split} \\
\begin{split}
[ \nabla_{{a}} ,~ \nabla_{{b}} \} &=  2 \, \e_{{a} {b} {c}} \, [~d \S^{\a {c}} + e \, \tfrac{2}{3} \, (\g^{{c}})^{\a \b} (\nabla_{\b} R) ~] \nabla_{\a} + \e_{{a} {b} {c}} \, {\cal F}^{{c} I}\,t_I  \\
& \quad + \e_{{a} {b} {c}} \,[~ {\widehat {\cal R}}{}^{{c} {d}} + \tfrac{2}{3} \, \eta^{{c} {d}} (f\nabla^2 R + g \, \tfrac{3}{2} \, R^2 )~] \,{\cal M}_{{d}} \,,
%
\end{split}
\end{align}
\end{subequations}
where
\begin{align}
&{\widehat {\cal R}}{}^{a b} - {\widehat {\cal R}}{}^{b a} = \eta_{{a} {b} } {\widehat {\cal R}}{}^{{a} {b} } = (\g_{{d} } )^{\a \b} \S_{\b}{}^{{d} } = 0 \,, \\
\intertext{and}
&\nabla_\a \S_\b^{~f} = -\tfrac{1}{4} \, (\g^e)_{\a\b}\widehat{\cal R}_e^{~f} + \tfrac{1}{6} \, [C_{\a\b}\h^{fd}+ \tfrac{1}{2} \, \e^{fde}(\g_e)_{\a\b}]\nabla_d R \,. \label{02-curl}
\end{align}
By checking the Bianchi identities,we will set the coefficients and derive constraints on the new superfield strengths as in \eqref{02-curl}. The Bianchi identity
\begin{equation}
[[\nabla_{(\a} , \nabla_\b \} , \nabla_{\g)}\} =0 \,,
\end{equation}
looks like
\begin{equation}
\begin{split}
[[\nabla_{(\a} ,\nabla_\b \},\nabla_{\g)} \} &= -(\g^c)_{\a\b}\Big \{ \big( \tfrac{1}{2} - a \big) (\g_c)_\g^{~\d}R\nabla_\d +{\cal F}_{\g c}^I\,t_I \\
& \quad +(\g_c)_\g^{~\d} \big[-2\S_\d^d + \tfrac{4}{3} \, b (\g^d)_\d^{~\e}(\nabla_\e R) \big] {\cal M}_d \Big \} \\
& \quad + (c-1)(\nabla_\g R){\cal M}_c + [\b\g\a ] + [\g\a\b ] \,.
\end{split}
\end{equation}
This equation is satisfied if $c=1$ and
\begin{equation}
(\g^{c})_{(\a\b}{\cal F}_{\g )c}^I=0 \quad \Rightarrow \quad {\cal F}_{\g c}^I= \tfrac{1}{3} \, (\g_{c})_\g^{~\a}W_\a^I \,.
\end{equation}
The identity
\begin{equation}
[\{\nabla_\a, \nabla_\b\} , \nabla_{c}\, ] + \{ [\nabla_{c}, \nabla_{(\a}],\nabla_{\b)}\} =0 \,,
\end{equation}
is quite complicated, so we restrict our attention to one algebra element at a time. The terms proportional to $t_I$ are
\begin{equation}
(\g^d)_{\a\b}\e_{dce}{\cal F}^{e I} - \tfrac{1}{3} \, (\g_c)_{(\a}^{~~\d}\nabla_{\b )}W_{\d}^I=0 \,.
\end{equation}
Multiplying by $(\g^{c})^{\a\b}$ implies $\nabla^\d W_\d^I=0$. Multiplying by $(\g_a)^{\a\b}$ and antisymmetrizing over $a$ and $c$ leads to
\begin{equation}
{\cal F}^{e I}= \tfrac{1}{3} \, (\g^{e})_\b^{~\d}\nabla^\b W_\d^I \,.
\end{equation}
Terms proportional to $\nabla_a$ are
\begin{equation}
-(\g^d)_{\a\b}\e_{cde}R\nabla^e+a(\g_c)_{(\a}^{~~\d}(\g^d)_{\b )\d}R\nabla_d=0 \,,
\end{equation}
which means $a=-\frac{1}{2}$. Continuing to the terms proportional to $\S_{\b c}\nabla_\a$
\begin{equation}
2d\e_{dce}(\g^d)_{\a\b}\S^{\d e}\nabla_\d +(\g_c)_{(\a}^{~~\d} (\g_d)_{\b)}^{~~\s}\S_\d^d\nabla_\s = 0 \,.
\end{equation}
Using \eqref{02-cons} and the fact that $\S_{\a c}$ is gamma traceless, we see that $d=-1$. The terms proportional to $R\,\nabla_\a$ are
\begin{equation}
\begin{split}
{\cal J}_{\a\b\g} \nabla^\g &= \Big\{ \tfrac{4}{3} \, e ~ C_{\a\g} (\g_c)_b^{~\r} (\nabla_\r R) + \tfrac{4}{3} \, e (\g_c)_{\g\a} (\nabla_\b R) + (\g_c)_{\g(\a}(\nabla_{\b)}R) \\
& \quad + \tfrac{4}{3} \, b ~ (\g_c)_{\a\b}(\nabla_\g R) + \tfrac{2}{3} \, b ~ (\g_c)_{\g(\a}(\nabla_{\b)}R) \Big\} ~ \nabla^\g  \,.
\end{split}
\end{equation}
Hence ${\cal J}_{\a\b\g}=0$. ${\cal J}_{\a\b\g}$ is symmetric on $\a\b$ and therefore it is the sum of two independent irreducible spin tensors corresponding to the completely symmetric \raisebox{-.2ex}{\usebox{\symmtiii}} and corner \raisebox{-1ex}{\usebox{\corner}} Young tableaux. Both of these should vanish separately. Taking ${\cal J}_{(\a\b\g)}=0$ we see that
\begin{equation}
4e+8b+6=0 \,.
\end{equation}
Then setting ${\cal J}^{\g}_{~\b\g}=0$ we have
\begin{equation}
-8e+2b-3=0 \,.
\end{equation}
Thus,
\begin{equation}
e=b=-\tfrac{1}{2} \,.
\end{equation}
We now turn to the last terms, they are proportional to the Lorentz generator. Looking at non-linear terms involving $R$ we have
\begin{subequations}
\begin{align}
&[g-2a] \, \e_c^{~fd}\,(\g_d)_{\a\b}R^2{\cal M}_f =0 \,, \\
&[f-4b] \, \tfrac{2}{3} \, (\g^d)_{\a\b}\, \e_{dc}^{~~f}\, \nabla^2 R {\cal M}_f=0 \,,
\end{align}
\end{subequations}
which implies that
\begin{subequations}
\begin{align}
g&=2a=-1  \,, \\
f&=4b=-2 \,.
\end{align}
\end{subequations}
We have used the following fact to extract these contributions
\begin{equation}
\nabla_\a\nabla_\b R = \tfrac{1}{2} \, (\g^d)_{\a\b}(\nabla_d R) - C_{\a\b}\nabla^2 R \,.
\end{equation}
The remaining terms in this Bianchi identity are
\begin{equation}
\begin{split}
\Big\{ (\g^d)_{\a\b} \, &\e_{dce} \widehat{\cal R}^{ef} +(\g^f)_{\a\b}(\nabla_c R) + 2 (\g_c)_{(\a}^{~~\d} \nabla_{\b)} \S_\d^{~f} \\
& - \tfrac{2}{3} \, b (\g_c\g^f\g^d)_{(\a\b)} (\nabla_d R) -(\g^d)_{\a\b} (\nabla_d R) \d_c^{~f} \Big\}{\cal M}_f=0 \,.
\end{split}
\end{equation}
After converting the free vector index into two symmetric spinor indices by contracting with $(\g^c)_{\g\d}$ we have an expression of the form ${\cal J}_{\a\b\g\d}^f{\cal M}_f=0$. This tensor is the product of two rank two symmetric spin tensors and has the following decomposition in terms of Young tableaux: \raisebox{-.2ex}{\usebox{\symmtii}} $\otimes$ \kern-.4em\raisebox{-.2ex}{\usebox{\symmtii}} $=$ \kern-.4em\raisebox{-.2ex}{\usebox{\symmtiii}} $\oplus$ \kern-.4em\raisebox{-1ex}{\usebox{\symmtiv}} $\oplus$ \kern-.4em\raisebox{-1ex}{\usebox{\gun}} . The completely symmetric term vanishes identically. The box diagram $\sim$ $C^{\g\a}C^{\d\b}{\cal J}_{\a\b\g\d}^f{\cal M}_f$ takes the form
\begin{equation}
0=\{-2\nabla^c R -12\nabla^\d \S_\d^{~c}+8~b~\nabla^c R+2\nabla^c R \}{\cal M}_c \,,
\end{equation}
which implies that
\begin{equation}
\nabla^\s \S_\s^{~f} = - \tfrac{1}{3} \, \nabla^f R \,.
\end{equation}
The gun diagram $\sim \, C^{\g\a}{\cal J}_{\a(\b\d)\g}^f{\cal M}_f$ takes the form
\begin{equation}
0=\{-4(\g^e)_{\b\d}\widehat {\cal R}_e^{~f}-8\nabla_{(\b}\S_{\d)}^{~~f} + (2+2-\tfrac{8}{3} \,) \, \e^{cde} \, (\g_e)_{\b\d} \, \nabla_d R \} \, {\cal M}_f \,,
\end{equation}
which implies that
\begin{equation}
\nabla_{(\b}\S_{\d)}^{~~f} = \tfrac{1}{6} \, \e^{fde} \, (\g_e)_{\d\b}\nabla_d R - \tfrac{1}{2} \, (\g^e)_{\b\d} \, \widehat{\cal R}_e^{~f} \,.
\end{equation}
Thus, the spinorial derivative of $\S_\a^{~f}$ takes the form
\begin{equation}
\nabla_\a \S_\b^{~f} = -\tfrac{1}{4} \, (\g^e)_{\a\b} \widehat{\cal R}_e^{~f} +\tfrac{1}{6} \, \big[ C_{\a\b}\h^{fd}+ \tfrac{1}{2} \, \e^{fde} \, (\g_e)_{\a\b} \big] \, \nabla_d R \,.
\end{equation}
This completes the analysis of the spin-spin-vector Bianchi identity. We now move on to the spin-vector-vector Bianchi identity
\begin{equation}
[[\nabla_\a ,\nabla_b \} , \nabla_c\}+[[\nabla_b , \nabla_c \} , \nabla_\a \} + [[\nabla_c , \nabla_\a \} , \nabla_b\}=0 \,.
\end{equation}
This identity is satisfied identically, yielding no further constraints. The final identity is all vector derivatives: $[[\nabla_{[a} ,\nabla_b \} , \nabla_{c]} \}=0$. This identity yields some more differential constraints which are of no consequence to the derivations.


\newpage

\chapter[Review of Spin(7) Holonomy Manifolds]{REVIEW OF SPIN(7) HOLONOMY MANIFOLDS}
\label{app-spin7}

\setcounter{equation}{0}
\renewcommand{\theequation}{\thechapter.\arabic{equation}}


This appendix contains a brief review of some of the relevant aspects of Spin(7) holonomy manifolds. An elegant discussion can be found in the book written by Joyce \cite{Joyce}. On an Riemannian manifold $X$ of dimension $n$, the spin connection ${\omega}$ is, in general, an $SO(n)$ gauge field. If we parallel transport a spinor $\psi$ around a closed path $\gamma$, the spinor comes back as $U\psi$, where $U=Pexp\int_{\gamma}{\omega}~dx$ is the path ordered exponential of $\omega$ around the curve $\gamma$. The set of $U$ transformations form a subgroup of $SO(n)$ called the holonomy group of the manifold $X$.

A compactification of M-theory (or string theory) on some internal manifold $X$ preserves some amount of supersymmetry if $X$ admits one (or more) covariantly constant spinors. Such spinors return upon parallel transport to their original values, i.e. they satisfy $U\psi=\psi$. The holonomy of the manifold is then a (proper) subgroup of $SO(n)$. A Spin(7) holonomy manifold is an eight-di\-men\-sio\-nal manifold, for which one such spinor exists. Therefore, if we compactify M-theory on these manifolds we obtain an ${\cal N}=1$ theory in three dimensions. Spin(7) is a subgroup of $GL(8,\IR)$ defined as follows. Introduce on ${\IR}^{\kern.05em 8}$ the coordinates $(x_1,\dots,x_8)$ and the four-form $dx_{ijkl}=dx_i\wedge dx_j\wedge dx_k\wedge dx_l$. Define a self-dual four-form $\Omega$ on ${\IR}^{\kern.05em 8}$ by
\begin{equation}
\begin{split}
\Omega &= dx_{1234}+dx_{1256}+dx_{1278}+dx_{1357}-dx_{1368} \\
& \quad -dx_{1458}-dx_{1467}-dx_{2358}-dx_{2367}-dx_{2457} \\
& \quad +dx_{2468}+dx_{3456}+dx_{3478}+dx_{5678} \,.
\end{split}
\end{equation}
The subgroup of $GL(8,\IR)$ preserving $\Omega$ is the holonomy group Spin(7). It is a compact, connected, simply connected, semisimple, twenty-one-di\-men\-sio\-nal Lie group, which is isomorphic to the double cover of $SO(7)$. Many of the mathematical properties of Spin(7) holonomy manifolds are discussed in detail in \cite{Joyce}. Let us here only mention that these manifolds are Ricci flat but are, in general, not K\"ahler.

The cohomology of a compact Spin(7) holonomy manifold can be decomposed into the following representations of Spin(7)
\begin{subequations}
\label{02-cohomology}
\begin{align}
H^0(X, \IR) & =  \IR \,, \\
H^1(X, \IR) & =  0 \,, \\
H^2(X, \IR) & =  H^2_{{\bf 21}}(X,\IR) \,, \\
H^3(X, \IR) & =  H^3_{{\bf 48}}(X, \IR) \,, \\
H^4(X, \IR) & =  H^4_{{\bf 1}^+} (X, \IR) \oplus H^4_{{\bf 27}^+} (X, \IR) \oplus H^4_{{\bf 35}^-} (X, \IR) \,, \\
H^5 (X, \IR) & =  H^5_{{\bf 48}} (X, \IR) \,, \\
H^6 (X, \IR) & =  H^6_{{\bf 21}} (X, \IR) \,, \\
H^7 (X, \IR) & =  0 \,,\\
H^8 (X,\IR) & = \IR \,,
\end{align}
\end{subequations}
where the label $``\pm"$ indicates self-dual and anti-self-dual four-forms, respectively, and the subindex indicates the representation. The Cayley calibration $\Omega$ belongs to the cohomology $H_{{\bf 1}^+}^4(X,\IR)$.

Next we will briefly discuss deformations of the Cayley calibration. More details can be found in \cite{Joyce} and \cite{Karigiannis}. The tangent space to the family of torsion-free Spin(7) structures, up to diffeomorphism is naturally isomorphic to the direct sum $H^4_{{{\bf 1}^+}}(X, \IR) \oplus H^4_{{\bf 35}^-} (X, \IR)$ if $X$ is compact and the holonomy is Spin(7) and not some proper subgroup. Thus, if the holonomy is Spin(7) the family has dimension $1+b_4^-$, and the infinitesimal variations in $\Omega$ are of the form $c\Omega+ \xi$, where $\xi$ is a harmonic anti-self-dual four-form and $c$ is a number.

When we are moving in moduli space along the ``radial direction'' $\phi$, the Cayley calibration deformation is
\begin{equation}
\delta\Omega=K \delta \phi \Omega\,,
\end{equation}
or in other words
\begin{equation}
\label{02-partial-0-omega}
\frac{\partial \Omega}{\partial \phi} = K \Omega \,.
\end{equation}
If we consider infinitesimal displacements in moduli space along the other $b_4^-$ directions, then the Cayley calibration deformation is
\begin{equation}
\delta\Omega=\delta \phi^A(\xi_A-K_A \Omega) \,,
\end{equation}
or in other words
\begin{equation}
\label{02-partail-a-omega}
\frac{\partial \Omega}{\partial \phi^A} = \xi_A-K_A \Omega \,,
\end{equation}
where $\xi_A$ are the anti-self-dual harmonic four-forms. If the movement in the moduli space is not along some particular direction then
\begin{equation}
\delta\Omega=\delta \phi^A \, \xi_A +(\delta \phi \, K - \delta \phi^A \, K_A )\Omega \,.
\end{equation}
We note that the potential
\begin{equation}
P= \frac{1}{2}\ln \Big( \int_{M_8} \Omega \wedge \star \Omega\,\Big) \,,
\end{equation}
generates $K=\partial P$ and $K_A=-\partial_A P$. The fact that
\begin{equation}
\int_{M_8} \Omega \wedge \star \Omega = 14 {\cal V}_{M_8} = e^{2P} \,,
\end{equation}
fixes $K=2$, where ${\cal V}_{M_8}$ is the volume of the internal manifold.




\bibliographystyle{hunsrt}
\bibliography{bibliography}

\begin{thebibliography}{10}

\bibitem{Joyce}
D.~D. Joyce.
\newblock {\em {\em ``Compact Manifolds with Special Holonomy''}}.
\newblock Oxford Science Publications, Oxford, U.K., (2000).

\bibitem{Shatashvili:1994zw}
S.~L. Shatashvili and C.~Vafa.
\newblock {\em ``Superstrings and Manifolds of Exceptional Holonomy''}.
\newblock (1994),
  \href{http://www.arxiv.org/abs/hep-th/9407025}{hep-th/9407025}.

\bibitem{Papadopoulos:1995da}
G.~Papadopoulos and P.~K. Townsend.
\newblock {\em ``Compactification of $D = 11$ Supergravity on Spaces of
  Exceptional Holonomy''}.
\newblock {\em Phys. Lett.}, {\bf B357}:300, (1995),
  \href{http://www.arxiv.org/abs/hep-th/9506150}{hep-th/9506150}.

\bibitem{Harvey:1982xk}
R.~Harvey and Jr. Lawson, H.~B.
\newblock {\em ``Calibrated Geometries''}.
\newblock {\em Acta Math.}, {\bf 148}:47, (1982).

\bibitem{Gukov:1999ya}
S.~Gukov, C.~Vafa, and E.~Witten.
\newblock {\em ``CFT's from Calabi-Yau Four-folds''}.
\newblock {\em Nucl. Phys.}, {\bf B584}:69, (2000),
  \href{http://www.arxiv.org/abs/hep-th/9906070}{hep-th/9906070}.

\bibitem{Gukov:1999gr}
S.~Gukov.
\newblock {\em ``Solitons, Superpotentials and Calibrations''}.
\newblock {\em Nucl. Phys.}, {\bf B574}:169, (2000),
  \href{http://www.arxiv.org/abs/hep-th/9911011}{hep-th/9911011}.

\bibitem{Taylor:1999ii}
T.~R. Taylor and C.~Vafa.
\newblock {\em ``RR Flux on Calabi-Yau and Partial Supersymmetry Breaking''}.
\newblock {\em Phys. Lett.}, {\bf B474}:130, (2000),
  \href{http://www.arxiv.org/abs/hep-th/9912152}{hep-th/9912152}.

\bibitem{Giddings:2001yu}
S.~B. Giddings, S.~Kachru, and J.~Polchinski.
\newblock {\em ``Hierarchies from Fluxes in String Compactifications''}.
\newblock {\em Phys. Rev.}, {\bf D66}:106006, (2002),
  \href{http://www.arxiv.org/abs/hep-th/0105097}{hep-th/0105097}.

\bibitem{Gates:2000fj}
S.~J. Gates, Jr., S.~Gukov, and E.~Witten.
\newblock {\em ``Two Two-dimensional Supergravity Theories from Calabi-Yau
  Four-folds''}.
\newblock {\em Nucl. Phys.}, {\bf B584}:109, (2000),
  \href{http://www.arxiv.org/abs/hep-th/0005120}{hep-th/0005120}.

\bibitem{Haack:2000di}
M.~Haack, J.~Louis, and M.~Marquart.
\newblock {\em ``Type IIA and Heterotic String Vacua in $D=2$''}.
\newblock {\em Nucl. Phys.}, {\bf B598}:30, (2001),
  \href{http://www.arxiv.org/abs/hep-th/0011075}{hep-th/0011075}.

\bibitem{Gukov:2002iq}
S.~Gukov and M.~Haack.
\newblock {\em ``IIA String Theory on Calabi-Yau Fourfolds with Background
  Fluxes''}.
\newblock {\em Nucl. Phys.}, {\bf B639}:95, (2002),
  \href{http://www.arxiv.org/abs/hep-th/0203267}{hep-th/0203267}.

\bibitem{Beasley:2002db}
C.~Beasley and E.~Witten.
\newblock {\em ``A Note on Fluxes and Superpotentials in M-theory
  Compactifications on Manifolds of $G(2)$ Holonomy''}.
\newblock {\em JHEP}, {\bf 07}:046, (2002),
  \href{http://www.arxiv.org/abs/hep-th/0203061}{hep-th/0203061}.

\bibitem{Dine:1985rz}
M.~Dine, R.~Rohm, N.~Seiberg, and E.~Witten.
\newblock {\em ``Gluino Condensation in Superstring Models''}.
\newblock {\em Phys. Lett.}, {\bf B156}:55, (1985).

\bibitem{Behrndt:2000zh}
K.~Behrndt and S.~Gukov.
\newblock {\em ``Domain Walls and Superpotentials from ${\cal M}$ theory on
  Calabi-Yau Three-folds''}.
\newblock {\em Nucl. Phys.}, {\bf B580}:225, (2000),
  \href{http://www.arxiv.org/abs/hep-th/0001082}{hep-th/0001082}.

\bibitem{Strominger:1986uh}
A.~Strominger.
\newblock {\em ``Superstrings with Torsion''}.
\newblock {\em Nucl. Phys.}, {\bf B274}:253, (1986).

\bibitem{deWit:1986xg}
B.~de~Wit, D.~J. Smit, and N.~D. Hari~Dass.
\newblock {\em ``Residual Supersymmetry of Compactified $D=10$ Supergravity''}.
\newblock {\em Nucl. Phys.}, {\bf B283}:165, (1987).

\bibitem{Becker:1996gj}
K.~Becker and M.~Becker.
\newblock {\em ``${\cal M}$-theory on Eight-Manifolds''}.
\newblock {\em Nucl. Phys.}, {\bf B477}:155, (1996),
  \href{http://www.arxiv.org/abs/hep-th/9605053}{hep-th/9605053}.

\bibitem{Sethi:1996es}
S.~Sethi, C.~Vafa, and E.~Witten.
\newblock {\em ``Constraints on Low-Dimensional String Compactifications''}.
\newblock {\em Nucl. Phys.}, {\bf B480}:213, (1996),
  \href{http://www.arxiv.org/abs/hep-th/9606122}{hep-th/9606122}.

\bibitem{Dasgupta:1999ss}
K.~Dasgupta, G.~Rajesh, and S.~Sethi.
\newblock {\em ``${\cal M}$ Theory, Orientifolds and G-flux''}.
\newblock {\em JHEP}, {\bf 08}:023, (1999),
  \href{http://www.arxiv.org/abs/hep-th/9908088}{hep-th/9908088}.

\bibitem{Haack:2001jz}
M.~Haack and J.~Louis.
\newblock {\em ``${\cal M}$-theory Compactified on Calabi-Yau Fourfolds with
  Background Flux''}.
\newblock {\em Phys. Lett.}, {\bf B507}:296, (2001),
  \href{http://www.arxiv.org/abs/hep-th/0103068}{hep-th/0103068}.

\bibitem{Berg:2002es}
M.~Berg, M.~Haack, and H.~Samtleben.
\newblock {\em ``Calabi-Yau Fourfolds with Flux and Supersymmetry Breaking''}.
\newblock {\em JHEP}, {\bf 04}:046, (2003),
  \href{http://www.arxiv.org/abs/hep-th/0212255}{hep-th/0212255}.

\bibitem{Witten:1994cg}
E.~Witten.
\newblock {\em ``Is Supersymmetry Really Broken?''}.
\newblock {\em Int. J. Mod. Phys.}, {\bf A10}:1247, (1995),
  \href{http://www.arxiv.org/abs/hep-th/9409111}{hep-th/9409111}.

\bibitem{Witten:1995rz}
E.~Witten.
\newblock {\em ``Strong Coupling and the Cosmological Constant''}.
\newblock {\em Mod. Phys. Lett.}, {\bf A10}:2153, (1995),
  \href{http://www.arxiv.org/abs/hep-th/9506101}{hep-th/9506101}.

\bibitem{Becker:1995sp}
K.~Becker, M.~Becker, and A.~Strominger.
\newblock {\em ``Three-dimensional Supergravity and the Cosmological
  Constant''}.
\newblock {\em Phys. Rev.}, {\bf D51}:6603, (1995),
  \href{http://www.arxiv.org/abs/hep-th/9502107}{hep-th/9502107}.

\bibitem{Vafa:1995fj}
C.~Vafa and E.~Witten.
\newblock {\em ``A One Loop Test of String Duality''}.
\newblock {\em Nucl. Phys.}, {\bf B447}:261, (1995),
  \href{http://www.arxiv.org/abs/hep-th/9505053}{hep-th/9505053}.

\bibitem{Duff:1995wd}
M.~J. Duff, J.~T. Liu, and R.~Minasian.
\newblock {\em ``Eleven-dimensional Origin of String / String Duality: A
  One-loop Test''}.
\newblock {\em Nucl. Phys.}, {\bf B452}:261, (1995),
  \href{http://www.arxiv.org/abs/hep-th/9506126}{hep-th/9506126}.

\bibitem{Witten:1997md}
E.~Witten.
\newblock {\em ``On Flux Quantization in ${\cal M}$-theory and the Effective
  Action''}.
\newblock {\em J. Geom. Phys.}, {\bf 22}:1, (1997),
  \href{http://www.arxiv.org/abs/hep-th/9609122}{hep-th/9609122}.

\bibitem{Becker:2000jc}
K.~Becker.
\newblock {\em ``A Note on Compactifications on $Spin(7)$-Holonomy
  Manifolds''}.
\newblock {\em JHEP}, {\bf 05}:003, (2001),
  \href{http://www.arxiv.org/abs/hep-th/0011114}{hep-th/0011114}.

\bibitem{Acharya:2002vs}
B.~Acharya, X.~de~la Ossa, and S.~Gukov.
\newblock {\em ``G-Flux, Supersymmetry and $Spin(7)$ Manifolds''}.
\newblock {\em JHEP}, {\bf 09}:047, (2002),
  \href{http://www.arxiv.org/abs/hep-th/0201227}{hep-th/0201227}.

\bibitem{Becker:2002jj}
M.~Becker and D.~Constantin.
\newblock {\em ``A Note on Flux Induced Superpotentials in String Theory''}.
\newblock {\em JHEP}, {\bf 08}:015, (2003),
  \href{http://www.arxiv.org/abs/hep-th/0210131}{hep-th/0210131}.

\bibitem{Gukov:2001hf}
S.~Gukov and J.~Sparks.
\newblock {\em ``${\cal M}$-Theory on $Spin(7)$ Manifolds. I''}.
\newblock {\em Nucl. Phys.}, {\bf B625}:3, (2002),
  \href{http://www.arxiv.org/abs/hep-th/0109025}{hep-th/0109025}.

\bibitem{Cvetic:2001pg}
M.~Cvetic, G.~W. Gibbons, H.~Lu, and C.~N. Pope.
\newblock {\em ``New Complete Non-Compact $Spin(7)$ Manifolds''}.
\newblock {\em Nucl. Phys.}, {\bf B620}:29, (2002),
  \href{http://www.arxiv.org/abs/hep-th/0103155}{hep-th/0103155}.

\bibitem{Gukov:2002zg}
S.~Gukov, J.~Sparks, and D.~Tong.
\newblock {\em ``Conifold Transitions and Five-brane Condensation in ${\cal
  M}$-theory on $Spin(7)$ Manifolds''}.
\newblock {\em Class. Quant. Grav.}, {\bf 20}:665, (2003),
  \href{http://www.arxiv.org/abs/hep-th/0207244}{hep-th/0207244}.

\bibitem{Acharya:2004qe}
B.~S. Acharya and S.~Gukov.
\newblock {\em ``${\cal M}$-Theory and Singularities of Exceptional Holonomy
  Manifolds''}.
\newblock {\em Phys. Rept.}, {\bf 392}:121, (2004),
  \href{http://www.arxiv.org/abs/hep-th/0409191}{hep-th/0409191}.

\bibitem{Becker:2003wb}
M.~Becker, D.~Constantin, S.~J. Gates, Jr., W.~D. Linch, III, W.~Merrell, and
  J.~Phillips.
\newblock {\em ``${\cal M}$-theory on $Spin(7)$ Manifolds, Fluxes and $3D$, $N
  = 1$ Supergravity''}.
\newblock {\em Nucl. Phys.}, {\bf B683}:67, (2004),
  \href{http://www.arxiv.org/abs/hep-th/0312040}{hep-th/0312040}.

\bibitem{Constantin:2004eg}
D.~Constantin.
\newblock {\em ``${\cal M}$-theory Vacua from Warped Compactifications on
  $Spin(7)$ Manifolds''}.
\newblock {\em Nucl. Phys.}, {\bf B706}:221, (2005),
  \href{http://www.arxiv.org/abs/hep-th/0410157}{hep-th/0410157}.

\bibitem{Strominger:1997eb}
A.~Strominger.
\newblock {\em ``Loop Corrections to the Universal Hypermultiplet''}.
\newblock {\em Phys. Lett.}, {\bf B421}:139, (1998),
  \href{http://www.arxiv.org/abs/hep-th/9706195}{hep-th/9706195}.

\bibitem{Peeters:2003ys}
K.~Peeters, P.~Vanhove, and A.~Westerberg.
\newblock {\em ``Towards Complete String Effective Actions Beyond Leading
  Order''}.
\newblock {\em Fortsch. Phys.}, {\bf 52}:630, (2004),
  \href{http://www.arxiv.org/abs/hep-th/0312211}{hep-th/0312211}.

\bibitem{Witten:1987kg}
L.~Witten and E.~Witten.
\newblock {\em ``Large Radius Expansion of Superstring Compactifications''}.
\newblock {\em Nucl. Phys.}, {\bf B281}:109, (1987).

\bibitem{Becker:2001pm}
K.~Becker and M.~Becker.
\newblock {\em ``Supersymmetry Breaking, ${\cal M}$-theory and Fluxes''}.
\newblock {\em JHEP}, {\bf 07}:038, (2001),
  \href{http://www.arxiv.org/abs/hep-th/0107044}{hep-th/0107044}.

\bibitem{Cremmer:1978km}
E.~Cremmer, B.~Julia, and J.~Scherk.
\newblock {\em ``Supergravity Theory in 11 Dimensions''}.
\newblock {\em Phys. Lett.}, {\bf B76}:409, (1978).

\bibitem{Tseytlin:2000sf}
A.~A. Tseytlin.
\newblock {\em ``$R^4$ Terms in $11$ Dimensions and Conformal Anomaly of
  $(2,0)$ Theory''}.
\newblock {\em Nucl. Phys.}, {\bf B584}:233, (2000),
  \href{http://www.arxiv.org/abs/hep-th/0005072}{hep-th/0005072}.

\bibitem{Martelli:2003ki}
D.~Martelli and J.~Sparks.
\newblock {\em ``G-Structures, Fluxes and Calibrations in ${\cal M}$-Theory''}.
\newblock {\em Phys. Rev.}, {\bf D68}:085014, (2003),
  \href{http://www.arxiv.org/abs/hep-th/0306225}{hep-th/0306225}.

\bibitem{deWit:1978sh}
B.~de~Wit.
\newblock {\em ``Properties of $SO(8)$ Extended Supergravity''}.
\newblock {\em Nucl. Phys.}, {\bf B158}:189, (1979).

\bibitem{Gross:1986iv}
D.~J. Gross and E.~Witten.
\newblock {\em ``Superstring Modifications of Einstein's Equations''}.
\newblock {\em Nucl. Phys.}, {\bf B277}:1, (1986).

\bibitem{Lu:2004ng}
H.~Lu, C.~N. Pope, K.~S. Stelle, and P.~K. Townsend.
\newblock {\em ``String and ${\cal M}$-theory Deformations of Manifolds with
  Special Holonomy}.
\newblock (2004),
  \href{http://www.arxiv.org/abs/hep-th/0410176}{hep-th/0410176}.

\bibitem{Banks:1998nr}
T.~Banks and M.~B. Green.
\newblock {\em ``Non-Perturbative Effects in $AdS_5 \times S^5$ String Theory
  and $d = 4$ SUSY Yang-Mills''}.
\newblock {\em JHEP}, {\bf 05}:002, (1998),
  \href{http://www.arxiv.org/abs/hep-th/9804170}{hep-th/9804170}.

\bibitem{Gubser:1998nz}
S.~S. Gubser, I.~R. Klebanov, and A.~A. Tseytlin.
\newblock {\em ``Coupling Constant Dependence in the Thermodynamics of $N = 4$
  Supersymmetric Yang-Mills Theory''}.
\newblock {\em Nucl. Phys.}, {\bf B534}:202, (1998),
  \href{http://www.arxiv.org/abs/hep-th/9805156}{hep-th/9805156}.

\bibitem{Isham:1988jb}
C.~J. Isham, C.~N. Pope, and N.~P. Warner.
\newblock {\em ``Nowhere Vanishing Spinors and Triality Rotations in Eight
  Manifolds''}.
\newblock {\em Class. Quant. Grav.}, {\bf 5}:1297, (1988).

\bibitem{Gibbons:1990er}
G.~W. Gibbons, D.~N. Page, and C.~N. Pope.
\newblock {\em ``Einstein Metrics on $S^3$, $R^3$ and $R^4$ Bundels''}.
\newblock {\em Commun. Math. Phys.}, {\bf 127}:529, (1990).

\bibitem{Lu:2003ze}
H.~Lu, C.~N. Pope, K.~S. Stelle, and P.~K. Townsend.
\newblock {\em ``Supersymmetric Deformations of $G(2)$ Manifolds from
  Higher-Order Corrections to String and ${\cal M}$-theory''}.
\newblock {\em JHEP}, {\bf 10}:019, (2004),
  \href{http://www.arxiv.org/abs/hep-th/0312002}{hep-th/0312002}.

\bibitem{Freeman:1986br}
M.~D. Freeman and C.~N. Pope.
\newblock {\em ``Beta Functions and Superstring Compactifications''}.
\newblock {\em Phys. Lett.}, {\bf B174}:48, (1986).

\bibitem{Salam:1989fm}
(Ed.) Salam, A. and (Ed.) Sezgin, E.
\newblock {\em {\em ``Supergravities in Diverse Dimensions. Vol. 1, 2''}}.
\newblock North-Holland, Amsterdam, Netherlands, (1989).

\bibitem{Rocek:1986bk}
M.~Rocek and P.~van Nieuwenhuizen.
\newblock {\em ``$N \geq 2$ Supersymmetric Chern-Simons Terms as $D = 3$
  Extended Conformal Supergravity''}.
\newblock {\em Class. Quant. Grav.}, {\bf 3}:43, (1986).

\bibitem{vanNieuwenhuizen:1985cx}
P.~van Nieuwenhuizen.
\newblock {\em ``$D = 3$ Conformal Supergravity and Chern-Simons Terms''}.
\newblock {\em Phys. Rev.}, {\bf D32}:872, (1985).

\bibitem{Karlhede:1987qd}
A.~Karlhede, U.~Lindstrom, M.~Rocek, and P.~van Nieuwenhuizen.
\newblock {\em ``On 3-D Nonlinear Vector-Vector Duality''}.
\newblock {\em Phys. Lett.}, {\bf B186}:96, (1987).

\bibitem{Karlhede:1987qf}
A.~Karlhede, U.~Lindstrom, M.~Rocek, and P.~van Nieuwenhuizen.
\newblock {\em ``Supersymmetric Vector-Vector Duality''}.
\newblock {\em Class. Quant. Grav.}, {\bf 4}:549, (1987).

\bibitem{Park:1999cw}
J.~H. Park.
\newblock {\em ``Superconformal Symmetry in Three-dimensions''}.
\newblock {\em J. Math. Phys.}, {\bf 41}:7129, (2000),
  \href{http://www.arxiv.org/abs/hep-th/9910199}{hep-th/9910199}.

\bibitem{Ivanov:2000tz}
E.~Ivanov, S.~Krivonos, and O.~Lechtenfeld.
\newblock {\em ``Double-vector Multiplet and Partially Broken $N = 4$, $D = 3$
  Supersymmetry''}.
\newblock {\em Phys. Lett.}, {\bf B487}:192, (2000),
  \href{http://www.arxiv.org/abs/hep-th/0006017}{hep-th/0006017}.

\bibitem{Zupnik:1999tf}
B.~M. Zupnik.
\newblock {\em ``Partial Spontaneous Breakdown of Three-dimensional $N = 2$
  Supersymmetry''}.
\newblock {\em Theor. Math. Phys.}, {\bf 123}:463, (2000),
  \href{http://www.arxiv.org/abs/hep-th/9905108}{hep-th/9905108}.

\bibitem{Zupnik:1997nn}
B.~M. Zupnik.
\newblock {\em ``Harmonic Superspaces for Three-Dimensional Theories''}.
\newblock (1997),
  \href{http://www.arxiv.org/abs/hep-th/9804167}{hep-th/9804167}.

\bibitem{Alexandre:2003qb}
J~Alexandre.
\newblock {\em ``From ${\cal N} = 1$ to ${\cal N} = 2$ Supersymmetries in $2+1$
  Dimensions''}.
\newblock {\em Phys. Rev.}, {\bf D67}:105017, (2003),
  \href{http://www.arxiv.org/abs/hep-th/0301034}{hep-th/0301034}.

\bibitem{deWit:1992up}
B.~de~Wit, A.~K. Tollsten, and H.~Nicolai.
\newblock {\em ``Locally Supersymmetric $D = 3$ Nonlinear Sigma Models''}.
\newblock {\em Nucl. Phys.}, {\bf B392}:3, (1993),
  \href{http://www.arxiv.org/abs/hep-th/9208074}{hep-th/9208074}.

\bibitem{deWit:2003ja}
B.~de~Wit, I.~Herger, and H.~Samtleben.
\newblock {\em ``Gauged Locally Supersymmetric $D = 3$ Nonlinear Sigma
  Models''}.
\newblock {\em Nucl. Phys.}, {\bf B671}:175, (2003),
  \href{http://www.arxiv.org/abs/hep-th/0307006}{hep-th/0307006}.

\bibitem{Brown:1979ma}
M.~Brown and S.~J. Gates, Jr.
\newblock {\em ``Superspace Bianchi Identities and the Supercovariant
  Derivative''}.
\newblock {\em Ann. Phys.}, {\bf 122}:443, (1979).

\bibitem{Grisaru:1997ub}
M.~T. Grisaru, M.~E. Knutt-Wehlau, and W.~Siegel.
\newblock {\em ``A Superspace Normal Coordinate Derivation of the Density
  Formula''}.
\newblock {\em Nucl. Phys.}, {\bf B523}:663, (1998),
  \href{http://www.arxiv.org/abs/hep-th/9711120}{hep-th/9711120}.

\bibitem{Gates:1997ag}
S.~J. Gates, Jr., M.~T. Grisaru, M.~E. Knutt-Wehlau, and W.~Siegel.
\newblock {\em ``Component Actions from Curved Superspace: Normal Coordinates
  and Ectoplasm''}.
\newblock {\em Phys. Lett.}, {\bf B421}:203, (1998),
  \href{http://www.arxiv.org/abs/hep-th/9711151}{hep-th/9711151}.

\bibitem{Gates:1998hy}
S.~J. Gates, Jr.
\newblock {\em ``Ectoplasm has no Topology''}.
\newblock {\em Nucl. Phys.}, {\bf B541}:615, (1999),
  \href{http://www.arxiv.org/abs/hep-th/9809056}{hep-th/9809056}.

\bibitem{Hawking:1998bg}
S.~W. Hawking and M.~M. Taylor-Robinson.
\newblock {\em ``Bulk Charges in Eleven Dimensions''}.
\newblock {\em Phys. Rev.}, {\bf D58}:025006, (1998),
  \href{http://www.arxiv.org/abs/hep-th/9711042}{hep-th/9711042}.

\bibitem{Isham:1987qe}
C.~J. Isham and C.~N. Pope.
\newblock {\em ``Nowhere Vanishing Spinors and Topological Obstructions to the
  Equivalence of the NSR and GS Superstrings''}.
\newblock {\em Class. Quant. Grav.}, {\bf 5}:257, (1988).

\bibitem{Gates:1983nr}
S.~J. Gates, Jr., M.~T. Grisaru, M.~Rocek, and W.~Siegel.
\newblock {\em {\em ``Superspace, or One Thousand and One Lessons in
  Supersymmetry''}}.
\newblock Benjamin/Cummings Publishing Company, Inc., Massachusetts 01867,
  U.S.A., (1983),
  \href{http://www.arxiv.org/abs/hep-th/0108200}{hep-th/0108200}.

\bibitem{Wess:1992cp}
J.~Wess and J.~Bagger.
\newblock {\em {\em ``Supersymmetry and Supergravity''}}.
\newblock Princeton University Press, Princeton, U.S.A., (1992).

\bibitem{Becker:1995kb}
K.~Becker, M.~Becker, and A.~Strominger.
\newblock {\em ``Five-branes, Membranes and Nonperturbative String Theory''}.
\newblock {\em Nucl. Phys.}, {\bf B456}:130, (1995),
  \href{http://www.arxiv.org/abs/hep-th/9507158}{hep-th/9507158}.

\bibitem{Marino:1999af}
M.~Marino, R.~Minasian, G.~W. Moore, and A.~Strominger.
\newblock {\em ``Nonlinear Instantons from Supersymmetric $p$-Branes''}.
\newblock {\em JHEP}, {\bf 01}:005, (2000),
  \href{http://www.arxiv.org/abs/hep-th/9911206}{hep-th/9911206}.

\bibitem{Witten:1985bz}
E.~Witten.
\newblock {\em ``New Issues in Manifolds of $SU(3)$ Holonomy''}.
\newblock {\em Nucl. Phys.}, {\bf B268}:79, (1986).

\bibitem{Candelas:1990pi}
P.~Candelas and X.~de~la Ossa.
\newblock {\em ``Moduli Space of Calabi-Yau Manifolds''}.
\newblock {\em Nucl. Phys.}, {\bf B355}:455, (1991).

\bibitem{Gates:1983py}
Jr. Gates, S.~J.
\newblock {\em ``Superspace Formulation of New Nonlinear Sigma Models''}.
\newblock {\em Nucl. Phys.}, {\bf B238}:349, (1984).

\bibitem{Becker:2003yv}
K.~Becker, M.~Becker, K.~Dasgupta, and P.~S. Green.
\newblock {\em ``Compactifications of Heterotic Theory on Non-K\"aehler Complex
  Manifolds. I''}.
\newblock {\em JHEP}, {\bf 04}:007, (2003),
  \href{http://www.arxiv.org/abs/hep-th/0301161}{hep-th/0301161}.

\bibitem{Kachru:2003aw}
S.~Kachru, R.~Kallosh, A.~Linde, and S.~P. Trivedi.
\newblock {\em ``De Sitter Vacua in String Theory''}.
\newblock {\em Phys. Rev.}, {\bf D68}:046005, (2003),
  \href{http://www.arxiv.org/abs/hep-th/0301240}{hep-th/0301240}.

\bibitem{Becker:2002nn}
K.~Becker, M.~Becker, M.~Haack, and J.~Louis.
\newblock {\em ``Supersymmetry Breaking and $\alpha'$-corrections to Flux
  Induced Potentials''}.
\newblock {\em JHEP}, {\bf 06}:060, (2002),
  \href{http://www.arxiv.org/abs/hep-th/0204254}{hep-th/0204254}.

\bibitem{Dundarer:1984fe}
R.~Dundarer, F.~Gursey, and C.-H. Tze.
\newblock {\em ``Generalized Vector Products, Duality and Octonionic Identities
  in $D = 8$ Geometry''}.
\newblock {\em J. Math. Phys.}, {\bf 25}:1496, (1984).

\bibitem{deWit:1984gs}
B.~de~Wit and H.~Nicolai.
\newblock {\em ``The Parallelizing $S^7$ Torsion in Gauged $N=8$
  Supergravity''}.
\newblock {\em Nucl. Phys.}, {\bf B231}:506, (1984).

\bibitem{Duff:1995an}
M.~J. Duff, R.~R. Khuri, and J.~X. Lu.
\newblock {\em ``String Solitons''}.
\newblock {\em Phys. Rept.}, {\bf 259}:213, (1995),
  \href{http://www.arxiv.org/abs/hep-th/9412184}{hep-th/9412184}.

\bibitem{Karigiannis}
Spiro Karigiannis.
\newblock {\em ``Deformations of $G_2$ and $Spin(7)$ Structures on
  Manifolds''}.
\newblock {\em Ph.D. Thesis - Harvard University}, (2003),
  \href{http://www.arxiv.org/abs/math.AG/0301218}{math.DG/0301218}.

\end{thebibliography}


\end{document}